\documentclass{article}
\usepackage{graphicx}
\usepackage{epstopdf, epsfig}
\usepackage{amssymb,amsmath}
\usepackage{mathtools}
\usepackage{tikz}
\usepackage{dsfont}
\usepackage{accents}
\usepackage{pgfplots}
\usepackage{hyperref}
\usepackage{arxiv,epsf}
\usepackage{caption, subcaption}
\usepackage[export]{adjustbox}
\usepackage{float}
\usepackage{orcidlink}
\pgfplotsset{compat=1.17}
\newlength{\dhatheight}
\usetikzlibrary{calc}
\usepackage{relsize}
\usepackage{newtxtext}
\usepackage{newtxmath}
\usepackage[authoryear,round]{natbib}
\usepackage{lineno}
\usepackage{xr-hyper}
\hypersetup{
	colorlinks = true,
	urlcolor   = blue,
	citecolor  = black,
}

\tikzset{fontscale/.style = {font=\relsize{#1}}}
\renewcommand{\exp}{\mathrm{e}}
\newcommand{\ci}{\mathrm{i}}
\newcommand{\im}{\mathrm{i}}
\newcommand{\wrt}{\mathrm{d}}
\newcommand{\rhoh}{\hat{\rho}}

\numberwithin{equation}{section}
\title{Wave propagation through periodic arrays of freely floating rectangular floes}
\author{Lloyd Dafydd \& Richard Porter}

\begin{document}
	
	\maketitle
\begin{abstract}
The two-dimensional propagation of small-amplitude waves through an infinite periodic array of freely-floating rectangular floes is considered under the assumptions of inviscid linearised wave theory. Fluid gaps between adjacent floes allow a complex interaction of the fluid with heave, surge and pitch motions. In particular, the presence of fluid resonance in the vertical channels between floes has a significant influence on wave propagation around certain critical frequencies. Bloch-Floquet theory is used and encodes the wavenumber for propagating waves into periodic boundary conditions. Solutions of the resulting boundary-value problem posed in a fundamental cell are formulated in terms of integral equations in which the three rigid body modes of the problem are treated individually. The dispersion relationship between frequency and wavenumber is expressed in terms of the vanishing of a $3 \times 3$ determinant which encodes the hydrodynamic coupling between the modes. Accurate numerical solutions are determined using Galerkin's method to approximate solutions to the integral equations. A particular focus of the paper is determining simple explicit approximations for the dispersion relation by assuming the gap between adjacent floes is small compared to the submerged draft of the floe. Approximations are shown to compare well to numerical results for a large range of gap sizes and some surprising results emerge for low-frequency wave propagation. This is particularly relevant to the application area that motivates this study: the modelling of wave propagation through broken ice.
\end{abstract}

\section{Introduction}\label{sec:introduction}

In two recent papers by the authors \citet{dafandpor24,dafandpor26}, the 
two-dimensional propagation of waves through broken floating ice 
of randomly-varying thickness was studied.
Their primary purpose was 
to consider whether randomness could provide an explanation for the 
observed frequency-dependent attenuation of ocean-ice coupled waves
within the Marginal Ice Zones: regions of fragmented sea ice on the 
boundary of the open ocean and the shore-fast sea ice. For a further 
description see the reviews of \citet{squire18}, \citet{meyetal18} and the many papers cited therein.
In the models of \citet{dafandpor24,dafandpor26} it is assumed 
that fragmented ice completely covers the surface of the ocean with no
fluid gaps between neighbouring floes which are constrained to
move vertically (in {\it heave}, in the language of ship hydrodynamics). 
This choice of model was made for simplicity and extends the widely-used 
mass-loading model (see \citet{weiandkel50}, \cite{mosetal17} 
and others) to ice of slowly-varying thickness. 
Explicit theoretical predictions of the attenuation due to wave scattering 
through random environments (often referred in other physical contexts 
as localisation) made by \citet{dafandpor24,dafandpor26} were confirmed by 
numerical simulations. Thus the dependence of attenuation with 
frequency, $f$, was shown to be proportional to $f^2$ in shallow water 
and to $f^8$ in deep water (both results applying for low frequencies).
However these results are not aligned with field measurements (see,
for example, \cite{meyetal14}, \cite{dobetal15},
\cite{rogetal16}, \cite{hosetal20}) 
which suggest the functional attenuation dependence on frequency of 
between $f^2$ and $f^4$. 
On the other hand some features of the theoretical results of 
\citet{dafandpor24,dafandpor26} compared more favourably with the data. 
Specifically, this included a quadratic relation between ice thickness 
and attenuation and the prediction of a high-frequency
`rollover effect' (see \citet{lietal17}), a persistent feature of 
datasets in many field measurements (see \citet{wadetal88}, 
\citet{dobetal15} for example) although recently theorised by 
\citet{thoetal21} to be a statistical effect produced by noisy data.

The purpose of the present paper is to consider how the introduction 
of gaps between adjacent floes in the broken ice cover might affect
wave propagation. Specifically, this paper aims to answer the question: 
is the zero-gap heave-constrained (or mass loading) model used in 
\citet{dafandpor24,dafandpor26} a reasonable approximation to the more 
physically realistic scenario of a broken ice cover which includes
small gaps ?
A positive answer to the question would confirm the validity of the 
earlier, simple, model.
A negative answer allows for the possibility that the theoretical prediction
of attenuation may be altered by the introduction of gaps between floes.

Adopting the notion of a slow spatial variation in the floe 
characteristics, a key assumption of the models used by 
\citet{dafandpor24,dafandpor26},
allows us to consider a simpler problem in which waves 
propagate locally as though they were within an infinite periodic array 
of identical floes with
equally-spaced gaps between neighbouring floes (see the arguments 
of \citet{allaire92} for example). This argument allows
us to simplify the problem, using Bloch-Floquet theory, to a single 
floe in a single fundamental cell with the extension to other periods 
in the array made through the introduction of periodic boundary 
conditions. The Bloch-Floquet wavenumber 
which is encoded into the lateral cell boundary conditions plays the role 
of the local wavenumber of propagating waves 
and its dependence upon frequency in this unforced (i.e. spectral) problem 
gives rise to a dispersion relation. Determining this dispersion relation 
and its relationship to the size and shape of the floes and the gaps between
them is our main goal which we approach both theoretically and numerically.

Within the fundamental cell the floating rectangular floe is allowed to
move freely in heave, surge and pitch motions. Each rigid-body mode of 
motion is coupled to the other two modes via the fluid and is manifested 
by (9 in total) fundamental hydrodynamic forces. The coupling is
encoded within the dynamic equations (as describe ship motions; see 
\citet{newman77} or \citet{meietal05})
of heave, surge and pitch and thus the dispersion relation is expressed
as the vanishing determinant of a $3 \times 3$ matrix. We establish
the governing system described above in Section~\ref{sec:definition} of the paper.

The generality of the current work appears to make it somewhat distinct 
from other problems in the literature.
In terms of considering two-dimensional wave propagation through floating
structures, \citet{chou98} considered waves through periodic arrays of 
flexible floating ice sheets, whilst
there are more examples of waves interacting with two-dimensional 
periodic arrays of fixed structures (for example, \citet{linton11}, 
\citet{huaandpor25}) and periodic beds (for example 
\citet{davandhea84}).
The majority of the mathematical models that approximate the
propagation of waves through regions of floating ice treat the ice 
as flexible but with zero draft; many examples are listed in the 
review of \citet{squire20}, but see \citet{pitandben26}
for a recent example. This description allows 
heave and pitch motions to be captured (in addition to any flexural waves 
through the ice/fluid coupling; only relevant for sufficiently large 
floes) but surge motions are not.
Notable exceptions
include the work of \citet{meyetal15}, \citet{sheandack91}.
Examples of work in the class of spectral problems which involve 
periodic arrays of freely-floating rigid bodies include 
\citet{diaandvid14} and \citet{carandmci13}.

The assumption of rectangular floes allow the solution to each of the
three independent components of the problem (forced heave, surge and pitch) to
be expressed in terms of integral equations which arise from matching
separation solutions at the common interface between two rectangular fluid 
regions. This is a non-trivial exercise especially in the case of forced 
pitch where both the sides and
the base of the floe are in prescribed motion. Because of this we outline
the solution procedure is detailed in Section~\ref{sec:solution-in-heave} of the paper for the
case of problem in which the floe moves in heave. Section~\ref{sec:solution-in-heave} also
includes the application of the numerical approximation method and 
asymptotic methods for small gaps for the case of heave-constrained
motion. To avoid swamping the paper with excessive complicated 
algebra, the detailed
solution for the remaining problems of surge and pitch are relegated
to the supplementary paper and include the associated numerical
methods and asymptotic approximations. In Section~\ref{sec:surge-pitch} we focus attention on the explicit 
expressions required to describe the dispersion for wave 
propagation in the presence of small gaps between neighbouring ice floes.
These asymptotic solutions are related to the full numerical solutions 
across a suite of results showing the different isolated and 
coupled effects of heave, surge and pitch. In Section~\ref{sec:fully-unconstrained} we 
present and discuss results for the fully unconstrained problem 
and we present our conclusions in Section~\ref{sec:conclusions}.

\section{Definition of the problem}\label{sec:definition}

\begin{figure}[!htbp]
		\def \globalscale {0.70000}
		\centering
		\begin{tikzpicture}[y=1cm, x=1cm, yscale=\globalscale,xscale=\globalscale, every node/.append style={scale=\globalscale}, inner sep=0pt, outer sep=0pt]
			\path[draw=black,even odd rule,rotate around={7.4:(0.0, 29.7)}] (7.4, 21.9) 
			rectangle (16.2, 16.8);

			\path[draw=black,<->] (18, 19.0) 
			-- (17.3, 24.1);

			\path[draw=black,<->] (9.1, 17.1) 
			-- (17.9, 18.2); 

			\path[draw=black, dashed] (4.5, 13.7) -- (4.5, 22.1);
			\path[draw=black, dashed] (7.4, 18.2) -- (7.4, 22.1);


			\path[draw=black, dashed] (3.9, 18.2) -- (18.1, 18.2);

			\path[draw=black,<->] (3.7, 18.2) 
			-- (3.7, 22.1);

			\path[draw=black, dashed] (3.9, 22.1) -- (18.1, 22.1);


			\path[draw=black,<->] (7.4, 17.5) 
			-- (4.5, 17.5);

			\path[draw=black,<->] (18.1, 14.5) 
			-- (4.5, 14.5);

			\path[draw=black,->] (4.5, 22.1) 
			-- (4.5, 23.2);

			\path[draw=black,->] (4.5, 22.1) 
			-- (5.4, 22.1);

			\path[draw=black] (13.0, 21.2) -- (13.1, 20.6);

			\path[draw=black] (12.7, 20.9) -- (13.4, 21.0);
			
			\path[draw=black, dashed] (12.75,19.75) -- (12.75,20.55);
			\path[draw=black, dashed] (12.35,20.15) -- (13.15,20.15);
			
			\path[draw=black,<->] (12.75,19.75) -- (13.15,19.75);
			\path[draw=black,<->] (12.35,20.15) -- (12.35,20.9);
			
			\node at (11.25,14) [fontscale=2] {$L$};
			\node at (5.95,17) [fontscale=2] {$\ell$};
			\node at (11.25,16.25) [fontscale=2] {$V^-$};
			\node at (5.95,20) [fontscale=2] {$V^+$};
			\node at (13.5,17.15) [fontscale=2] {$L-\ell$};
			\node at (18.05,21.55) [fontscale=2] {$d$};	
			\node at (5.5,22.3) [fontscale=2] {$x$};
			\node at (4.6,23.3) [fontscale=2] {$z$};
			\node at (3.3,20.15) [fontscale=2] {$\rhoh d$};
			\node at (12.05,20.525) [fontscale=2] {$\zeta$};
			\node at (12.95,19.25) [fontscale=2] {$\xi$};
			\node at (14.4,21) [fontscale=2] {$\circlearrowright\theta$};
		\end{tikzpicture}
\caption{\label{fig1} Definition of the coordinate system and variables used}
\end{figure}
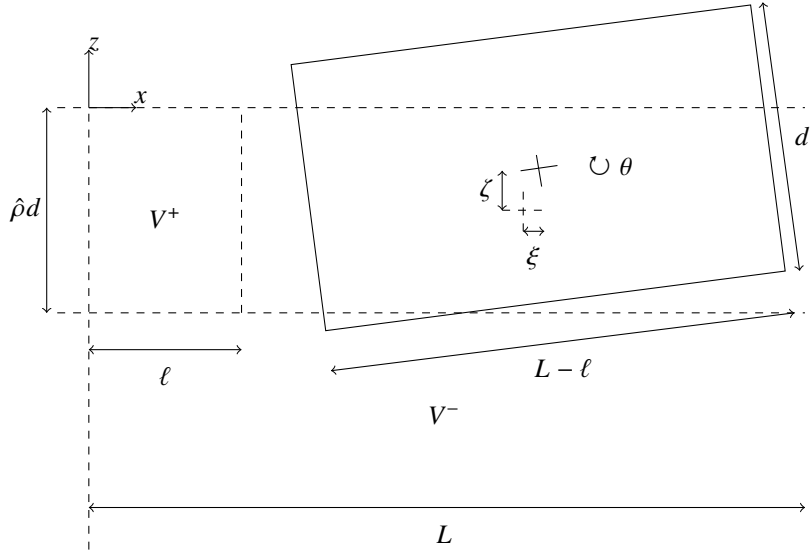

Without loss of generality we arrange the coordinate system 
such that the $x$ axis is aligned with the rest position of the 
fluid surface and the $z$ axis is directed vertically upwards in 
line with the reference position of the right side of a floe in the
periodic array (see Fig. \ref{fig1}). The width of the period is defined by $L$
and $\ell < L$ is the width of the fluid between opposing parallel sides of 
adjacent floes. Within the fundamental cell, defined as $0 < x < L$, 
the reference position describing floe/fluid boundaries include: 
the floe base along $z = -\hat{\rho}d$, $\ell < x < L$; its left side 
$x=\ell$, $-\hat{\rho} d < z < 0$ and the right-hand side of floe in the 
neighbouring period, floe is $x = 0$, $-\hat{\rho} d < z < 0$.
Here $\hat{\rho} = \rho_f/\rho < 1$ represents the ratio of floe and
fluid densities. The off-centre choice of geometry is made to simplify the
fluid domain in a fundamental cell so that it is comprised of the union 
of two rectangular domains labelled $V^\pm$ in Fig. \ref{fig1}.

Inviscid potential theory is used to
describe the assumed small-amplitude irrotational motion of an incompressible
fluid in which the fluid velocity equals $\Re \{ \nabla \phi
\exp^{-\ci \omega t} \}$ having assumed a harmonic time dependence of 
angular frequency $\omega$. The floe in $0 < x < L$ moves in response 
to the fluid with given oscillatory heave and surge displacements 
$\Re \{ \zeta \exp^{-\ci \omega t} \}$ and $\Re \{ \xi \exp^{-\ci \omega t} \}$
and a (clockwise) pitch angle $\Re \{ \theta \exp^{-\ci \omega t} \}$. 
These oscillations are made about the reference position of the floe 
stated previously.

In the fluid
\begin{equation}
	\nabla^2\phi = 0 
 \label{eqn:2.1}
\end{equation}
with $|\nabla \phi| \rightarrow 0$ as $z\rightarrow -\infty$ (it is assumed the fluid is infinitely deep). The combined linearised kinematic and dynamic free surface conditions can be expressed as
\begin{equation}
	\phi_z - K\phi = 0,\quad\text{on}\quad z=0,\;0<x<\ell 
 \label{eqn:2.2}
\end{equation}
where $K=\omega^2/g$. The linearised kinematic condition on the base of the floe is given by
\begin{equation}
	\phi_z = -\im\omega\left(\zeta - (x-x^c)\theta\right),\quad\text{on}\quad z=-\hat{\rho}d,\;\ell<x<L
 \label{eqn:2.3}
\end{equation}
where $x^c = \frac12 (\ell+L)$ denotes the centre of the floe. We also have Bloch-Floquet boundary conditions
\begin{equation}
	\phi(L,z) = \exp^{\im k L}\phi(0,z),\quad	\phi_x(L,z) = \exp^{\im k L}\phi_x(0,z)
 \label{eqn:2.4}
\end{equation}
for $z<-\rhoh d$ where $kL$ is the dimensionless Bloch wavenumber implying that the solution may be extended beyond $0 < x < L$ using $\phi(x + mL,z) = \exp^{\ci m k L} \phi(x,z)$ for $m \in \mathbb{Z}$. On the sidewalls of the floe, the linearised kinematic conditions are
\begin{equation}
	\phi_x(\ell,z) = -\im\omega(\xi + (z-z^c)\theta),\quad\text{on}\quad -\rhoh d<z<0
 \label{eqn:2.5}
\end{equation}
and
\begin{equation}
	\phi_x(0,z) = -\im\omega\exp^{\im k L}(\xi + (z-z^c)\theta),\quad\text{on}\quad -\rhoh d<z<0
 \label{eqn:2.6}
\end{equation}
where $z^c = (\frac12 - \rhoh) d$ is the vertical coordinate of the rest position of the centre of the floe, determined by Archimedes' law. The dynamic equations, measured in relation to the centre of mass, $(x^c,z^c)$, for heave, surge and pitch are (see \citet{newman77})
\begin{equation}
	-\omega^2 \rho_f \zeta (L-\ell)d = -\rho g (L-\ell)\zeta + \im\omega\rho\int_{\ell}^{L}\phi(x,-\rhoh d)\;\wrt x \label{eqn:2.7}
\end{equation}
and
\begin{equation}
	-\omega^2 \rho_f \xi (L-\ell)d  = \im\omega\rho\int_{-\rhoh d}^{0}\phi(\ell,z) - \exp^{\im k L}\phi(0,z)\;\wrt z \label{eqn:2.8}
\end{equation}
and
\begin{multline}
		-\omega^2 \frac{1}{12} \rho_f (L-\ell)d((L-\ell)^2+d^2)\theta = -\frac{1}{12}\rho g (L-\ell)^3 \theta + \frac12 \rho_f g d^2 (1-\rhoh)(L-\ell)\theta\\ - \im\omega\rho\int_{\ell}^{L}\left(x-x^c\right)\phi(x,-\rhoh d)\;\wrt x + \im\omega\rho\int_{-\rhoh d}^{0}(z-z^c)\left(\phi(\ell,z) - \exp^{\im k L}\phi(0,z)\right)\;\wrt z
 \label{eqn:2.9}
\end{multline}
where periodicity has been used to relate forces and moments
along $x=L$ to the boundary along $x=0$ lying within the fundamental period.

The first two terms on the right-hand side of (\ref{eqn:2.9})
are contributions to hydrostatic restoring forces from the
base and sidewalls of the floe respectively. These first-order
terms arise, like the first term in (\ref{eqn:2.7}), from integrating 
the hydrostatic component of the pressure over the instantaneous 
wetted surface of the floe in motion. All first-order hydrodynamic 
contributions to the pressure are integrated over the mean wetted surface.

The problem set out above is homogeneous and requires us to establish 
the relationship between the wavenumber, $k$, and the frequency 
parameter $K = \omega^2/g$. To do this we separate the contributions
from the three motions by writing
\begin{equation}
		\phi(x,z) = -\im \omega \left(\zeta\phi^{(h)}(x,z) + \xi \phi^{(s)}(x,z) + (L-\ell)\theta \phi^{(p)}(x,z)\right)
 \label{eqn:2.10}
\end{equation}
where $\phi^{(h,s,p)}$ are required to satisfy (\ref{eqn:2.1}), (\ref{eqn:2.2}) and (\ref{eqn:2.4}) along with the additional boundary conditions on the floes consisting of
\begin{equation}
		\phi_z^{(h)}(x,-\rhoh d) = 1,\quad \phi_z^{(s)}(x,-\rhoh d) = 0,\quad \phi_z^{(p)}(x,-\rhoh d) = -(x-x^c)/(L-\ell)
 \label{eqn:2.11}
\end{equation}
for $\ell<x<L$ and
\begin{equation}
		\phi_x^{(h)}(\ell,z) = 0, \quad \phi_x^{(s)}(\ell,z) = 1,\quad \phi_x^{(p)}(\ell,z) = (z-z^c)/(L-\ell)
 \label{eqn:2.12}
\end{equation}
with $\phi_x^{(h,s,p)}(0,z) = \exp^{-\im k L}\phi_x^{(h,s,p)}(\ell,z)$ for $-\rhoh d < z < 0$. 

Quantities of interest that will be sought from the solution
to the problems for $\phi^{(h,s,p)}$ are components proportional 
to the dynamic heave and surge forces and pitch moments 
on the base and the sidewalls due to fluid motion. We
define the normalised dynamic force in mode $b \in \{h,s,p\}$
due to forced motion in mode $a \in \{h,s,p\}$ by
\begin{eqnarray}
		F^{(a,b)} &=& \frac{1}{d(L-\ell)}\int_{\ell}^{L}\phi^{(a)}(x,-\rhoh d)\overline{\phi_z^{(b)}}(x,-\rhoh d)\;\wrt x
 \nonumber \\
		& &+ \frac{1}{d(L-\ell)}\int_{-\rhoh d}^{0}\phi^{(a)}(\ell,z)\overline{\phi_x^{(b)}}(\ell,z) - \phi^{(a)}(0,z)\overline{\phi_x^{(b)}}(0,z)\; \wrt z.
	\label{eqn:2.13}
\end{eqnarray}

For example, the heave induced force due to forced 
heave, surge or pitch motions is
\begin{equation}
		F^{(a,h)} = \frac{1}{d(L-\ell)}\int_{\ell}^{L}\phi^{(a)}(x,-\rhoh d) \; \wrt x
 \label{eqn:2.14}
\end{equation}
where $a \in \{ h,s,p \}$.

Thus, the strategy is to solve each hydrodynamic problem for 
$\phi^{(h,s,p)}$ in order to compute these quantities and then, 
using (\ref{eqn:2.10}) in (\ref{eqn:2.7})--(\ref{eqn:2.9}) with
the definition (\ref{eqn:2.13}), we have the system of equations
\begin{equation}
	\begin{pmatrix}
		1- Kd(\rhoh + F^{(h,h)}) & -KdF^{(s,h)} & -KdF^{(p,h)} \\
		-KdF^{(h,s)} & -Kd(\rhoh + F^{(s,s)}) & -KdF^{(p,s)} \\
		-KdF^{(h,p)} & -KdF^{(s,p)} & \frac{1}{12}-\frac12\rhoh(1-\rhoh)\hat{d}^2 - \frac{1}{12}K\rhoh d(1+\hat{d}^2)-KdF^{(p,p)} 
	\end{pmatrix}\begin{pmatrix}
		\zeta \\
		\xi \\
		(L-\ell)\theta 
	\end{pmatrix}  = 0
 \label{eqn:2.15}
\end{equation}
where $\hat{d} = d/(L-\ell)$ is the aspect ratio of the floe.
Eigensolutions of (\ref{eqn:2.15}) determine $k$ as a 
function of $K$ and the corresponding floe motions.
The matrix in (\ref{eqn:2.15}) is often presented as the sum of three
matrices representing the solid inertia, added fluid inertia (both 
proportional to $\omega^2$) and hydrostatic restoring forces. In this 
respect the absence of the usual exciting forces and radiation damping 
components is expected since the fluid motion is unforced and energy is
confined within the finite domain.
Note that we require $\hat{d} < 1/\sqrt{6\rhoh(1-\rhoh)}$ for static stability of the rectangular floe.

If we have two functions $\varphi$ and $\chi$ that are harmonic in the domain $V$, Green's identity states that
\begin{equation}
	\int_{S}\frac{\partial \varphi}{\partial n}\chi - \frac{\partial \chi}{\partial n}\varphi\;\wrt s = 0
 \label{eqn:2.16}
\end{equation}
where $s$ measures the arclength on $S$, which is a closed surface around $V$, and the normal derivative is directed outwards from $V$.
Now if, for example, we let $\varphi=\phi^{(a)}$ and $\chi = \phi^{(b)}$ in the domain occupied by the fluid and apply (\ref{eqn:2.2}), (\ref{eqn:2.4}), (\ref{eqn:2.11}) and (\ref{eqn:2.12}) we get the reciprocity relation
\begin{equation}
		F^{(a,b)} = \overline{F^{(b,a)}}.
 \label{eqn:2.17}
\end{equation}
Clearly, this implies for all $a\in {h,s,p}$ we have that $F^{(a,a)}$ is real and the matrix in (\ref{eqn:2.15}) is Hermitian (assuming $K \in \mathbb{R}$) and therefore has real eigenvalues. It is the vanishing of one of those eigenvalues that determines solutions. The corresponding eigenvectors provide information about the motion itself.

\section{Solution in heave}\label{sec:solution-in-heave}

A general expression for $\phi^{(h)}(x,z)$ in the fluid region $V^+ :=\left\{0<x<\ell,-\rhoh d<z<0\right\}$ (see Fig.~\ref{fig1}) satisfying both (\ref{eqn:2.1}) and (\ref{eqn:2.11}) can be written in terms of the series
\begin{equation}
		\phi^{(h)}(x,z) = \sum_{m=0}^{\infty}a_m Z_m(z) \cos{\alpha_m x}
 \label{eqn:3.1}
\end{equation}
where $\alpha_m = m\pi/\ell$ and
\begin{equation}
	Z_0(z) = 1+Kz,\quad Z_m(z) = \cosh{\alpha_m z} + (K/\alpha_m)\sinh{\alpha_m z}\label{eqn:3.2}
\end{equation}
are chosen to satisfy (\ref{eqn:2.2}).
Using the orthogonality result
\begin{equation}
		\frac{1}{\ell}\int_{0}^{\ell}\cos{\alpha_m x}\cos{\alpha_n x}\;\wrt x = \begin{cases}
			1,\quad m = n = 0,\\
			\frac12, \quad m=n\neq 0,\\
			0,\quad\text{otherwise}
 \label{eqn:3.3}
	\end{cases}
\end{equation}
we can express the unknown coefficients $a_m$ in (\ref{eqn:3.1}) as
\begin{equation}
		a_0 = \frac{1}{K\ell}\int_{0}^{\ell}W^{(h)}(x)\;\wrt x \label{eqn:3.4}
\end{equation}
and 
\begin{equation}
		\left(K\cosh\left(\alpha_m \rhoh d\right) - \alpha_m\sinh\left(\alpha_m \rhoh d\right)\right)a_m = \frac{2}{\ell}\int_{0}^{\ell}W^{(h)}(x)\cos{\alpha_m x}\;\wrt x
 \label{eqn:3.5}
\end{equation}
for $m \geq 1$ in terms of $W^{(h)}(x) \equiv\phi_z^{(h)}(x,-\rhoh d)$, defined over $0<x<\ell$ to be the vertical component of the fluid velocity across the interface between the two rectangular fluid sub-domains 
$V^+$ and $V^-:=\left\{z<-\rhoh d, 0<x<L\right\}$. 

In $V^-$, we require solutions to satisfy (\ref{eqn:2.1}), (\ref{eqn:2.4}) and (\ref{eqn:2.12}) and find a general solution can be written
\begin{equation}
	\phi^{(h)}(x,z) = \varphi^{(h)}(x,z) + \sum_{m=-\infty}^{\infty} b_m\exp^{\im \beta_m x}\exp^{\beta_m(z+\rhoh d)}
 \label{eqn:3.6}
\end{equation}
where $\beta_m = k + 2\pi m/L$. Here, we define $\varphi^{(h)}(x,z)$ to be a harmonic function in the region $z<-\rhoh d$, $0<x<L$ satisfying the periodicity conditions (\ref{eqn:2.4}) and decaying as $z \to -\infty$ which is used to absorb the inhomogeneous condition (\ref{eqn:2.11}) by defining
\begin{equation}
		\varphi^{(h)}_z(x,-\rhoh d) = \begin{cases}
			1, \quad \ell<x<L\\
			0, \quad 0<x<\ell
		\end{cases}
 \label{eqn:3.7}
\end{equation}
leaving the second term in (\ref{eqn:3.6}) to satisfy a homogeneous Neumann condition on $z=-\hat{\rho} d$, $\ell<x<L$. From this prescription for $\varphi^{(h)}(x,z)$ is straightforward to determine that
\begin{equation}
	\varphi^{(h)}(x,z) = \ell \sum_{m=-\infty}^{\infty} \frac{{g_m}^{(h)} \exp^{\im \beta_m x}}{\beta_m L}\exp^{\beta_m(z+\rhoh d)}
 \label{eqn:3.8}
\end{equation}
where
\begin{equation}
	g_m^{(h)} = \frac{1}{\ell}\int_{\ell}^{L}\exp^{-\im\beta_m x}\;\wrt x = \frac{1}{\im \beta_m \ell}\left(\exp^{-\im \beta_m \ell} - \exp^{-\im \beta_m L}\right).
 \label{eqn:3.9}
\end{equation}
In deriving the result above we have used the orthogonality relation
\begin{equation}
	\frac{1}{L}\int_{0}^{L}\exp^{\im \beta_m  x}\exp^{-\im \beta_n x}\;\wrt x = \begin{cases}
		1,\quad m=n\\
		0,\quad \text{otherwise}.
	\end{cases}
 \label{eqn:3.10}
\end{equation}
Returning to (\ref{eqn:3.6}), it is clear we can now also express the coefficients in the series in terms of the function $W^{(h)}(x)$ so that
\begin{equation}
	b_m = \frac{1}{\beta_m L}\int_{0}^{\ell}W^{(h)}(x)\exp^{-\im \beta_m x}\;\wrt x
 \label{eqn:3.11}
\end{equation}
for all $m$. 

The only remaining condition of the problem is to ensure the continuity of 
$\phi^{(h)}$ across the interface $z=-\rhoh d$, $0<x<\ell$. Thus, from 
(\ref{eqn:3.1}) and (\ref{eqn:3.6}) and definitions for $a_m$ and $b_m$ we arrive at 
the integral equation
\begin{equation}
	(\mathcal{T} W^{(h)})(x) =\frac{1}{\ell}\int_{0}^{\ell}W^{(h)}(x')T(x,x')\;\wrt x' = G^{(h)}(x),\quad 0<x<\ell
 \label{eqn:3.12}
\end{equation}
for the function $W^{(h)}(x)$ where
\begin{equation}
	T(x,x') = \frac{K\rhoh d-1}{K\ell} + 2 \sum_{m=1}^{\infty}\frac{\left(K\tanh\left(\alpha_m\rhoh d\right)-\alpha_m\right)\cos{\alpha_m x}\cos{\alpha_m x'}}{\alpha_m \ell\left(K- \alpha_m\tanh\left(\alpha_m\rhoh d\right)\right)} + \sum_{m=-\infty}^{\infty}\frac{\exp^{\im \beta_m (x-x')}}{\beta_m L}
 \label{eqn:3.13}
\end{equation}
defines the kernel of the integral operator $\mathcal{T}$ and where
\begin{equation}
	G^{(h)}(x) = -(1/\ell)\varphi^{(h)}(x,-\rhoh d) = - \sum_{m=-\infty}^{\infty}\frac{g_m^{(h)}\exp^{\im\beta_m x}}{\beta_m L}
\label{eqn:3.14}
\end{equation}
represents the forcing term.
We note that $T(x,x') = \overline{T(x',x)}$ is Hermitian and hence $\mathcal{T}$ is self-adjoint. The solution of (\ref{eqn:3.12}) can be used to determine $\phi$ everywhere in the fluid. However, our primary goal is to compute $F^{(h,h)}$ as defined by (\ref{eqn:2.14}) and we consider this next.

Using Green's identity with $\phi^{(h)}(x,z)$ and $\overline{\varphi^{(h)}}(x,z)$, in the domain $V^-$ with boundary $\partial V^-$, gives us
\begin{align}
		0 &= \int_{\partial V^-} \frac{\partial \phi^{(h)}}{\partial n} \overline{\varphi^{(h)}} -  \frac{\partial \overline{\varphi^{(h)}}}{\partial n} {\phi^{(h)}}\;\wrt s\\
		&= \int_{0}^{\ell} \left(\frac{\partial \phi^{(h)}}{\partial z} \overline{\varphi^{(h)}} -  \frac{\partial \overline{\varphi^{(h)}}}{\partial z} {\phi^{(h)}}\right)_{z=-\rhoh d}\;\wrt x + \int_{\ell}^{L} \left(\frac{\partial \phi^{(h)}}{\partial z} \overline{\varphi^{(h)}} -  \frac{\partial \overline{\varphi^{(h)}}}{\partial z} {\phi^{(h)}}\right)_{z=-\rhoh d}\;\wrt x\\
		&= \int_{0}^{\ell}W^{h}(x)\overline{\varphi^{(h)}}(x,-\rhoh d)\;\wrt x + \int_{\ell}^{L} \overline{\varphi^{(h)}}(x,-\rhoh d) - \phi^{(h)}(x,-\rhoh d)\;\wrt x
 \label{eqn:3.17}
\end{align}
using conditions satisfied by $\phi^{(h)}(x,z)$ and $\varphi^{(h)}(x,z)$
and so it follows that
	\begin{equation}
		\int_{\ell}^{L}\phi^{(h)}(x,-\rhoh d)\;\wrt x = \int_{0}^{\ell}W^{(h)}(x)\overline{\varphi^{(h)}}(x,-\rhoh d)\;\wrt x + \int_{\ell}^{L} \overline{\varphi^{(h)}}(x,-\rhoh d) \;\wrt x.
 \label{eqn:3.18}
	\end{equation}
Using the definition (\ref{eqn:2.13}), the relation (\ref{eqn:3.14}) and (\ref{eqn:3.18})
results in 
	\begin{equation}
		F^{(h,h)} = -\frac{\ell^2}{(L-\ell)d} A^{(h,h)} + C^{(h,h)}
\label{eqn:3.19}
	\end{equation}
where, for convenience, here we have defined
\begin{equation}
		A^{(h,h)} = \frac{1}{\ell} 
\int_{0}^{\ell}W^{(h)}(x)\overline{G^{(h)}}(x)\;\wrt x
\label{eqn:3.20}
\end{equation}
and
\begin{equation}
 C^{(h,h)} = 
\frac{\ell^2}{(L-\ell)d}\sum_{m=-\infty}^{\infty}\frac{|{g_m}^{(h)}|^2}{\beta_m L} = \frac{(L-\ell)}{2 d} \cot (kL/2)
 \label{eqn:3.21}
\end{equation}
where the cot term arises from evaluating the infinite sum using contour integration (see equation (S.A.13) in Appendix A of the supplementary material).
The manipulations above have allowed us to express $F^{(h,h)}$, as required by
(\ref{eqn:2.15}), directly in terms of the solution $W^{(h)}(x)$ of the integral equation (\ref{eqn:3.12}). This formulation, along with the explicit evaluation of the series defining $C^{(h,h)}$ in (\ref{eqn:3.21}), is crucial for approximating solutions as we shall show in Section~\ref{sec:heave-small-gaps}.
	
\subsection{Numerical solution}\label{sec:numerical-solution}

In order to approximate (\ref{eqn:3.19}) to arbitrary precision we use Galerkin's method which involves approximating the solution of the integral equation (\ref{eqn:3.12}) as an expansion
\begin{equation}
	W^{(h)}(x)\approx\sum_{n=0}^{N} A_n u_n(x)
 \label{eqn:3.22}
\end{equation}
in terms of a set of basis functions $u_n(x)$ (to be defined in a moment) and unknown coefficients $A_n$; $N$ is a truncation parameter. After substituting (\ref{eqn:3.22}) into (\ref{eqn:3.12}), multiplying by $\overline{u_m(x)}$ and integrating over $0<x<\ell$, a process which characterises Galerkin's method, one obtains the linear system of equations
\begin{equation}
	\sum_{n=0}^N A_n T_{mn} = E_{m}^{(h)}
\label{eqn:3.23}
\end{equation}
defining $A_n$ where
\begin{equation}
		T_{mn} = \langle\mathcal{T}u_n,u_m\rangle
 \label{eqn:3.24}
\end{equation}
and
\begin{equation}
		E_m^{(h)} = \langle G^{(h)},u_m \rangle
 \label{eqn:3.25}
\end{equation}
are prescribed, the notation $\langle u,v \rangle$ being employed here as shorthand for the inner product $\ell^{-1} \int_{0}^{\ell} u(x)\overline{v(x)}\;\wrt x$.
	
Using (\ref{eqn:3.22}) in (\ref{eqn:3.20}) gives
\begin{equation}
		A^{(h,h)} \approx \sum_{n=0}^{N} A_n \overline{E_n^{(h)}}.
 \label{eqn:3.26}
\end{equation}
	
If we define
\begin{equation}
		\mathcal{I}_{mn} = \left\langle \cos{\alpha_n x}, u_m \right\rangle
\qquad \mbox{and} \qquad
		\mathcal{J}_{mn} = \left\langle \exp^{\im \beta_n x}, u_m \right\rangle 
 \label{eqn:3.27}
\end{equation}
it follows from (\ref{eqn:3.24}) that
\begin{equation}
		T_{mn} =  \frac{K\rhoh d-1}{K\ell}{\mathcal{I}_{m0}\overline{\mathcal{I}_{n0}}} + 2\sum_{r=1}^{\infty}\frac{K\tanh\left(\alpha_r\rhoh d\right) - \alpha_r  }{\alpha_r \ell \left(K-\alpha_r\tanh\left(\alpha_r \rhoh d\right)\right)}\mathcal{I}_{mr}\overline{\mathcal{I}_{nr}}+ \sum_{r=-\infty}^{\infty}\frac{1}{\beta_r L}\mathcal{J}_{mr}\overline{\mathcal{J}_{nr}}
 \label{eqn:3.28}
\end{equation}
and, from (\ref{eqn:3.14}), that
\begin{equation}
		E_m^{(h)} = -\sum_{r=-\infty}^{\infty}\frac{g_r^{(h)}}{\beta_r L}{\mathcal{J}_{mr}}.
 \label{eqn:3.29}
\end{equation}
With the knowledge that $W^{(h)}(x)$ is proportional to the vertical velocity of the flow across the opening $z = -\rhoh d$, $0 < x < \ell$, $W^{(h)}(x) \sim \mathcal{C}_0 x^{-1/3}$ as $x \to 0^+$ and $W^{(h)}(x) \sim \mathcal{C}_\ell (\ell-x)^{-1/3}$ as $x \to \ell^-$ ($\mathcal{C}_0$, $\mathcal{C}_\ell$ constants) corresponding to the local flow of an ideal fluid around a right-angled corner. 
To preserve a simple form of $T_{mn}$, we choose (see \citet{evaandfer95}, for example)
\begin{equation}
		u_m(x) = \frac{\im^m(m!)\Gamma(\frac16) \ell^{2/3}}{\sqrt{2}\pi\Gamma(m+\frac13) x^{1/3}(\ell-x)^{1/3}}C_m^{1/6}\left(\frac{2x-\ell}{\ell}\right)
 \label{eqn:3.30}
\end{equation}
where $C_n^\upsilon$ represents the orthogonal ultraspherical Gegenbauer polynomials. With the help of \citet{evaandfer95} the following results can be derived:
\begin{equation}
		\mathcal{I}_{m0} = \frac{6}{2^{1/6}\Gamma(\frac16)}\delta_{m0};
 \label{eqn:3.31}
\end{equation}
\begin{equation}
		\mathcal{I}_{mn}= (-\im)^m\cos\left(\left(m+n\right)\frac{\pi}{2}\right) \frac{J_{m+1/6}(n\pi/2)}{{\left(n\pi/2\right)}^{1/6}}
 \label{eqn:3.32}
\end{equation}
	for $n \geq 1$; and
\begin{equation}
	{\mathcal{J}_{mn}} = 
		(-1)^{m} \exp^{\im\beta_n \ell/2}{J_{m+1/6}(\beta_n \ell/2)}/{{\left(\beta_n \ell/2\right)}^{1/6}}.
 \label{eqn:3.33}
\end{equation}

\subsection{Heave-constrained motion and small gaps}\label{sec:heave-small-gaps}

If we assume the floe motion is constrained to move in heave only then
from (\ref{eqn:2.15}) the condition for propagating waves is reduced to 
\begin{equation}
  Kd (\hat{\rho} + F^{(h,h)}) = 1.
 \label{eqn:3.34}
\end{equation}
Let us consider $\epsilon = \ell/d \ll 1$. Further, it will be assumed 
that $|K\rhoh d-1| \gg \epsilon$; this avoids fluid resonance in narrow 
fluid channels which is associated with the condition $K \hat{\rho} d = 1$. 
It follows from (\ref{eqn:3.13}), after using (S.B.3), that
\begin{equation}
 T(x,x') \approx \frac1\epsilon \widetilde{T}(x,x'), \qquad
 \mbox{where} ~~\widetilde{T}(x,x') = \frac{K\rhoh d-1}{Kd} + \frac12\epsilon\cot\left(\frac{kL}{2}\right)
 \label{eqn:3.35}
\end{equation}
and tildes are used henceforth to denote terms of $O(1)$ with respect
to $\epsilon$. 
The second term is included since it can also contribute at $O(1)$ in the wavenumber intervals $kL \lesssim O(\epsilon)$ and $2\pi - kL \lesssim O(\epsilon)$ where $\cot(kL/2) = O(1/\epsilon)$. This turns out to have a significant effect on some of the asymptotic results we encounter whose dispersion curves cross, or lie within, the small wavenumber intervals close to $kL = 0$ and $kL=2\pi$. In other cases it has less of an effect. For example, the result (\ref{eqn:3.44}) below does not change when the second term in (\ref{eqn:3.35}) is removed.

We also find that, since $0 < x < \ell$,
\begin{equation}
 G^{(h)}(x) \approx \frac1\epsilon \widetilde{G}^{(h)},
 \label{eqn:3.36}
\end{equation}
where, from (\ref{eqn:3.9}), (\ref{eqn:3.14}), we have
\begin{equation}
 \widetilde{G}^{(h)} = \im \frac{L}{d} \left(1-\exp^{-\im k L} \right)
 \sum_{m=-\infty}^\infty \frac{1}{(\beta_m L)^2}
 = \im \frac{L}{4 d} \left(1-\exp^{-\im k L}\right) \text{cosec}^2\left(\frac{kL}{2}\right)
 \label{eqn:3.37}
\end{equation}
after using \citet{graandryz81} \S1.422(4).

It follows upon using (\ref{eqn:3.35}) and (\ref{eqn:3.36}) in (\ref{eqn:3.12}) that
\begin{equation}
 \frac{1}{\ell} \int_0^\ell W^{(h)}(x') \; \wrt x' \approx \widetilde{G}^{(h)}/\widetilde{T}
 \label{eqn:3.38}
\end{equation}
and then, together with (\ref{eqn:3.19}), that
\begin{equation}
 F^{(h,h)} \approx  C^{(h,h)} -\epsilon \frac{d}{L-\ell}
 |\widetilde{G}^{(h)}|^2/\widetilde{T}.
 \label{eqn:3.39}
\end{equation}
Full simplification of the result above results in
\begin{equation}
 F^{(h,h)} \approx F^{(h,h)}_0 + \epsilon F_1^{(h,h)}
 \label{eqn:3.40}
\end{equation}
where
\begin{equation}
 F_0^{(h,h)} = \frac{L}{2 d} \cot (kL/2), \qquad
 F_1^{(h,h)} = -\frac12\cot (kL/2) - 
 \frac{{KL\;}\text{cosec}^2\left(\frac{kL}{2}\right)}{4({K\rhoh d -1}) + 2 \epsilon Kd\cot(kL/2)} .
 \label{eqn:3.41}
\end{equation}
Using this in (\ref{eqn:3.34}) gives the approximate dispersion relation
for heave-constrained motion with small gaps as
\begin{equation}
		\frac{1-K\rhoh d}{Kd} \approx \frac{L\cot\left(kL/2\right)}{2d} - \epsilon\left(\frac{1}{2}\cot\left(\frac{kL}{2}\right) + \frac{{KL\;}\text{cosec}^2\left(\frac{kL}{2}\right)}{4({K\rhoh d -1}) + 2 \epsilon Kd\cot(kL/2)}\right)
 \label{eqn:3.42}.
\end{equation}
We can solve this for $k$ in terms of $K$ by letting 
$\cot (kL/2) \equiv X \approx X_0 + \epsilon X_1 +\ldots$ in (\ref{eqn:3.42}) 
and matching term-by-term to get
\begin{equation}
 X_0 = \kappa \equiv \frac{2(1-K \hat{\rho} d)}{KL},
  \qquad
 X_1 = -(d/L)/\kappa 
 \label{eqn:3.43}
\end{equation}
and so
\begin{equation}
 k \approx \frac{2}{L}\cot^{-1}\left( \kappa - \epsilon \frac{d}{L \kappa} \right).
 \label{eqn:3.44}
\end{equation}
This result ensures that $kL$ is $2 \pi$-periodic as
required by the problem whereas further Taylor expansion of the 
inverse cot function destroys this property.

We note that when $\epsilon \to 0$, (\ref{eqn:3.44}) reduces to
\begin{equation}
 k \approx \frac{2}{L}\cot^{-1}\left(\frac{2\left(1-K\rhoh d\right)}{KL}\right)
 \label{eqn:3.45}
\end{equation}
which is therefore the exact dispersion relation for heave-constrained motion in the case that there are no fluid gaps. In the case the underlying wavelength is long compared to the width of the floes ($KL \ll 1$) (\ref{eqn:3.45}) reduces further to
\begin{equation}
	k \approx \frac{K}{\left(1-K\rhoh d\right)},
 \label{eqn:3.46}
\end{equation}
now independent of $L$, and coincides with the mass loading model of \citet{weiandkel50} and the dispersion relation used in \citet{dafandpor26} when $d$ is taken to be a slowly-varying function of $x$.

\subsection{Results for heave constrained motion}\label{sec:heave-results}

\begin{figure}[!htbp]
		\centering
		\begin{subfigure}[t]{0.03\textwidth}
			\text{(a)}
		\end{subfigure}
		\begin{subfigure}[t]{0.45\textwidth}        \centering
		\includegraphics[width=\linewidth,trim={0cm 0cm 0cm 0cm}]{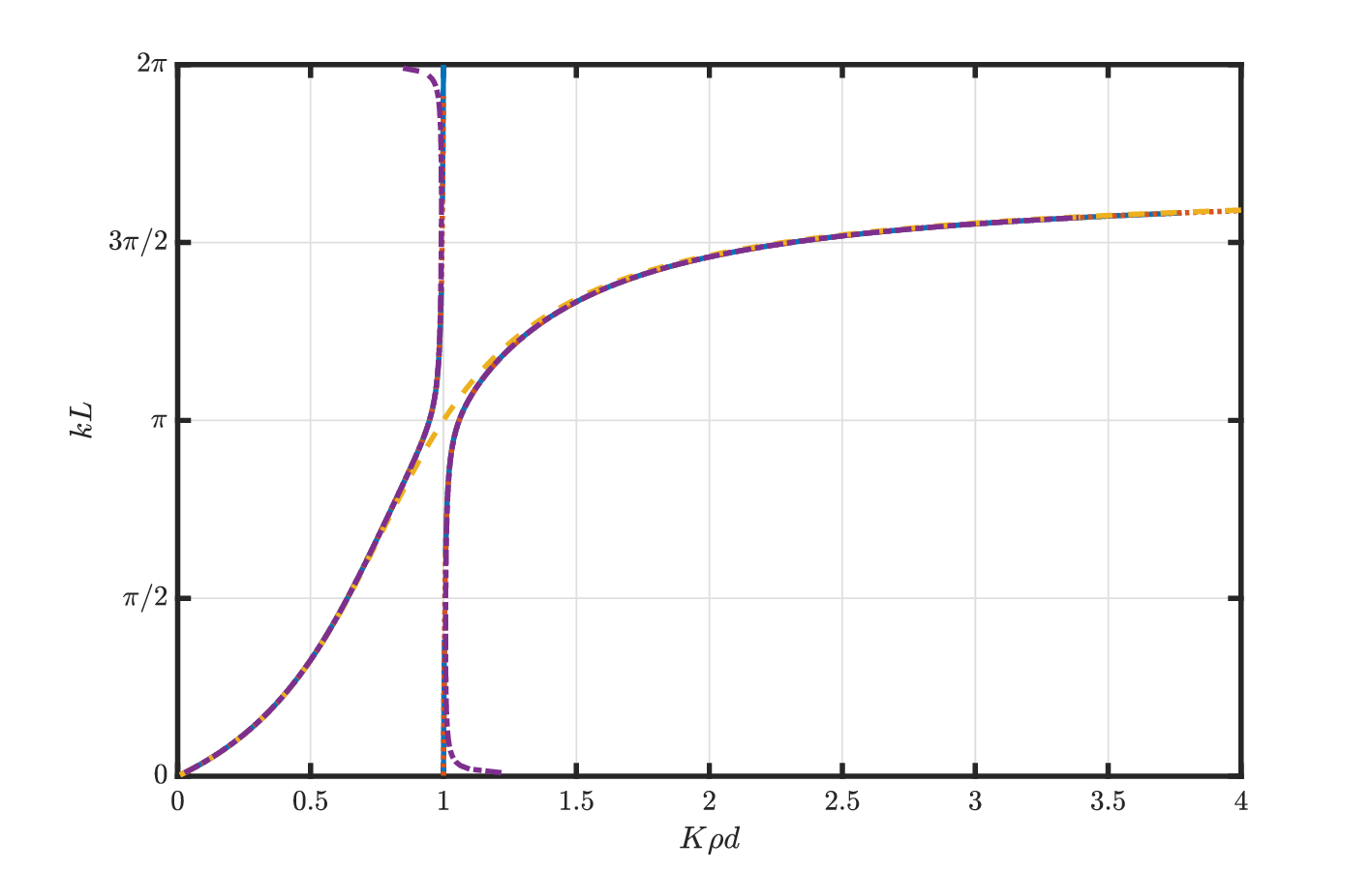}
		\end{subfigure}\hfill
		\begin{subfigure}[t]{0.03\textwidth}
			\text{(b)}
		\end{subfigure}
		\begin{subfigure}[t]{0.45\textwidth}        \centering
		\includegraphics[width=\linewidth,trim={0cm 0cm 0cm 0cm}]{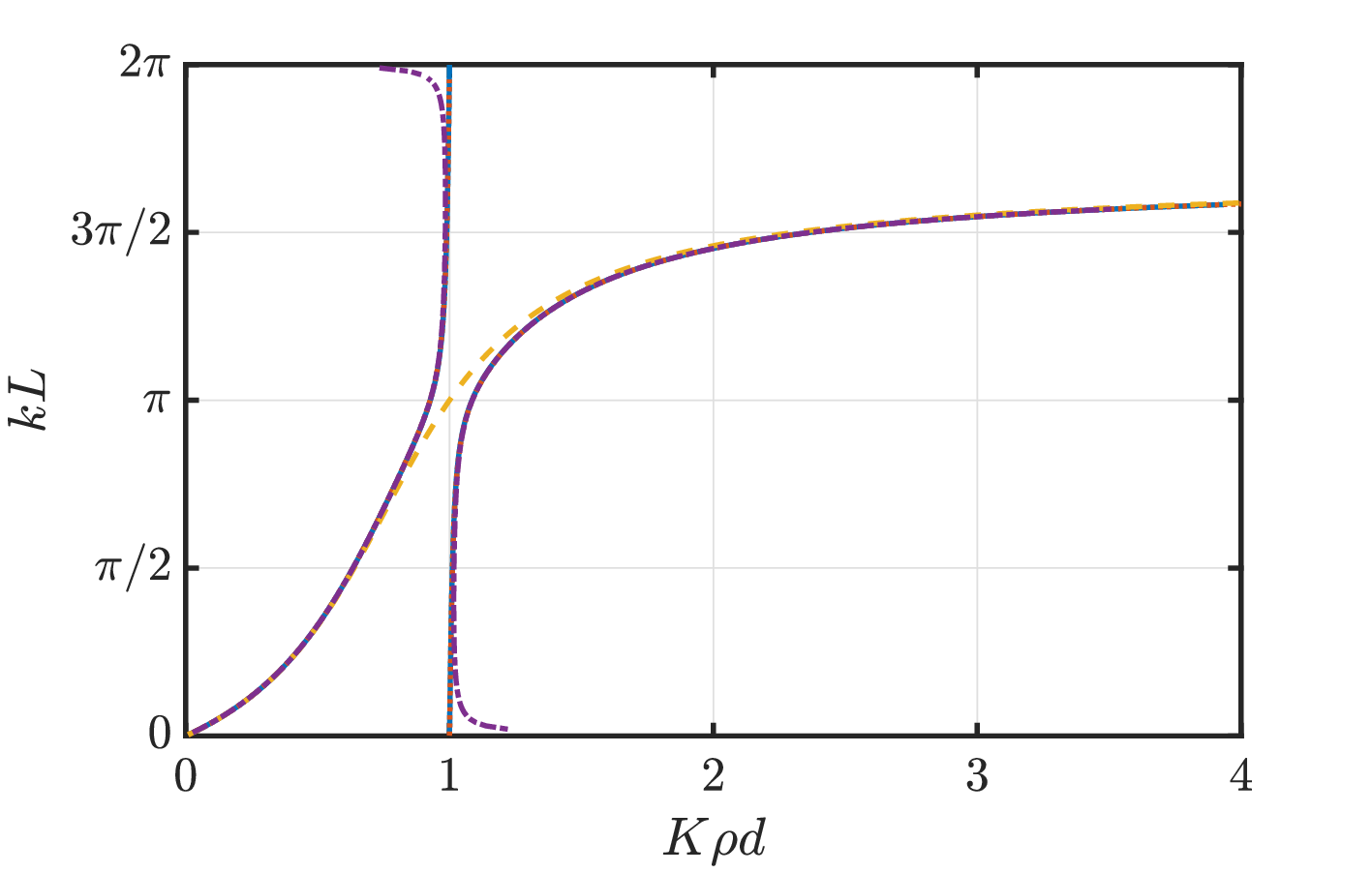}
		\end{subfigure}                               
		\begin{subfigure}[t]{0.03\textwidth}
			\text{(c)}
		\end{subfigure}
		\begin{subfigure}[t]{0.45\textwidth}        \centering
		\includegraphics[width=\linewidth,trim={0cm 0cm 0cm 0cm}]{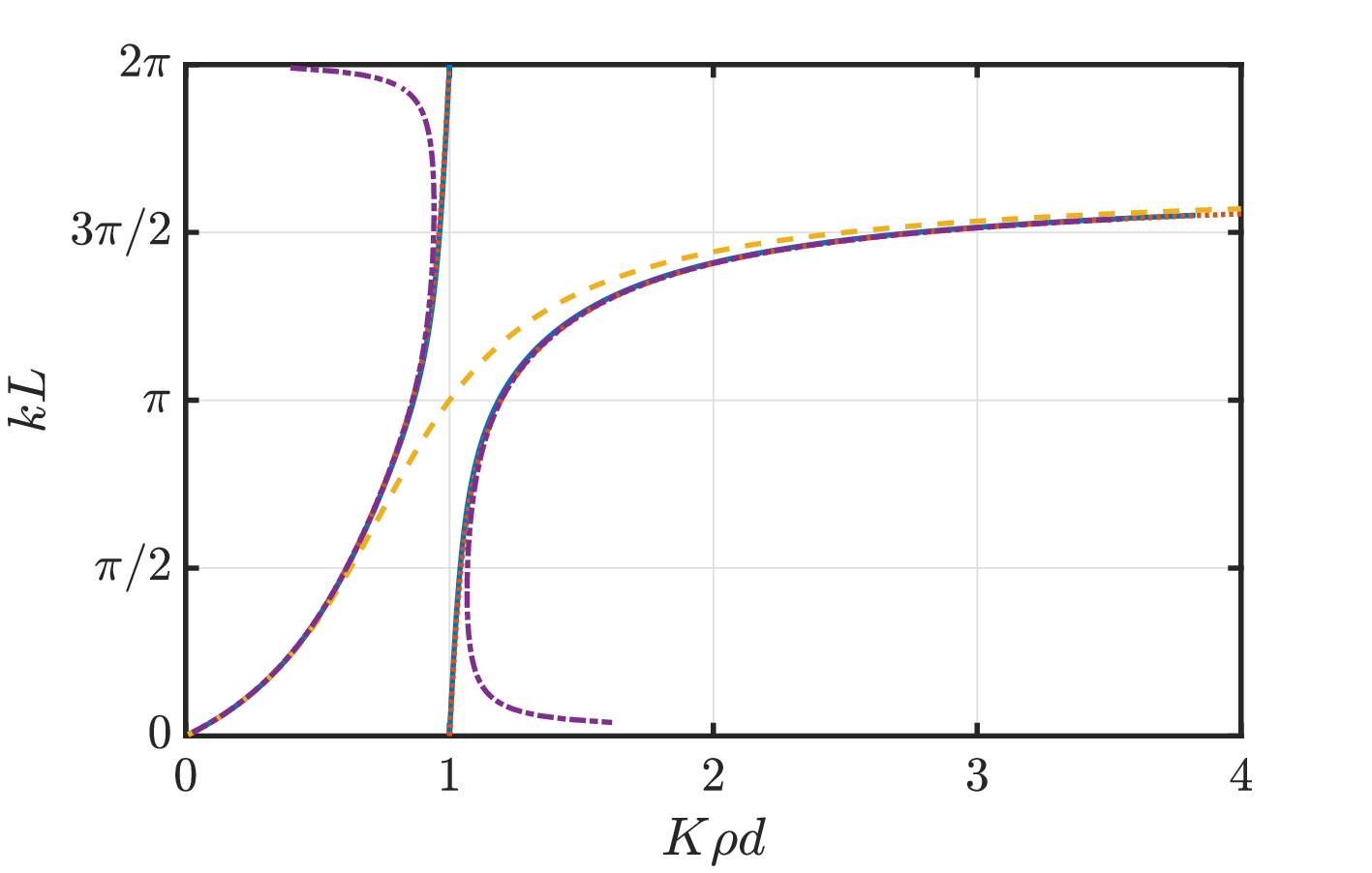}
		\end{subfigure}\hfill
		\begin{subfigure}[t]{0.03\textwidth}
			\text{(d)}
		\end{subfigure}
		\begin{subfigure}[t]{0.45\textwidth}        \centering
		\includegraphics[width=\linewidth,trim={0cm 0cm 0cm 0cm}]{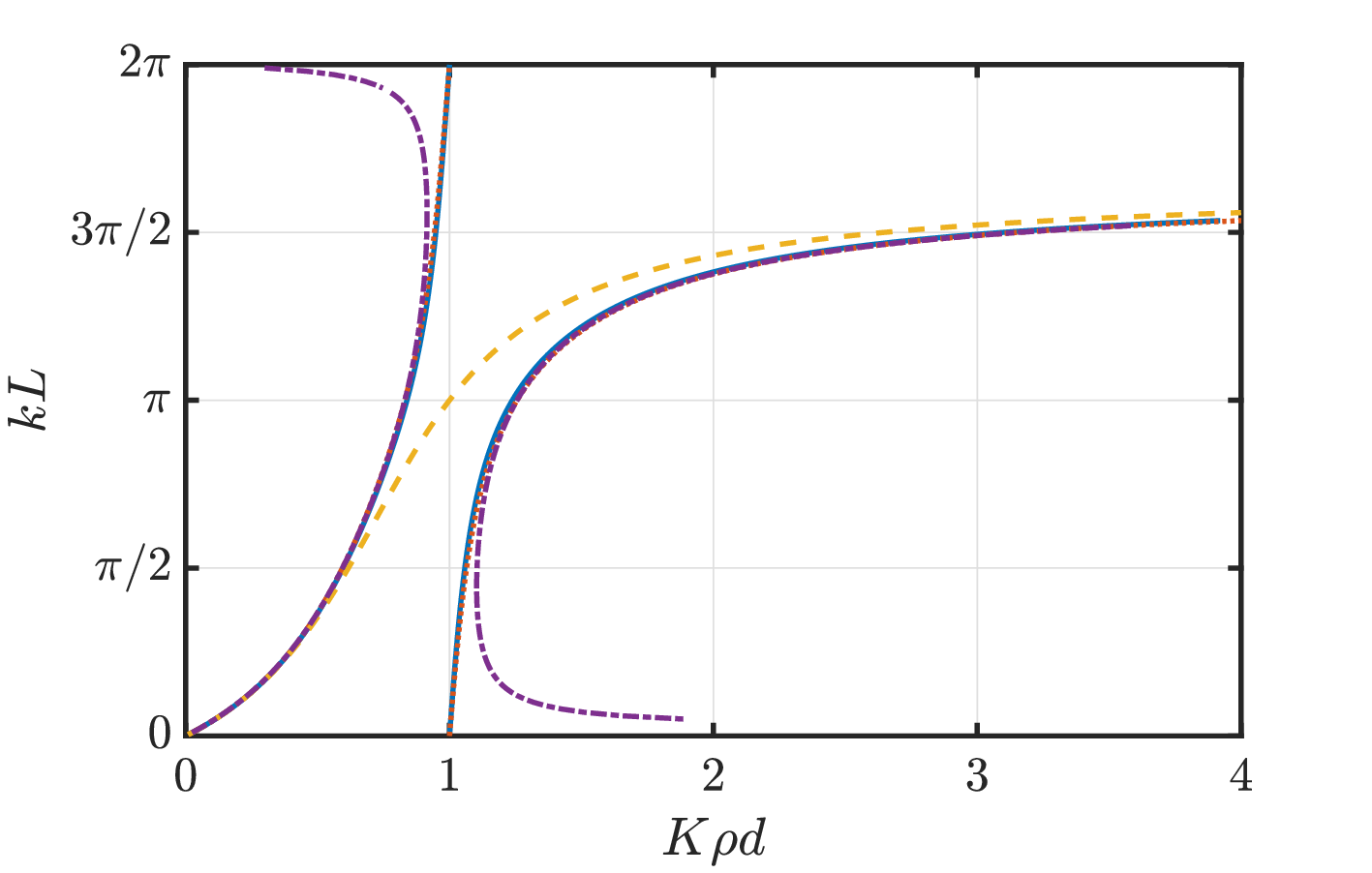}
		\end{subfigure}
		\caption{Dispersion curves ($kL$ versus $K \rhoh d$) for heave-constrained motions in the case $\rhoh=0.9$, $L/d = 1 + \epsilon$ (square ice floes). Exact results (blue solid curves), explicit $O(\epsilon)$ small-gap approximation (orange, dotted),leading order small-gap approximation (yellow, dashed) and small gap root-finding (purple, dot-dashed) for gap sizes: (a) $\epsilon = 0.01$, (b) $\epsilon = 0.02$, (c) $\epsilon = 0.08$, (d) $\epsilon = 0.12$.}
 \label{fig2}
	\end{figure}

In Fig.~\ref{fig2} we provide a comparison between the full numerical 
results based on the methods described in Section~\ref{sec:numerical-solution} with the various
asymptotic results derived in Section~\ref{sec:heave-small-gaps}. We have chosen
to focus on the case where $(L-\ell) = d$, and $\hat{\rho} = 0.9$
corresponding to a square ice block. The effect of the increasing 
the size of the fluid gap between adjacent ice blocks is
considered by setting $\epsilon = \ell/d = 0.01,~0.02,~0.08,~0.12$.
Each subplot shows: the full numerical solution of the dispersion relation;
the leading-order approximation, as given by (\ref{eqn:3.45}), which is 
independent of $\epsilon$; the explicit $O(\epsilon)$ approximation
given by (\ref{eqn:3.44}); and a solution which simply root-finds solutions to (\ref{eqn:3.42}). 
The explicit $O(\epsilon)$ solution provides a very good fit to the full numerical results for all
gap sizes considered and across all frequencies, including close to resonance ($K\rhoh d =1$),
to the extent that the curves are near indistinguishable in Figure \ref{fig2}. Pure root-finding
introduces discrepancies near $kL, 2\pi-kL \approx 0$ due to inconsistencies introduced when expanding
to $O(\epsilon)$; it is important to note that these are not associated with the retention of the
second term in (\ref{eqn:3.35}). These issues will not arise in surge and pitch constrained results
where asymptotic expansions will be limited to leading order, although these effects will reemerge
when other modes are coupled to heave.

\section{Solutions in surge and pitch}\label{sec:surge-pitch}

We can continue with a similar analysis for the other two modes of motion,
surge and pitch, and hence determine forces which are associated with 
these motions.
Since the details are long and algebraically complicated, we have relegated 
them to a supplementary material document. In summary, for each problem we can 
formulate the solution in terms of an 
integral equation for an unknown vertical velocity across the 
horizontal gap aligned with the lower edge of the floe. It turns out
that this integral equation always has the same kernel, $T(x,x')$, as
defined by (\ref{eqn:3.13}) and therefore the main difference is that the 
right-hand side
forcing term differs according to the problem being solved. This 
can be traced back to the way each problem is set up: a 
bespoke particular solution is needed to account for the 
inhomogeneous boundary conditions associated with the imposed 
velocity on the sides of the rectangular floe in different
modes of motion.
The other main difference is that the formulae that define the hydrodynamic forces
vary since these also depend upon the particular solution. The 
details do become complicated particularly when we consider pitch
motions and pitch-induced hydrodynamic forces since the imposed
conditions on the boundary have components in both domains 
$V^+$ and $V^-$.

The method for producing numerical approximations is also similar to
what has been described previously and details are given
in the supplementary material. 

In the remaining part of the paper we concentrate on the explicit
approximations derived for small gaps assuming frequencies that are
away from resonance, which is the main range of interest
for the application area in mind: the low frequency wave propagation 
through broken ice.

\subsection{Surge constrained motions for small gaps}\label{sec:surge-small-gaps}

From the solution of the surge problem 
we are able to determine the leading order approximation to the 
surge-induced surge component of the force in terms of the
dimensionless gap size $\epsilon$ as
\begin{equation}
    F^{(s,s)} \approx \frac{1}{\epsilon}\frac{4\rhoh^2}{KL}\sin^2\left(\frac{kL}{2}\right)\left(\frac13K\rhoh d-1 - \frac{\left( K\rhoh d -2\right)^2}{4 (K\rhoh d-1) + 2 \epsilon Kd \cot(kL/2)}\right).\label{eqn:4.1}
\end{equation}
(see equation (S.1.20) in the supplementary material).
We may immediately consider surge-constrained motions
by extracting from (\ref{eqn:2.15}) the condition 
\begin{equation}
	F^{(s,s)} + \rhoh = 0.
 \label{eqn:4.2}
\end{equation}
As this is clearly a transcendental equation, we may use a root-finding algorithm to find solutions--noting that we do not have the aforementioned issues as this form is only leading order.

When $kL \gg O(\epsilon)$ or $2\pi-kL \gg O(\epsilon)$, we note that $\cot(kL/2) = O(1)$ and we can neglect the $O(1)$ term $\cot(kL/2)$ in (\ref{eqn:4.1}), simplifying it to

\begin{equation}
	F^{(s,s)} \approx \frac{1}{\epsilon}\frac{\rhoh^3d(4-K\rhoh d)}{3L(1-K\rhoh d)}\sin^2\left(\frac{kL}{2}\right)
 \label{eqn:4.3}
\end{equation}
resulting in the dispersion relation
\begin{equation}
	\sin^2\left(\frac{kL}{2}\right) \approx \epsilon\frac{3L(1-K\rhoh d)}{\rhoh^2 d(K\rhoh d-4)}
 \label{eqn:4.4}
\end{equation}
which is $2 \pi$-periodic in $kL$.
This reduction is useful since it shows from (\ref{eqn:4.4}) that solutions are predicted to exist for 
$1 \leq K \hat{\rho} d \leq 4$ and that there is symmetry in $kL$ about $\pi$ in the surge-only case.

\begin{figure}[!htbp]
		\centering
		\begin{subfigure}[t]{0.03\textwidth}
			\text{(a)}
		\end{subfigure}
		\begin{subfigure}[t]{0.45\textwidth}        \centering
		\includegraphics[width=\linewidth,trim={0cm 0cm 0cm 0cm}]{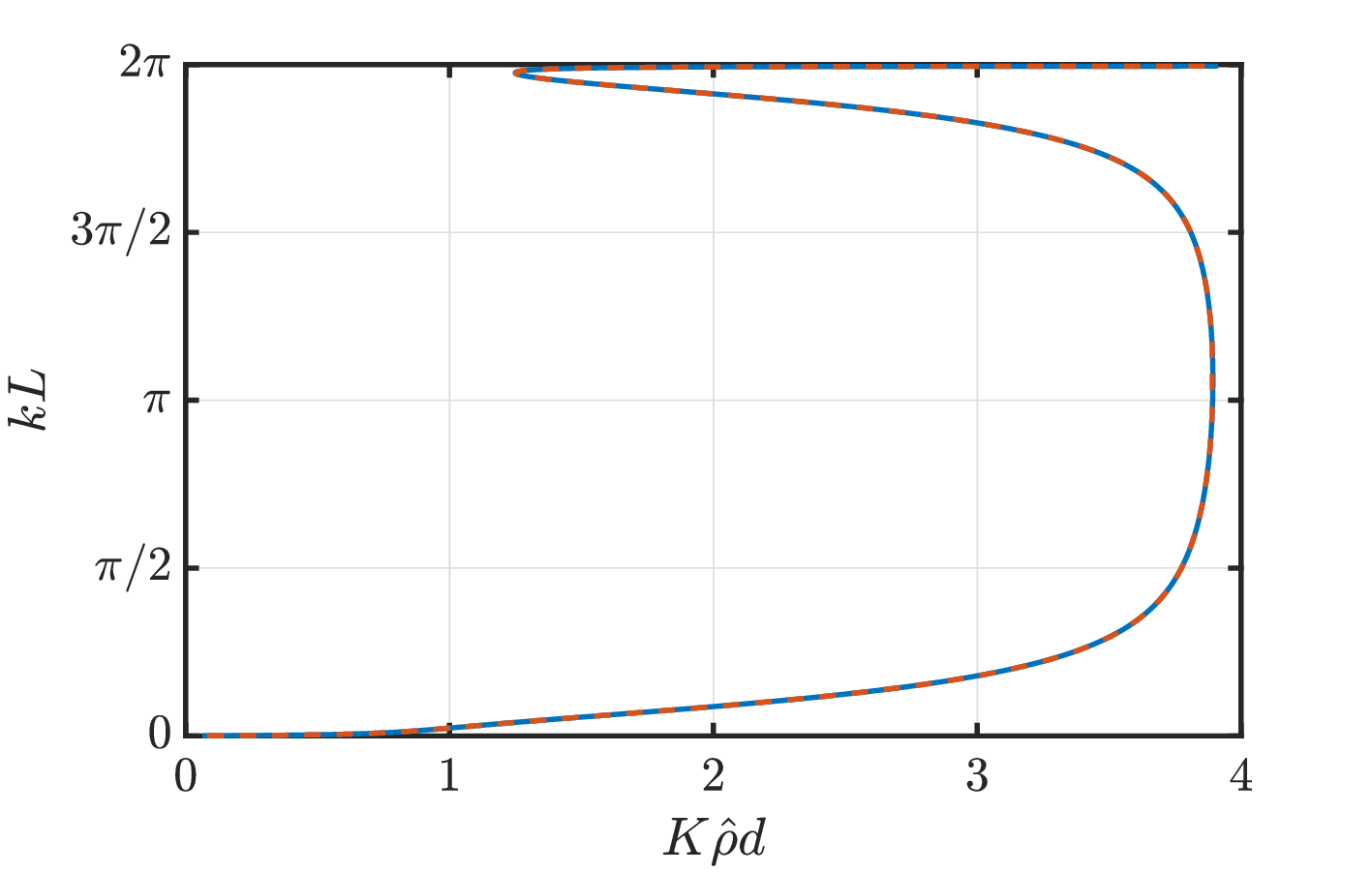}
		\end{subfigure}\hfill
		\begin{subfigure}[t]{0.03\textwidth}
			\text{(b)}
		\end{subfigure}
		\begin{subfigure}[t]{0.45\textwidth}        \centering
		\includegraphics[width=\linewidth,trim={0cm 0cm 0cm 0cm}]{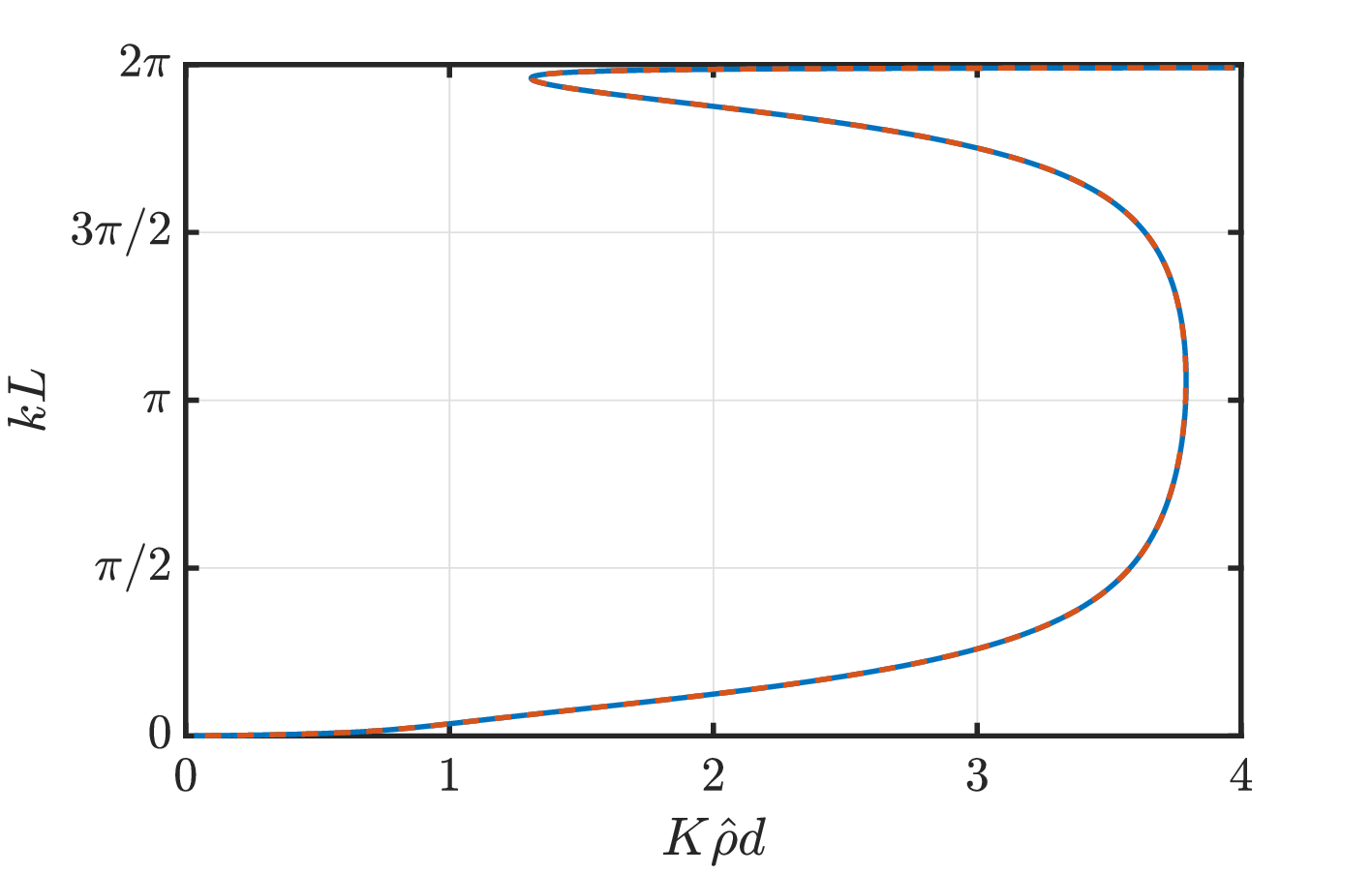}
		\end{subfigure}
		\begin{subfigure}[t]{0.03\textwidth}
			\text{(c)}
		\end{subfigure}
		\begin{subfigure}[t]{0.45\textwidth}        \centering
		\includegraphics[width=\linewidth,trim={0cm 0cm 0cm 0cm}]{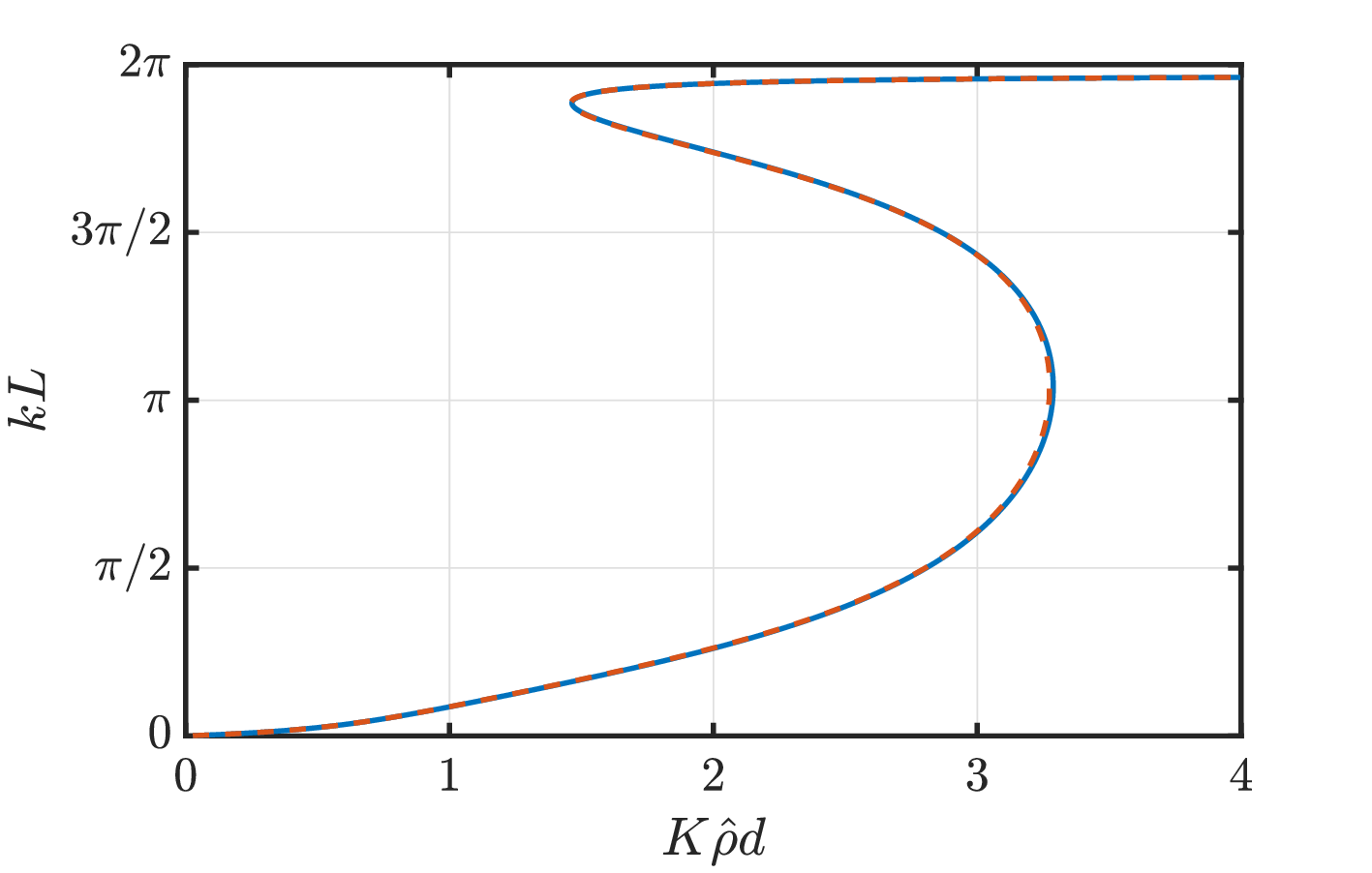}
		\end{subfigure}\hfill
		\begin{subfigure}[t]{0.03\textwidth}
			\text{(d)}
		\end{subfigure}
		\begin{subfigure}[t]{0.45\textwidth}        \centering
		\includegraphics[width=\linewidth,trim={0cm 0cm 0cm 0cm}]{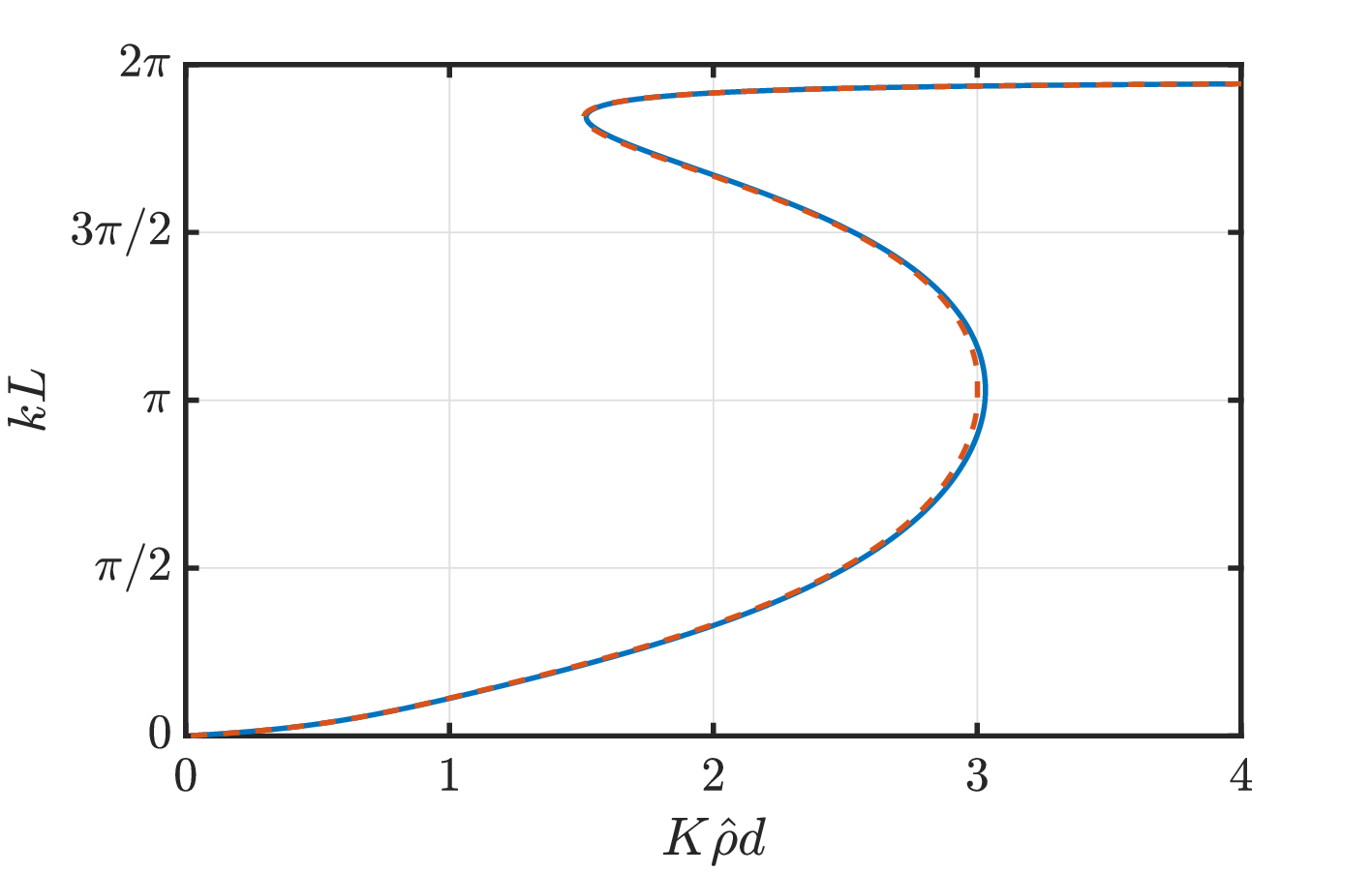}
		\end{subfigure}
		\caption{Dispersion curves ($kL$ versus $K \rhoh d$) for surge-constrained motions in the case $\rhoh=0.9$, $L/d = 1 + \epsilon$ (square ice floes). Exact results (blue solid curves) and small-gap approximations (orange, dashed) for gap sizes: (a) $\epsilon = 0.01$, (b) $\epsilon = 0.02$, (c) $\epsilon = 0.08$, (d) $\epsilon = 0.12$.}
\label{fig3}
	\end{figure}

The solution of (\ref{eqn:4.2}) using (\ref{eqn:4.1}) via root-finding results in the curves labelled in Fig.~\ref{fig3} as asymptotic results. They are shown to agree very well with the accurate full numerical results across all frequencies and improve as $\epsilon$ is reduced in size, as expected. As predicted, the dispersion curves fall mainly within the region $1 \leq K\hat{\rho} d \leq 4$. However, note the vital role that the extra term $\epsilon \cot(kL/2)$ in (\ref{eqn:4.1}) has in allowing the asymptotically-derived results to follow the exact results beyond this range as $kL$ tends to $0$ and $2\pi$ as the aforementioned symmetry is broken. Since we are motivated by results at low frequencies, we note the importance of retaining the $\epsilon \cot(kL/2)$ term in (\ref{eqn:4.1}).

\subsection{Heave-surge coupled motions for small gaps}\label{sec:heave-surge-coupled}

An asymptotic analysis for small gaps, away from resonance
(see supplementary material), leads to the result
\begin{equation}
	F^{(h,s)} = i\hat{\rho} \frac{\frac{1}{2}K\hat{\rho} d - 1}{K\hat{\rho} d - 1 + \frac{1}{2}\epsilon Kd\cot\left(\frac{kL}{2}\right)} = -F^{(s,h)} \label{eqn:4.5}
\end{equation}
confirming the reciprocal relation (\ref{eqn:2.17}). Again, we note the inclusion of a term proportional to $\epsilon \cot(kL/2)$ originating from (\ref{eqn:3.35}) which is $O(1)$ in the vicinity of $kL=0$ and $kL=2\pi$.

If the motion is constrained so that floes move freely in heave and surge
without pitching then the last element of (\ref{eqn:2.15}) is suppressed so that the motion is defined by the $2 \times 2$ upper-left system of equations and the dispersion relation is defined by
\begin{equation}
	Kd\left(\rhoh + F^{(s,s)}\right)\left(Kd\left(\rhoh + F^{(h,h)}\right)-1\right) - \left|KdF^{(h,s)}\right|^2 = 0.
 \label{eqn:4.6}
\end{equation}
We can make accurate numerical calculations of solutions of this 
dispersion relation using the numerical methods that define the
hydrodynamic forces.

Assuming small gaps and using the expressions (\ref{eqn:3.41}), (\ref{eqn:3.42})
(\ref{eqn:4.1}) and (\ref{eqn:4.5}), we see 
that all forces are $O(1)$ except for $F^{(s,s)} = O(1/\epsilon)$ and thus 
it is clear that, at leading order, (\ref{eqn:4.6}) reduces to 
\begin{equation}
	F_0^{(h,h)} \approx \frac{1-K\rhoh d}{Kd}
 \label{eqn:4.7}
\end{equation}
which is the leading-order approximation corresponding to heave-constrained motion. 
That is, solutions of (\ref{eqn:4.7}) are just
\begin{equation}
	k \approx \frac{2}{L} \cot^{-1} \left(\frac{2(1-K\rhoh d)}{KL} \right)
 \label{eqn:4.8}
\end{equation}
the same as (\ref{eqn:3.44}).
This is perhaps not surprising since as the gaps shrink, the 
effect of surge is reduced. Since we only have leading order estimates for the forces $F^{(s,h)}$ and $F^{(s,s)}$ (unlike for $F^{(h,h)}$) we are unable to perform a consistent expansion of (\ref{eqn:4.6}) in higher order powers of $\epsilon$ that allows us to deduce an improved explicit estimate of $k$ dependent upon $\epsilon$ as we did in the heave- and surge-constrained cases.

\begin{figure}[!htbp]
	\centering
		\begin{subfigure}[t]{0.03\textwidth}
			\text{(a)}
		\end{subfigure}
		\begin{subfigure}[t]{0.45\textwidth}        \centering
			\includegraphics[width=\linewidth,trim={0cm 0cm 0cm 0cm}]{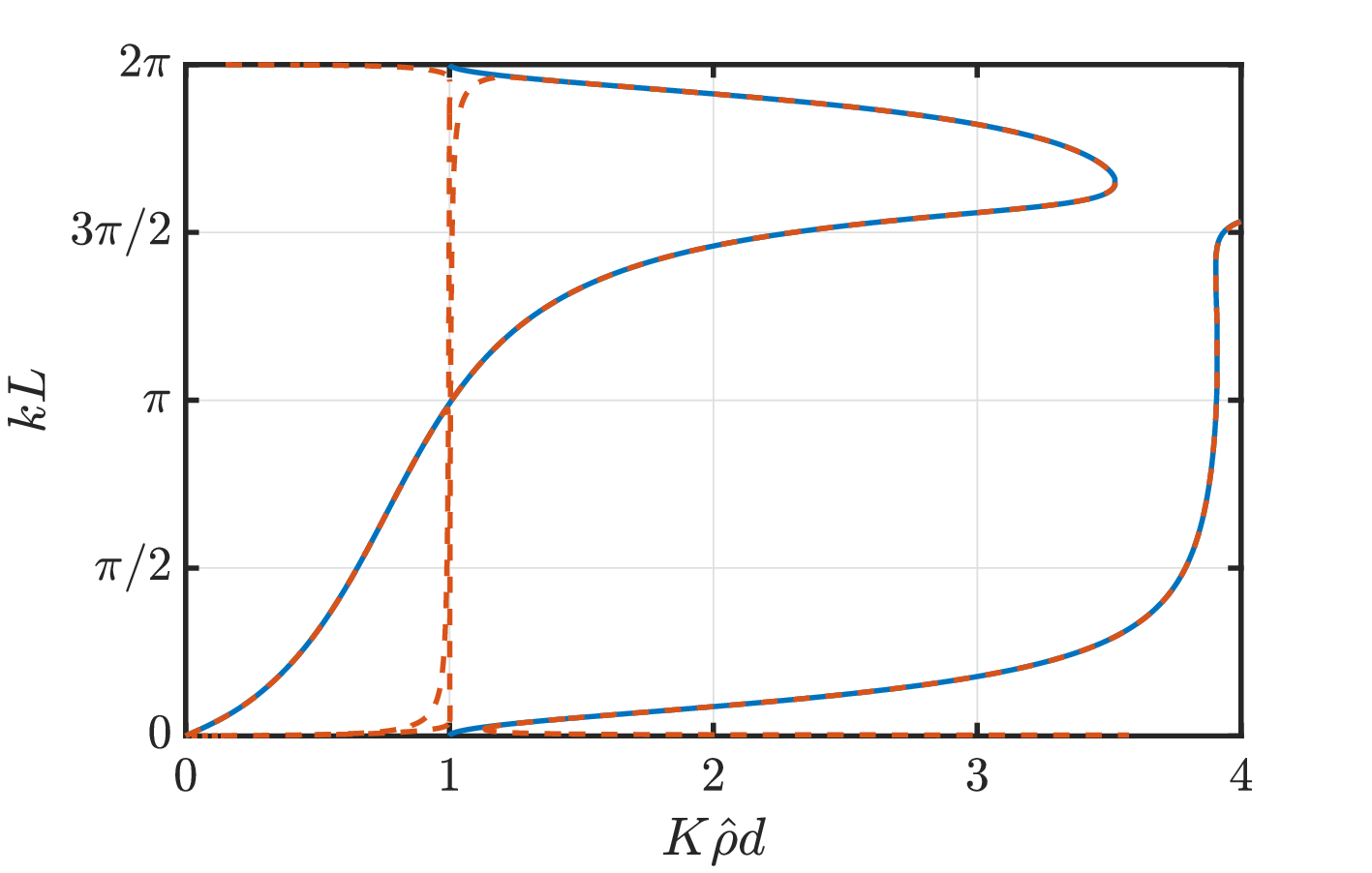}
		\end{subfigure}\hfill
		\begin{subfigure}[t]{0.03\textwidth}
			\text{(b)}
		\end{subfigure}
		\begin{subfigure}[t]{0.45\textwidth}        \centering
			\includegraphics[width=\linewidth,trim={0cm 0cm 0cm 0cm}]{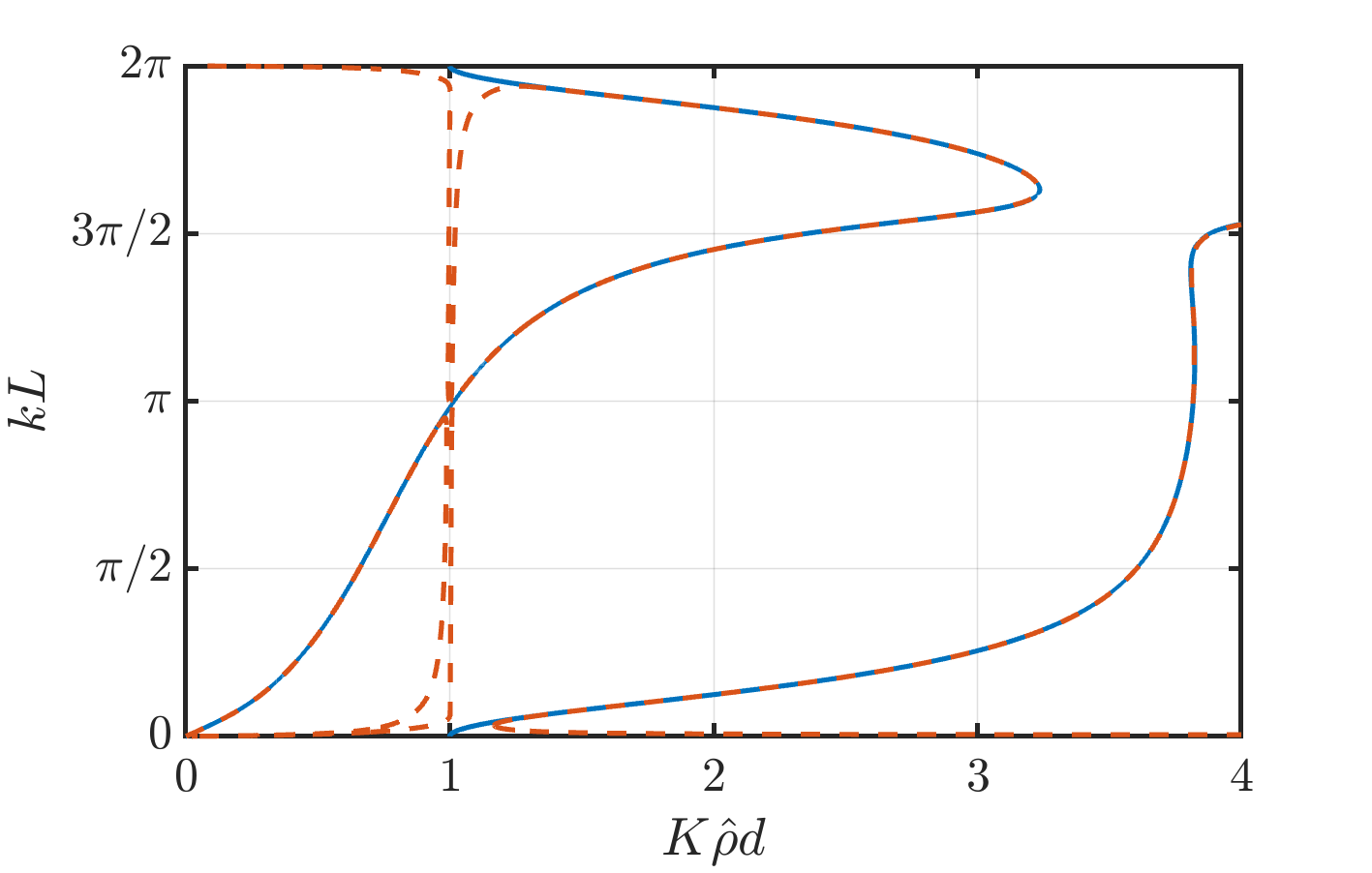}
		\end{subfigure}
		\begin{subfigure}[t]{0.03\textwidth}
			\text{(c)}
		\end{subfigure}
		\begin{subfigure}[t]{0.45\textwidth}        \centering
			\includegraphics[width=\linewidth,trim={0cm 0cm 0cm 0cm}]{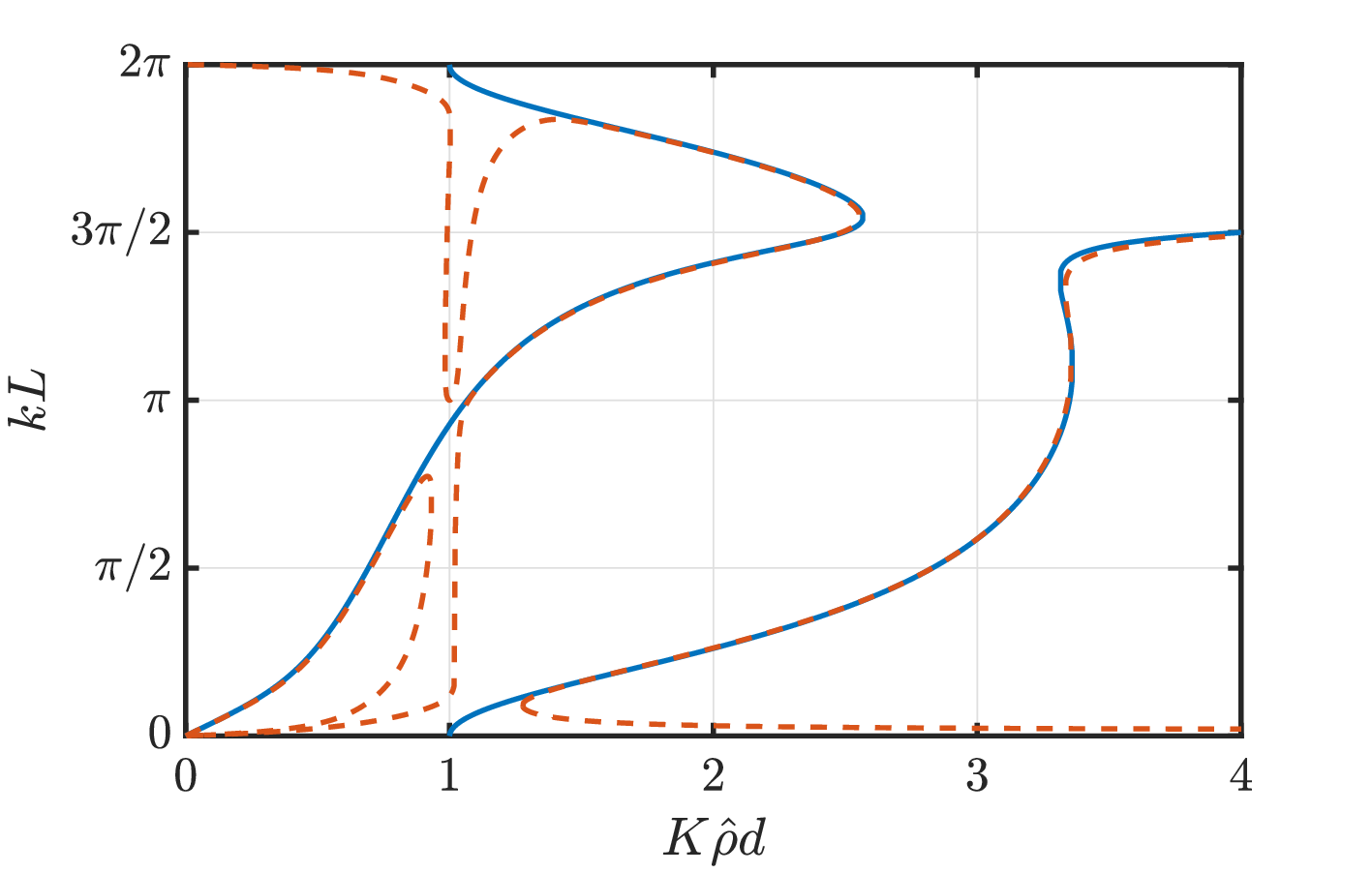}
		\end{subfigure}\hfill
		\begin{subfigure}[t]{0.03\textwidth}
			\text{(d)}
		\end{subfigure}
		\begin{subfigure}[t]{0.45\textwidth}        \centering
			\includegraphics[width=\linewidth,trim={0cm 0cm 0cm 0cm}]{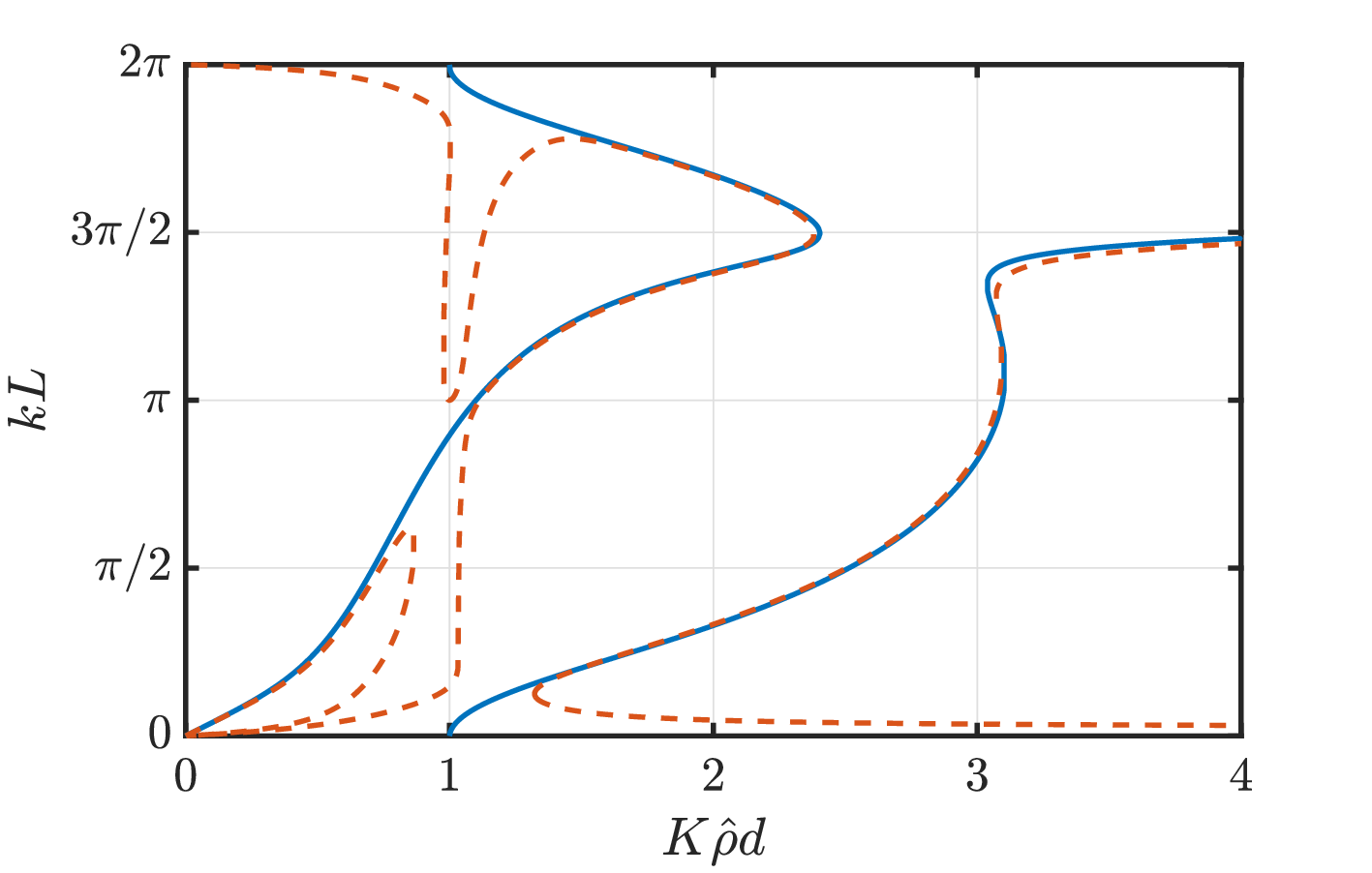}
		\end{subfigure}
		\caption{Dispersion curves ($kL$ versus $K \rhoh d$) for heave and surge-constrained motions in the case $\rhoh=0.9$, $L/d = 1 + \epsilon$ (square ice floes). Exact results (blue solid curves) and small-gap approximations (orange, dashed) for gap sizes: (a) $\epsilon = 0.01$, (b) $\epsilon = 0.02$, (c) $\epsilon = 0.08$, (d) $\epsilon = 0.12$.}
 \label{fig4}
	\end{figure}

Instead, Fig.~\ref{fig4} shows a comparison of the full numerical 
solutions of the coupled heave-surge motions alongside 
curves calculated from using the small gap approximations 
(\ref{eqn:3.40}), (\ref{eqn:3.41}), (\ref{eqn:4.1}) and 
(\ref{eqn:4.5}) in (\ref{eqn:4.6}) with increasing values of $\epsilon$ for the same
square floe configuration used in the previous results.

In the low frequency region $K \rhoh d < 1$ exact results correspond 
closely to the heave-constrained approximation consistent with the
result (\ref{eqn:4.8}). The influence of gap resonance and standing wave solutions on the results using the leading-order asymptotic theory are significant as $\epsilon$ moves away from zero. Additional solution branches, not associated with exact solutions, emerge from the narrow strips in the dispersion diagram where the analysis is invalid, exaggerated, perhaps, because of the $2\pi$-periodicity constraint on solutions.

\subsection{Pitch constrained motions for small gaps}\label{sec:pitch-small-gaps}

We consider floe motions which are constrained to the pitching mode only
(i.e. the heave/surge is suppressed). According to (\ref{eqn:2.15}) the condition
for such modes to exist is
\begin{equation}
 \frac{1}{12}-\frac12\rhoh(1-\rhoh)\hat{d}^2 - \frac{1}{12} K \rhoh d (1+\hat{d}^2)-Kd F^{(p,p)} = 0.
 \label{eqn:4.9}
\end{equation}
For small gaps we can use the approximation
\begin{equation}
    F^{(p,p)} = \frac{1}{\epsilon} \hat{\rho}^2 \left(\frac{d}{L}\right)^3 \left[\frac{\hat{\rho}^3}{5} - \frac{\hat{\rho}^2}{2} + \frac{\hat{\rho}}{3} - \frac{(1-\hat{\rho})^2}{Kd} - \frac{\left(\frac{\hat{\rho}^2}{3} - \frac{\hat{\rho}}{2} + \frac{1-\hat{\rho}}{Kd}\right)^2}{\frac{K\hat{\rho} d - 1}{Kd} + \frac{1}{2}\epsilon\cot\left(\frac{kL}{2}\right)}\right] \sin^2\left(\frac{kL}{2}\right)\label{eqn:4.10}
\end{equation}
which is derived in the supplementary material. For values of $kL$ not close to $0$ or $2\pi$ we may neglect the $\epsilon \cot(kL/2)$ term, simplifying the above to
\begin{equation}
 F^{(p,p)} 
 \approx \frac{1}{\epsilon} \frac{d^3}{L^3} \rhoh^2 
 \sin^2(kL/2) \left( \frac{Kd}{1-K \rhoh d} \left(
 \frac{\rhoh^2}{3}
 -\frac{\rhoh}{2}
 +\frac{(1-\rhoh)}{Kd}
 \right)^2
 +
 \frac{\rhoh^3}{5}
 -\frac{\rhoh^2}{2}
 +\frac{\rhoh}{3}
 -\frac{(1-\rhoh)^2}{Kd}
 \right).
 \label{eqn:4.11}
\end{equation}

\begin{figure}[!htbp]
		\centering
		\begin{subfigure}[t]{0.03\textwidth}
			\text{(a)}
		\end{subfigure}
		\begin{subfigure}[t]{0.45\textwidth}        \centering
			\includegraphics[width=\linewidth, valign=t,trim={0cm 0cm 0cm 0cm}]{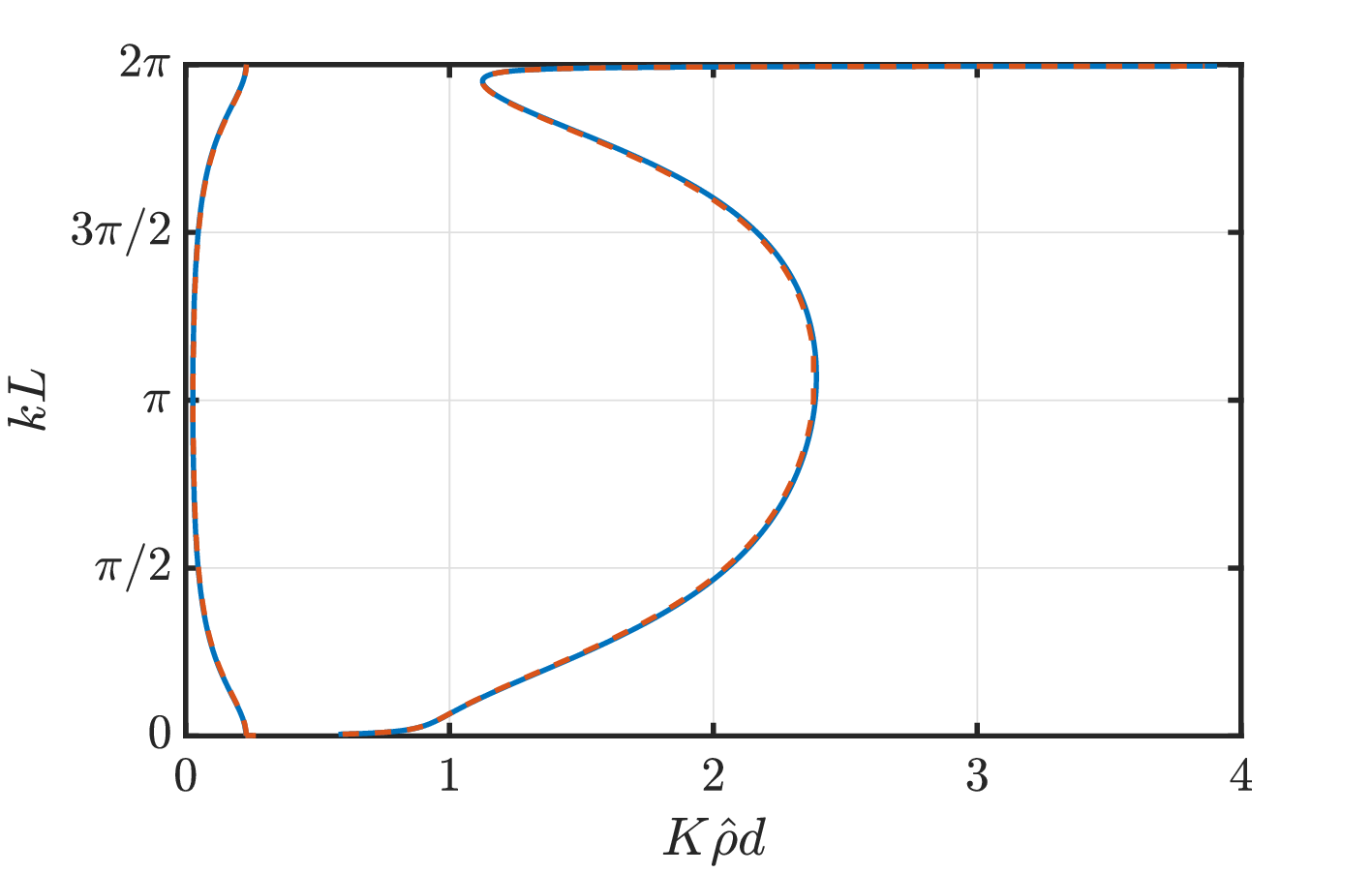}
		\end{subfigure}\hfill
		\begin{subfigure}[t]{0.03\textwidth}
			\text{(b)}
		\end{subfigure}
		\begin{subfigure}[t]{0.45\textwidth}        \centering
			\includegraphics[width=\linewidth, valign=t,trim={0cm 0cm 0cm 0cm}]{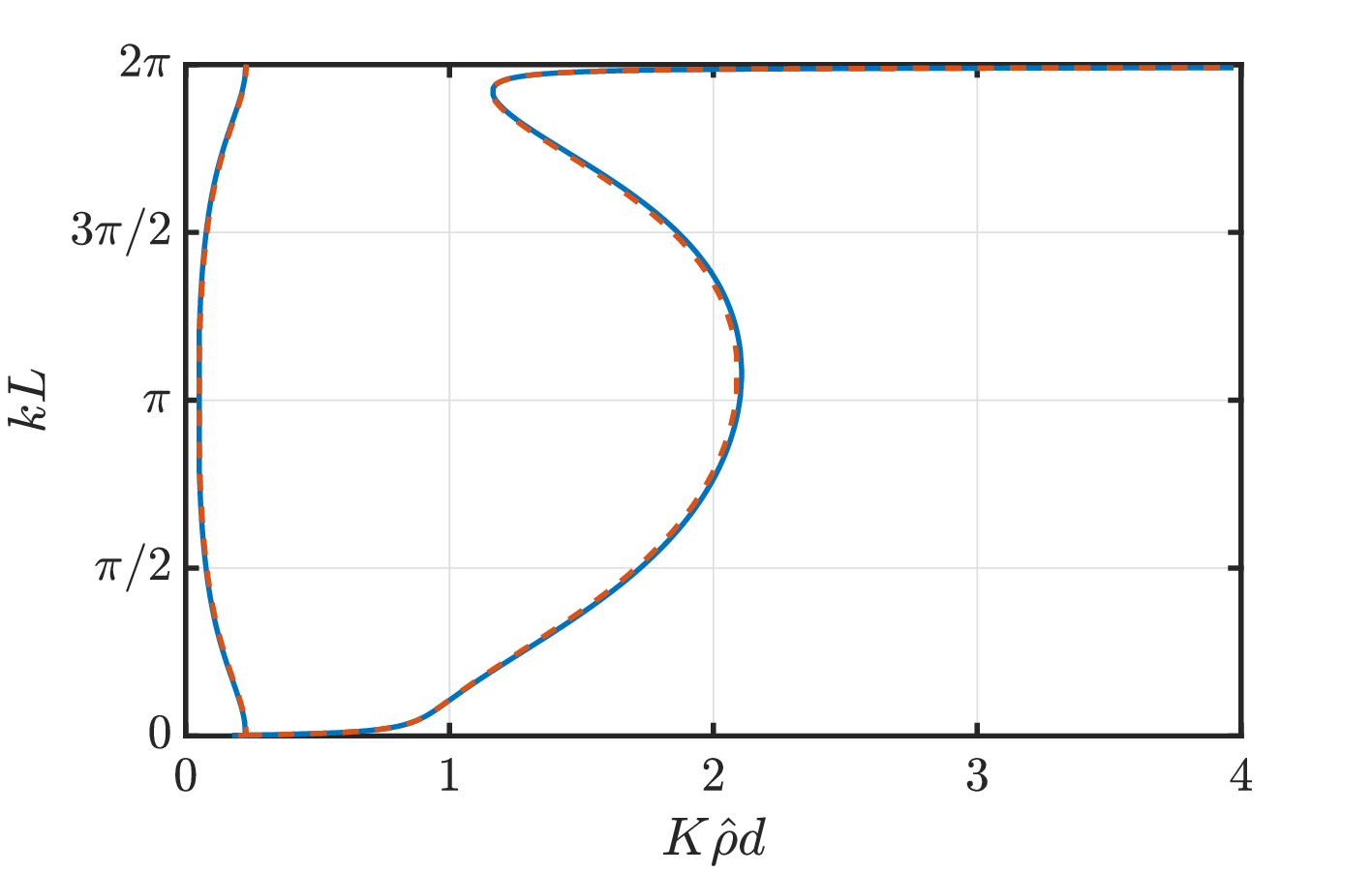}
		\end{subfigure}
		\begin{subfigure}[t]{0.03\textwidth}
			\text{(c)}
		\end{subfigure}
		\begin{subfigure}[t]{0.45\textwidth}        \centering
			\includegraphics[width=\linewidth, valign=t,trim={0cm 0cm 0cm 0cm}]{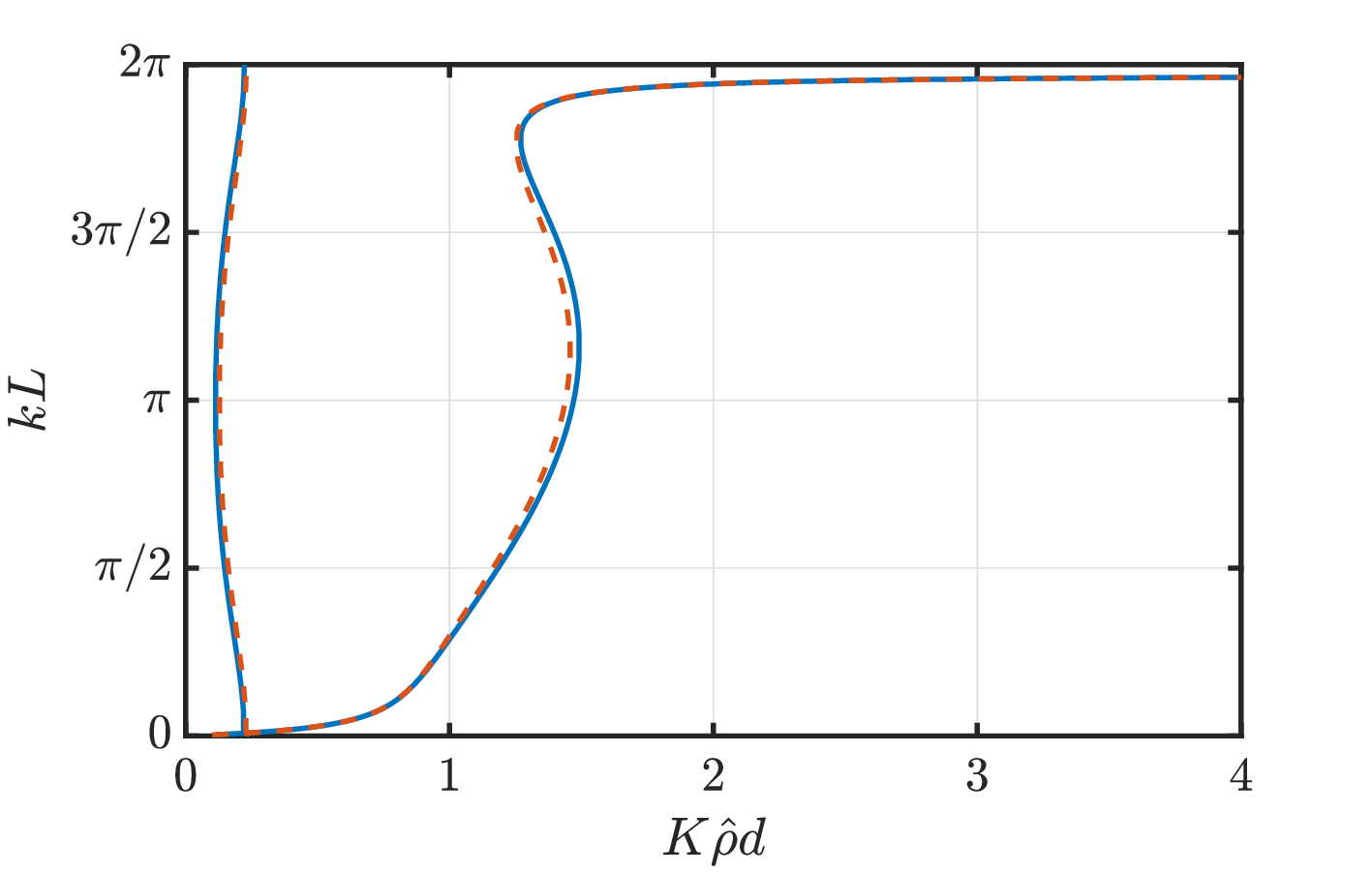}
		\end{subfigure}\hfill
		\begin{subfigure}[t]{0.03\textwidth}
			\text{(d)}
		\end{subfigure}
		\begin{subfigure}[t]{0.45\textwidth}        \centering
			\includegraphics[width=\linewidth, valign=t,trim={0cm 0cm 0cm 0cm}]{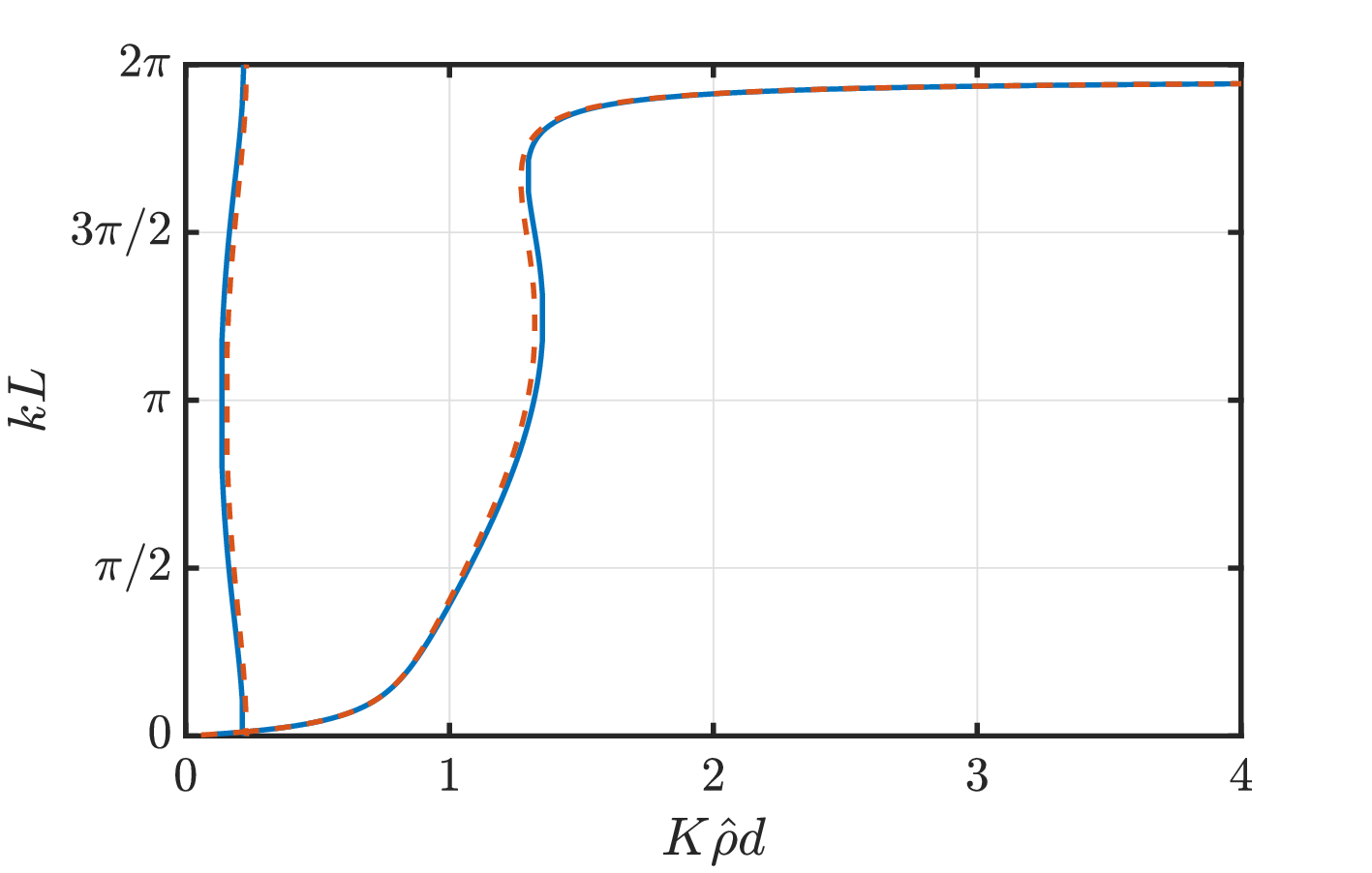}
		\end{subfigure}
		\caption{ \label{fig5} Dispersion curves ($kL$ versus $K \rhoh d$) for pitch-constrained motions in the case $\rhoh=0.9$, $L/d = 1 + \epsilon$ (square ice floes). Exact results (blue solid curves) and small-gap approximations (orange, dashed) for gap sizes: (a) $\epsilon = 0.01$, (b) $\epsilon = 0.02$, (c) $\epsilon = 0.08$, (d) $\epsilon = 0.12$.}
	\end{figure}

\begin{figure}[!htbp]
		\centering
		\begin{subfigure}[t]{0.03\textwidth}
			\text{(a)}
		\end{subfigure}
		\begin{subfigure}[t]{0.45\textwidth}        \centering
			\includegraphics[width=\linewidth, valign=t,trim={0cm 0cm 0cm 0cm}]{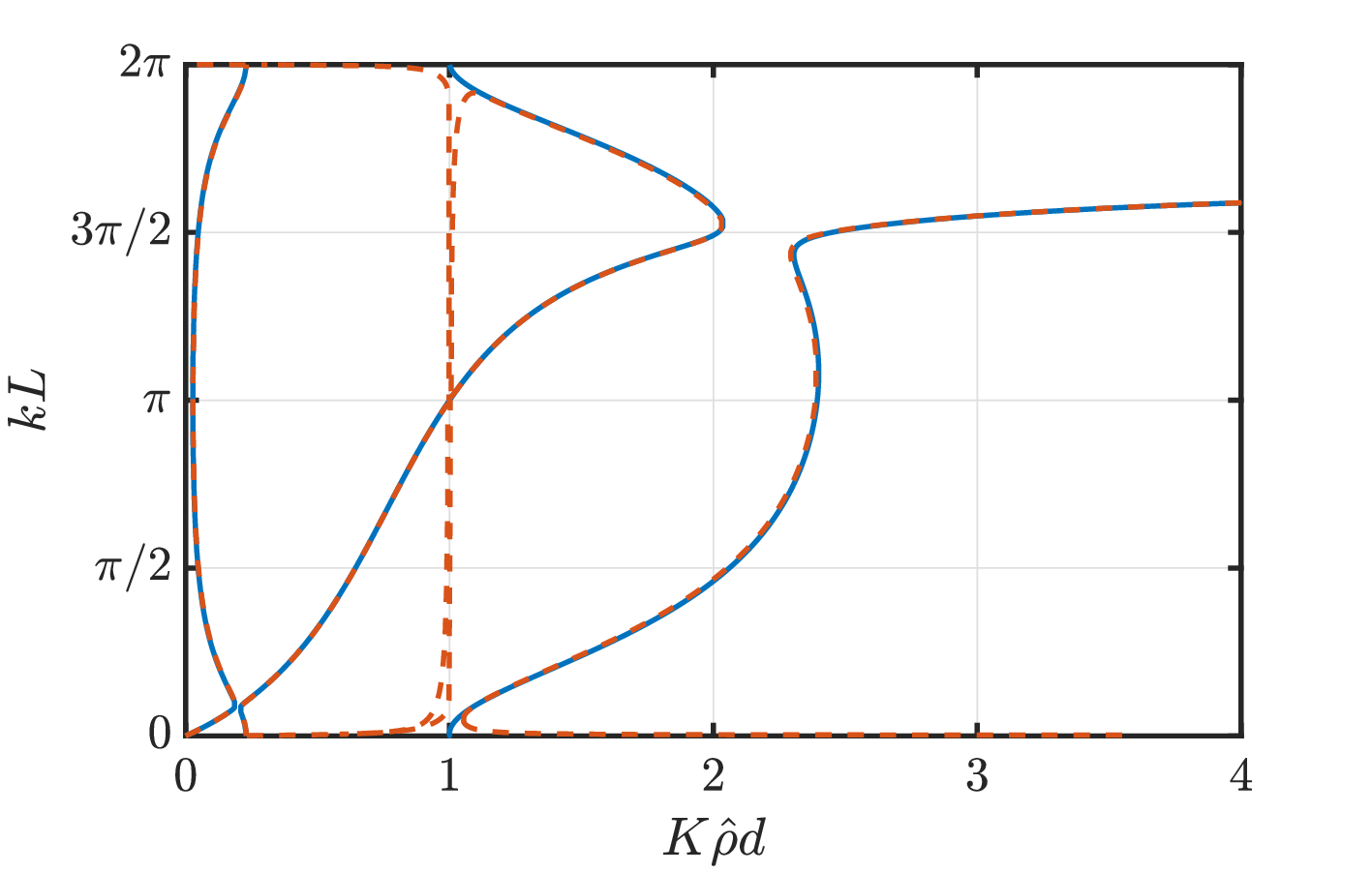}
		\end{subfigure}\hfill
		\begin{subfigure}[t]{0.03\textwidth}
			\text{(b)}
		\end{subfigure}
		\begin{subfigure}[t]{0.45\textwidth}        \centering
			\includegraphics[width=\linewidth, valign=t,trim={0cm 0cm 0cm 0cm}]{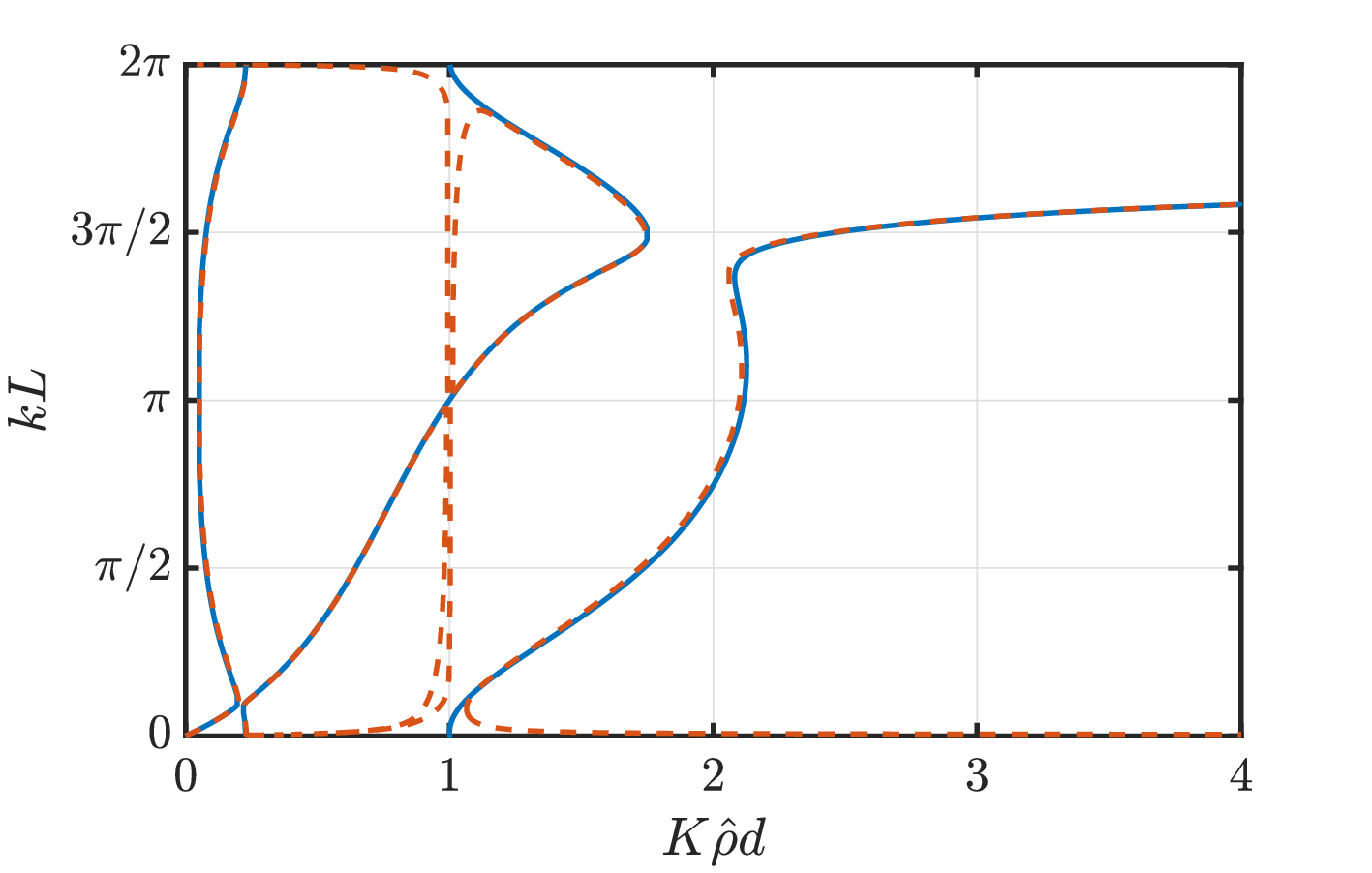}
		\end{subfigure}
		\begin{subfigure}[t]{0.03\textwidth}
			\text{(c)}
		\end{subfigure}
		\begin{subfigure}[t]{0.45\textwidth}        \centering
			\includegraphics[width=\linewidth, valign=t,trim={0cm 0cm 0cm 0cm}]{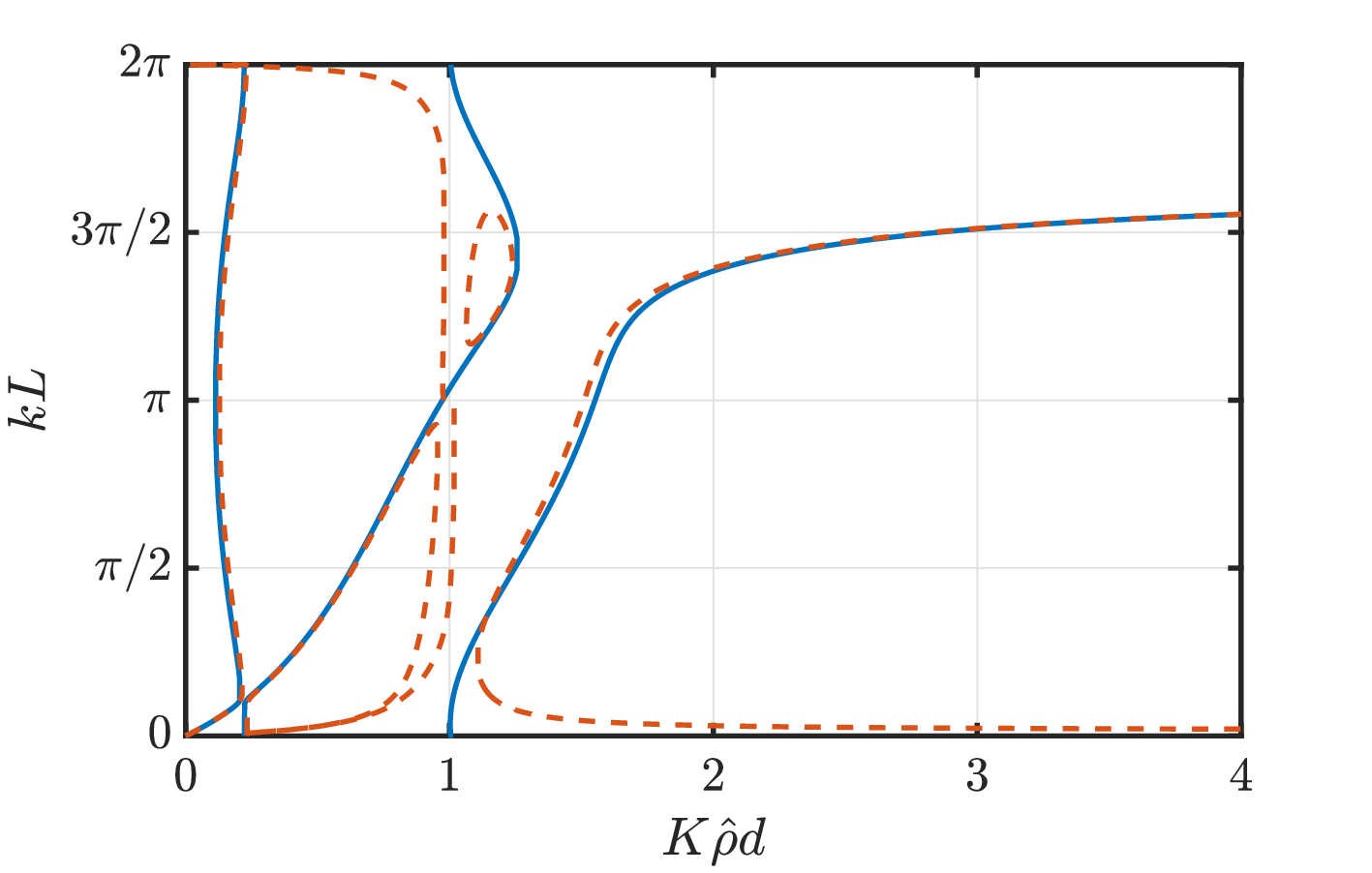}
		\end{subfigure}\hfill
		\begin{subfigure}[t]{0.03\textwidth}
			\text{(d)}
		\end{subfigure}
		\begin{subfigure}[t]{0.45\textwidth}        \centering
			\includegraphics[width=\linewidth, valign=t,trim={0cm 0cm 0cm 0cm}]{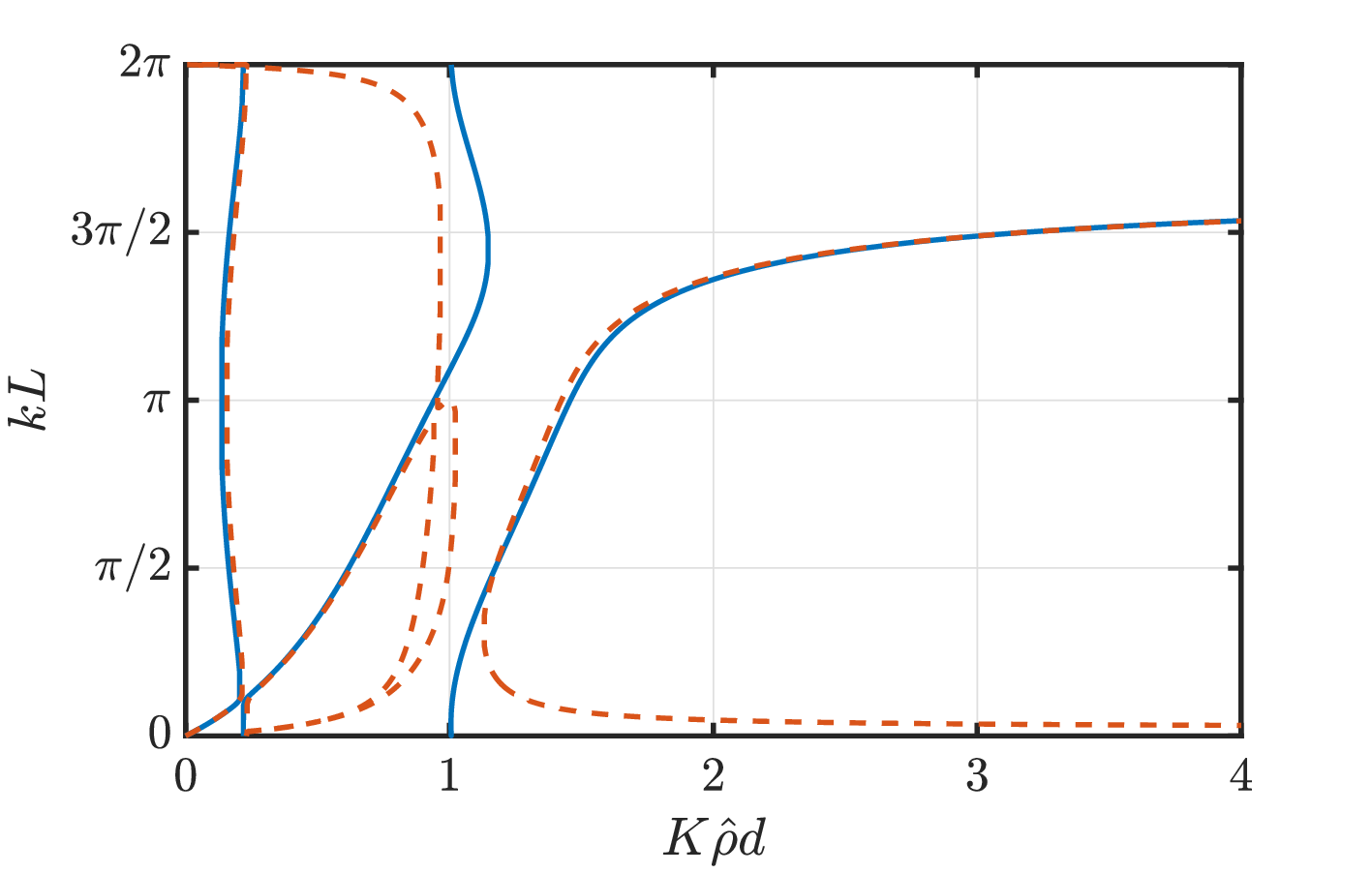}
		\end{subfigure}
		\caption{\label{fig6} Dispersion curves ($kL$ versus $K \rhoh d$) for heave/pitch-constrained motions in the case $\rhoh=0.9$, $L/d = 1 + \epsilon$ (square ice floes). Exact results (blue solid curves) and small-gap approximations (orange, dashed) for gap sizes: (a) $\epsilon = 0.01$, (b) $\epsilon = 0.02$, (c) $\epsilon = 0.08$, (d) $\epsilon = 0.12$.}
	\end{figure}

It can be shown from (\ref{eqn:4.11}) that as $Kd \to 0$
\begin{equation}
 F^{(p,p)} \to \frac{1}{\epsilon} \frac{d^3}{L^3}
 \left( \frac{28}{15}\rhoh^3-\frac{25}{6}\rhoh^2 + 
 \frac{7}{3}\rhoh \right) \sin^2 (kL/2).
 \label{eqn:4.13}
\end{equation}
That is, terms proportional to $1/Kd$ in (\ref{eqn:4.11}) cancel.
As a result there can be no solutions, $kL$, to (\ref{eqn:4.9}) in the zero frequency limit away from $(0,0)$. However, we note that for small $\epsilon$, (\ref{eqn:4.9}) has solutions, $kL$, for values of $Kd$ in proportion to $\epsilon$. This means that as $\epsilon \to 0$, solutions exist as $Kd \to 0$. 
This behaviour can be observed in the dispersion diagram of Fig.~\ref{fig5}, there being a low-frequency branch of solutions running from $kL=0$ to $kL = 2 \pi$. The small-gap approximation without the $\epsilon \cot(kL/2)$ term defined by (\ref{eqn:4.11}) is symmetric with respect to $kL = \pi$ when used in (\ref{eqn:4.9}). The exact results are not symmetric, as shown in Fig.~\ref{fig5}, once again showing the value of including the additional term. The comments made about surge motion close to $kL=0,~2\pi$ apply here also. The importance of the $\epsilon \cot(kL/2)$ term in (\ref{eqn:4.10}) in computing the small-gap approximation is evident in Fig.~\ref{fig6} where (\ref{eqn:4.10}) with (\ref{eqn:4.9}) are shown to produce accurate approximations in all cases.

To further illustrate the interaction between different modes we briefly consider
the effect of coupling pitch to heave and surge motions. As with coupled
heave-surge motion previously, this means
selectively restricting (\ref{eqn:2.15}) to the appropriate $2\times 2$ subsystem.
When small gaps are assumed, we use the approximation (see supplementary material)
\begin{equation}
    F^{(h,p)} = i\left(\frac{L}{12d} + \frac{\hat{\rho}d}{2L} \frac{\frac{\hat{\rho}^2}{3} - \frac{\hat{\rho}}{2} + \frac{1-\hat{\rho}}{Kd}}{\frac{K\hat{\rho} d - 1}{Kd} + \frac{1}{2}\epsilon\cot\left(\frac{kL}{2}\right)}\right) = -F^{(p,h)}\label{eqn:4.14}.
\end{equation}

In Fig.~\ref{fig6} we show the dispersion curves for motions constrained
to pitch/heave motions, i.e.\ surge motion is suppressed. Exact computations are
shown against approximations for small gaps. We see a superposition of the 
effects seen in Figs.~\ref{fig2} and \ref{fig5}. The low-frequency branch
from the pitch-constrained solution exists alongside the heave-constrained
mass-loading curve, but we note that it is not simly a superposition of heave and pitch. The heave resonance that manifests as an asymptote at $K\rhoh d = 1$ in the heave-constrained is suppressed when heave is coupled with either surge, pitch or both. Instead, exact results follow the leading-order heave solution in these coupled cases. 
Beyond low frequencies, we see similar effects to those observed for coupled heave-surge motions in relation to how approximate solutions diverge from the exact counterparts as frequencies tend to resonance ($K\rhoh d = 1$) or wave motion tends to standing waves ($kL = 0,~2\pi$). The distortions to approximate solutions that are constrained to be $2\pi$-periodic in $kL$ lead to the emergence of solution branches which are not associated with exact solutions.

\begin{figure}[!htbp]
		\centering
		\begin{subfigure}[t]{0.03\textwidth}
			\text{(a)}
		\end{subfigure}
		\begin{subfigure}[t]{0.45\textwidth}        \centering
			\includegraphics[width=\linewidth, valign=t,trim={0cm 0cm 0cm 0cm}]{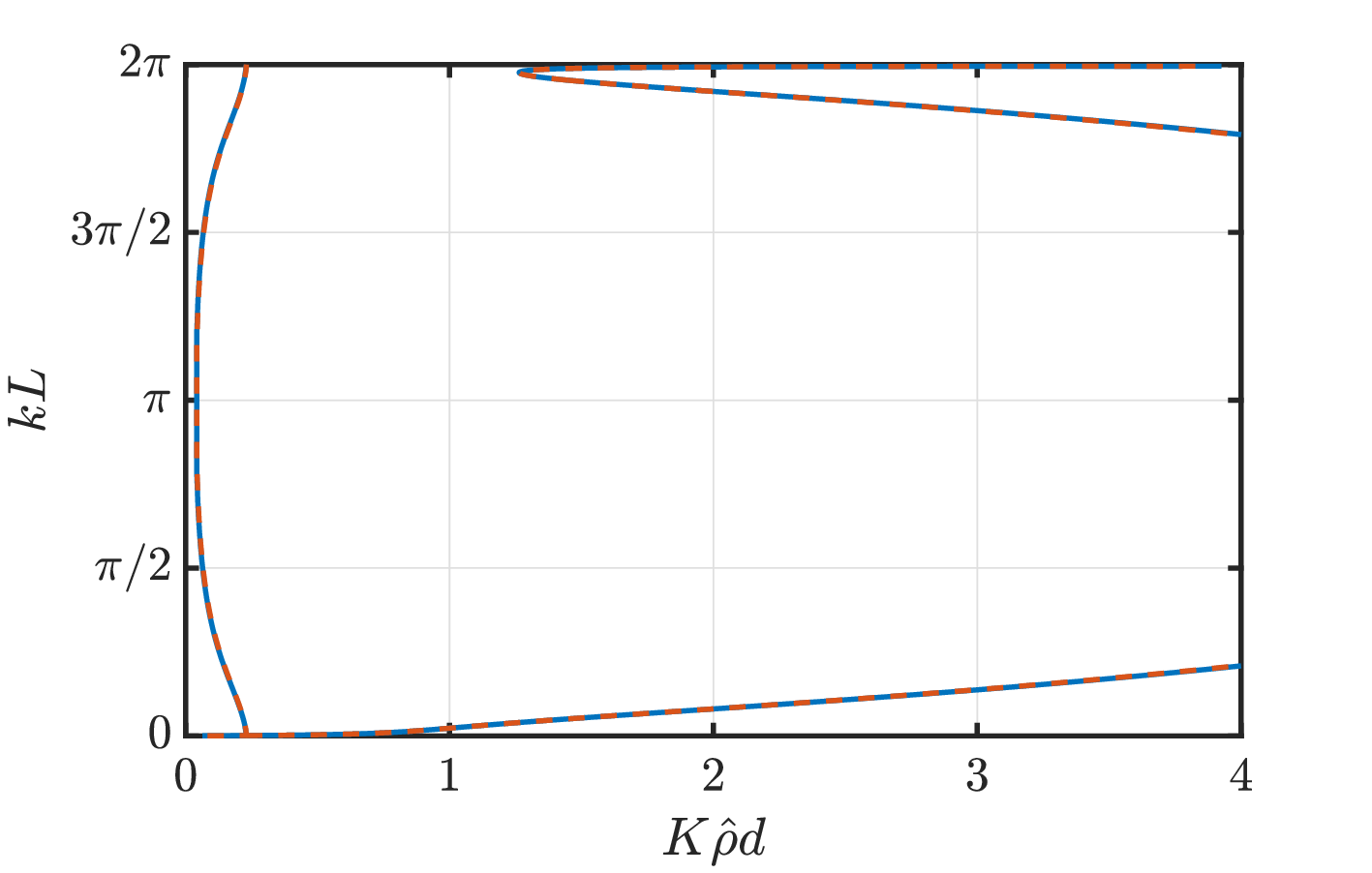}
		\end{subfigure}\hfill
		\begin{subfigure}[t]{0.03\textwidth}
			\text{(b)}
		\end{subfigure}
		\begin{subfigure}[t]{0.45\textwidth}        \centering
			\includegraphics[width=\linewidth, valign=t,trim={0cm 0cm 0cm 0cm}]{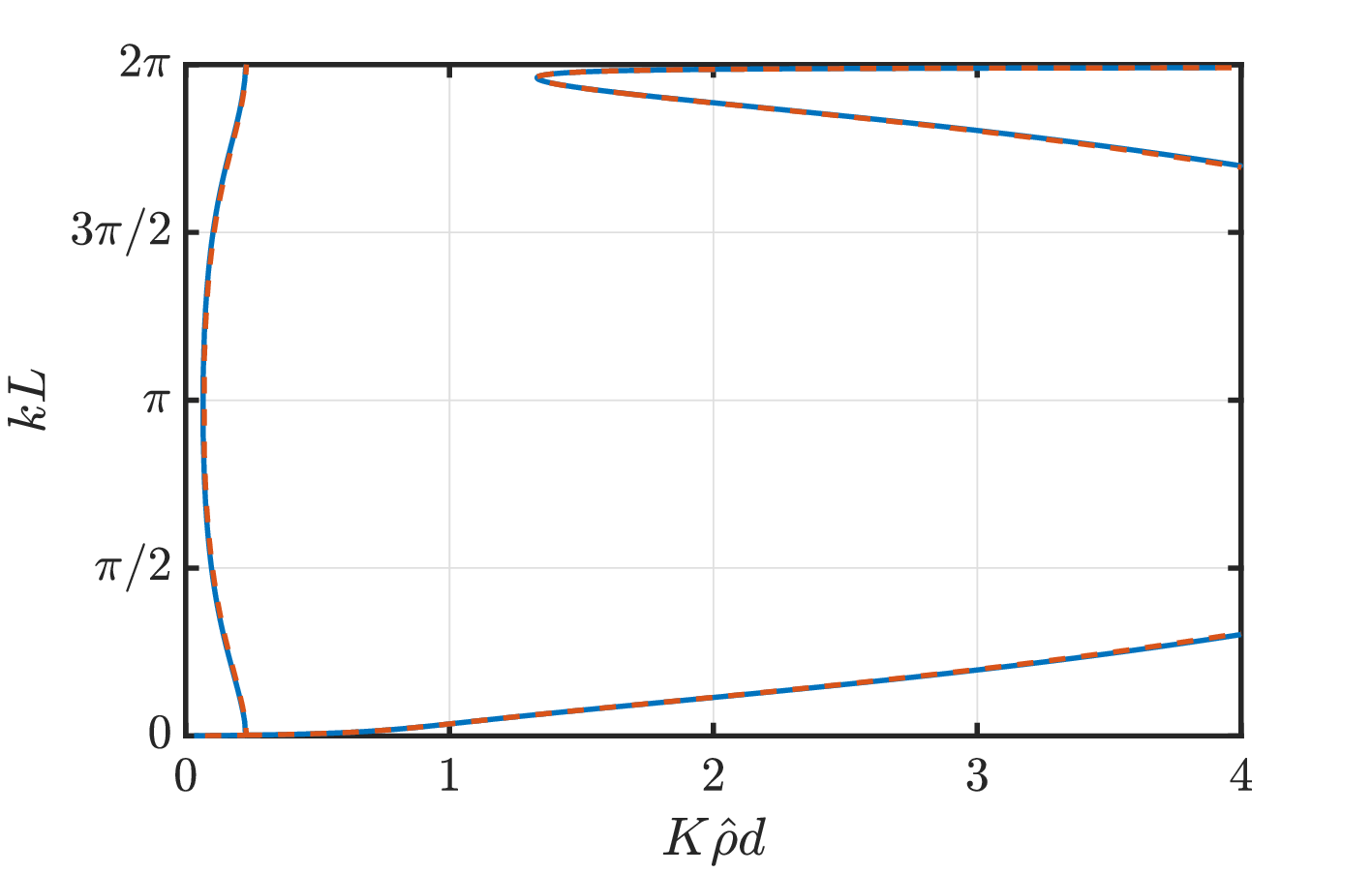}
		\end{subfigure}
		\begin{subfigure}[t]{0.03\textwidth}
			\text{(c)}
		\end{subfigure}
		\begin{subfigure}[t]{0.45\textwidth}        \centering
			\includegraphics[width=\linewidth, valign=t,trim={0cm 0cm 0cm 0cm}]{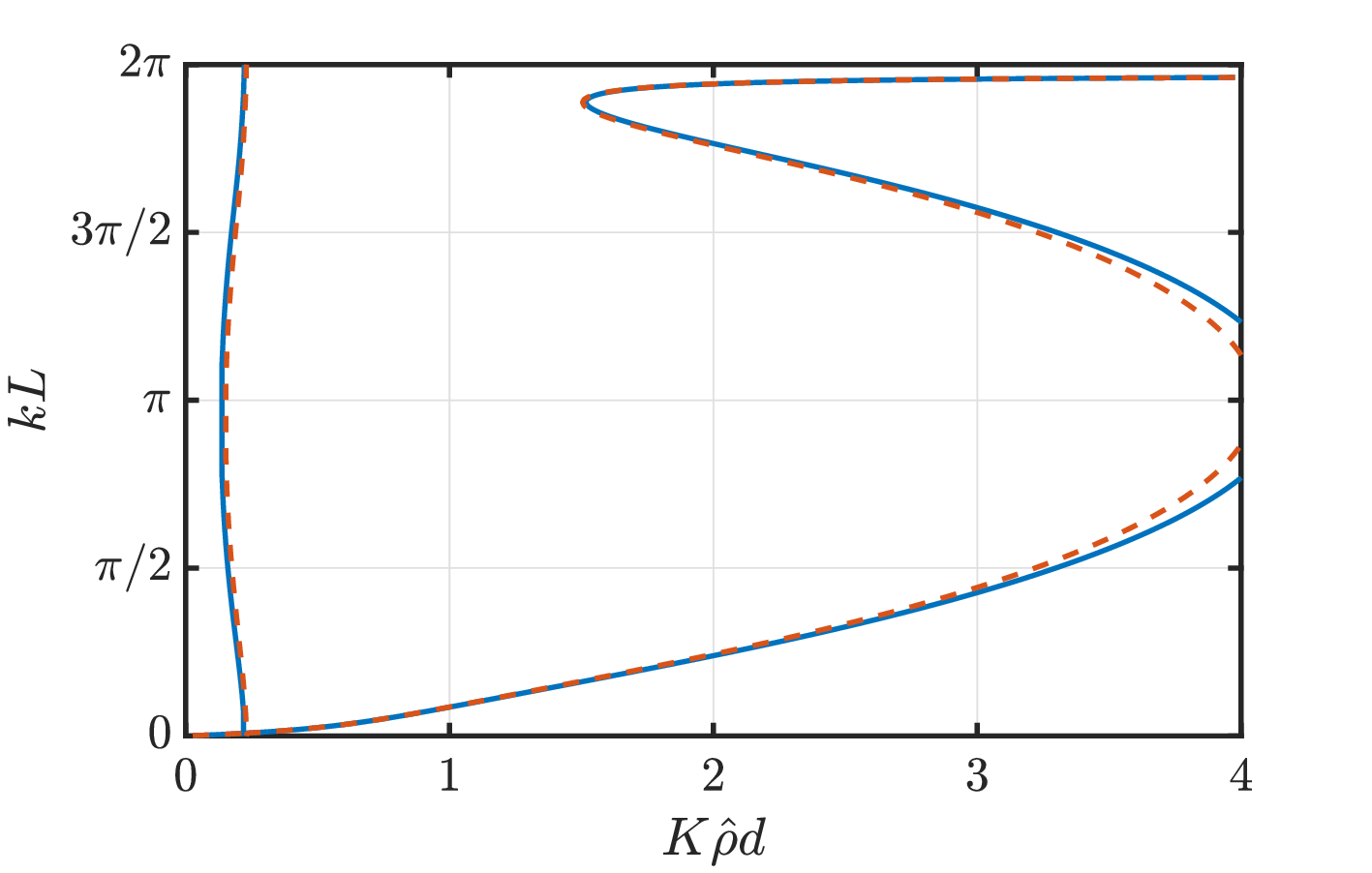}
		\end{subfigure}\hfill
		\begin{subfigure}[t]{0.03\textwidth}
			\text{(d)}
		\end{subfigure}
		\begin{subfigure}[t]{0.45\textwidth}        \centering
			\includegraphics[width=\linewidth, valign=t,trim={0cm 0cm 0cm 0cm}]{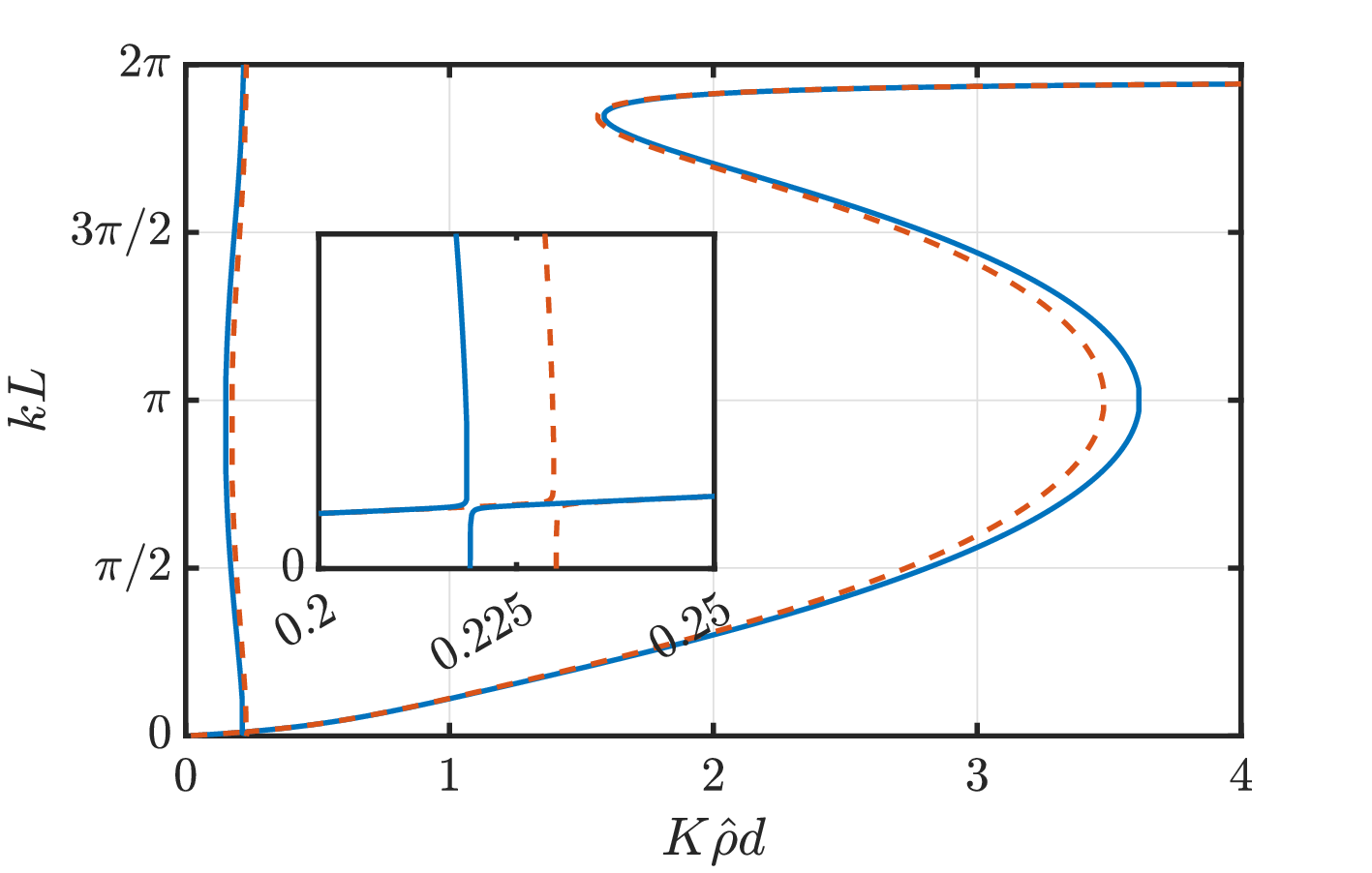}
		\end{subfigure}
		\caption{\label{fig7} Dispersion curves ($kL$ versus $K \rhoh d$) for surge/pitch-constrained motions in the case $\rhoh=0.9$, $L/d = 1 + \epsilon$ (square ice floes). Exact results (blue solid curves) and small-gap approximations (orange, dashed) for gap sizes: (a) $\epsilon = 0.01$, (b) $\epsilon = 0.02$, (c) $\epsilon = 0.08$, (d) $\epsilon = 0.12$.}
	\end{figure}

The same general observations can be made regarding Fig.~\ref{fig7}, where we 
consider pitch/surge coupled motions.
In the previous results where two modes have been coupled (heave/surge in Fig.~\ref{fig4} and heave/pitch in Fig.~\ref{fig6}) the resulting dispersion diagrams have, to a good approximation, appeared as a superposition of the two sets of dispersion curves from each component mode. In Fig.~\ref{fig7} we see that the dispersion curves are related to those in Figs.~\ref{fig3} and \ref{fig5}. However, only the low frequency branch of the pitch mode is represented in Fig.~\ref{fig7} and the branch which extends into $K\rhoh d > 1$ is missing. We cannot determine a physical or mathematical reason for why this might be. It is worth remarking that the results obtained using the asymptotic results for coupled surge/pitch motions match the exact results well across all frequencies. That is, there appears to be no issue with frequencies close to resonance, as experienced in the results involving coupling with heave.

Here, the small-gap approximation is derived in the supplementary material to be
\begin{equation}
    F^{(s,p)} = \frac{1}{\epsilon} \frac{\hat{\rho}^2 d}{KL^2} \left(2Kd\left(\frac{\hat{\rho}^2}{4} - \frac{\hat{\rho}}{3} + \frac{1-\hat{\rho}}{Kd}\right) - \frac{K\hat{\rho} d - 2}{\frac{K\hat{\rho} d - 1}{Kd} + \frac{1}{2}\epsilon\cot\left(\frac{kL}{2}\right)} \left(\frac{\hat{\rho}^2}{3} - \frac{\hat{\rho}}{2} + \frac{1-\hat{\rho}}{Kd}\right)\right) \sin^2\left(\frac{kL}{2}\right)\label{eqn:4.15}
\end{equation}

\section{Fully unconstrained problem}\label{sec:fully-unconstrained}

The fully unconstrained problem treats the rectangular floe as a rigid body
free to heave, surge and pitch. The dispersion relation is expressed by the vanishing of the $3 \times 3$ determinant in (\ref{eqn:2.15}), coupling all three rigid-body modes. Numerical solutions are obtained using the Galerkin method described in Section~\ref{sec:solution-in-heave}, extended to include the surge and pitch subproblems. Approximate solutions for small gaps use the asymptotic force expressions derived in the preceding sections.

	\begin{figure}[!htbp]
		\centering
		\begin{subfigure}[t]{0.03\textwidth}
			\text{(a)}
		\end{subfigure}
		\begin{subfigure}[t]{0.45\textwidth}        \centering
			\includegraphics[width=\linewidth,trim={0cm 0cm 0cm 0cm}]{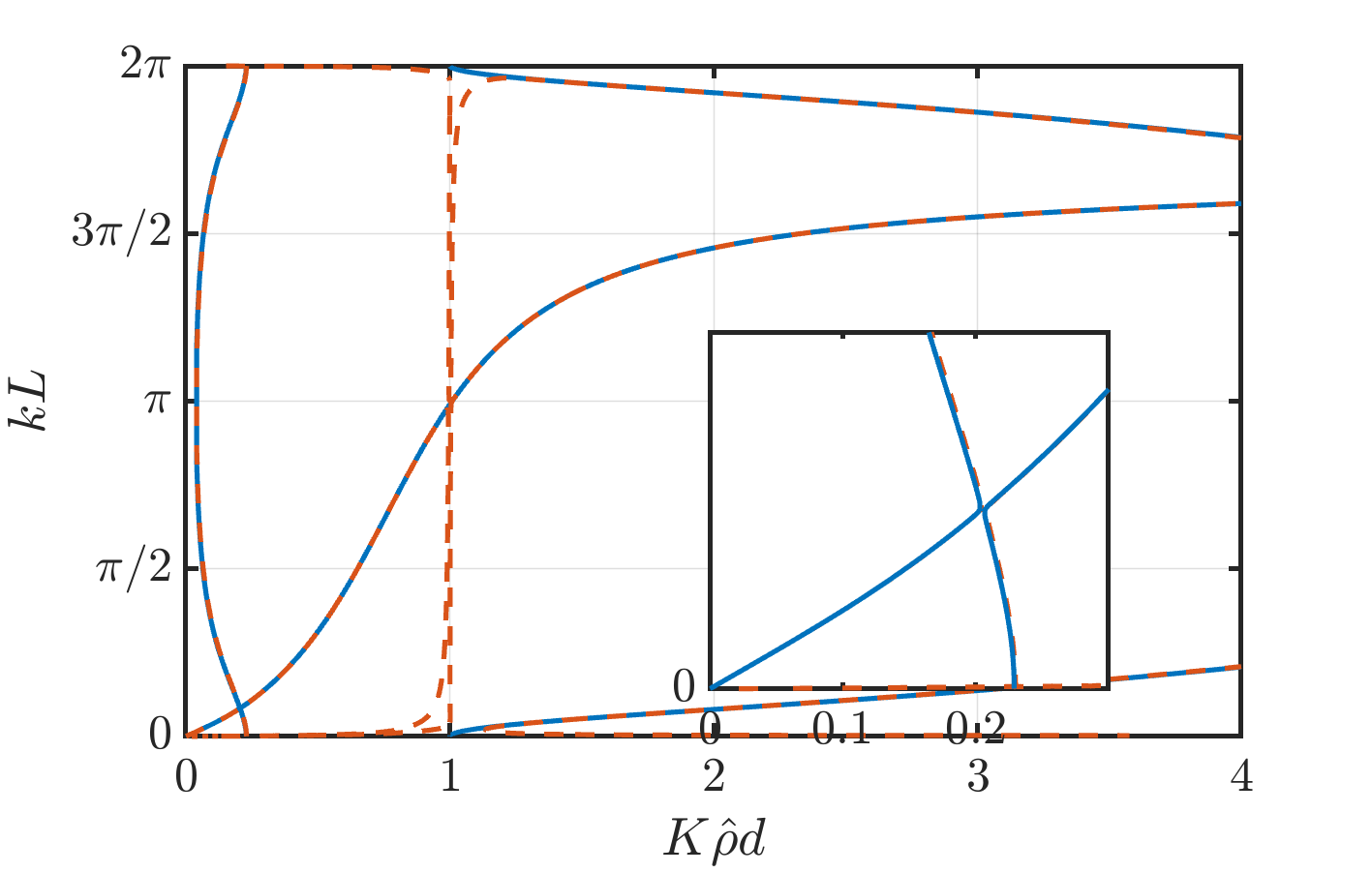}
		\end{subfigure}\hfill
		\begin{subfigure}[t]{0.03\textwidth}
			\text{(b)}
		\end{subfigure}
		\begin{subfigure}[t]{0.45\textwidth}        \centering
			\includegraphics[width=\linewidth,trim={0cm 0cm 0cm 0cm}]{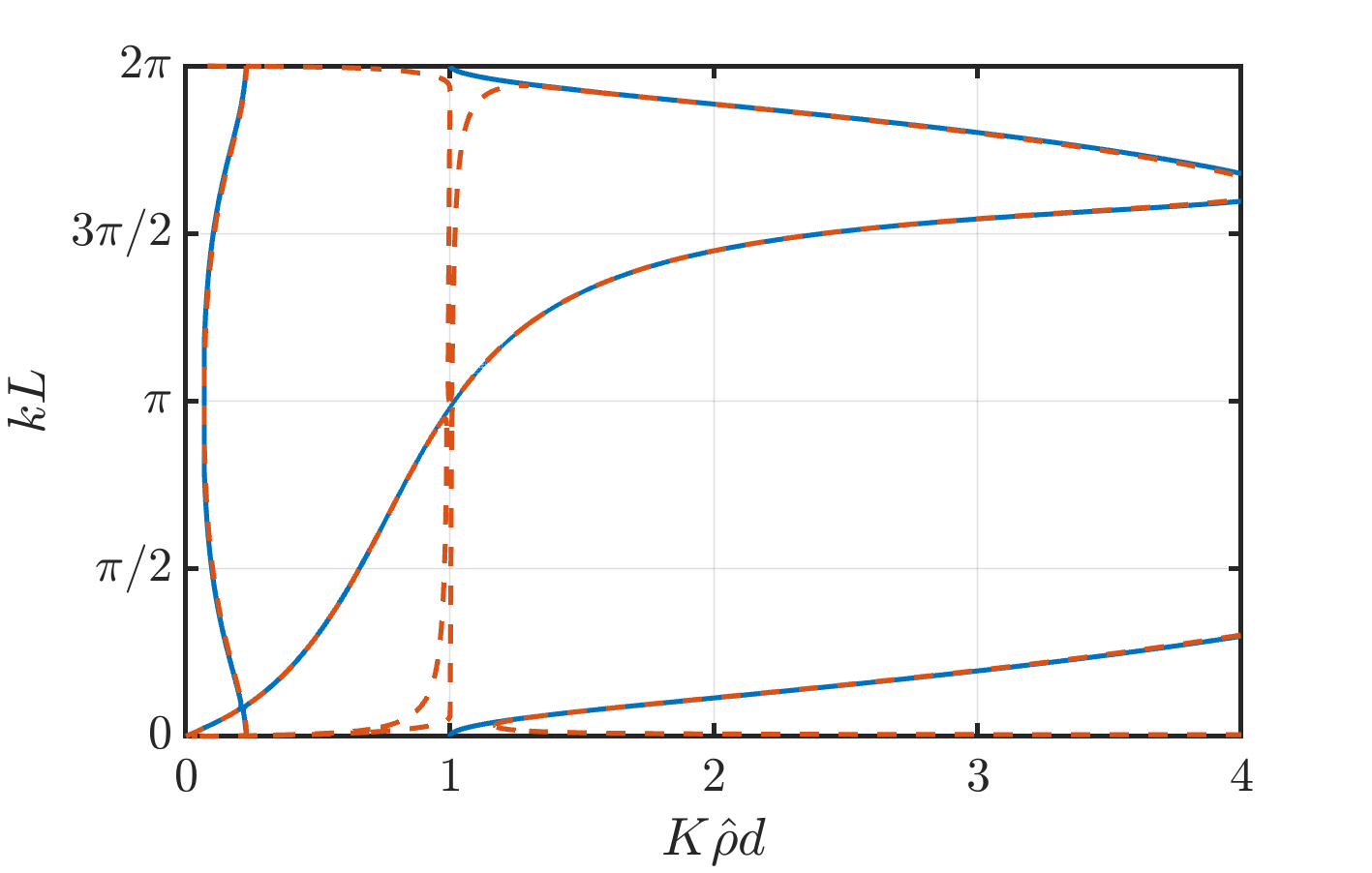}
		\end{subfigure}
		\begin{subfigure}[t]{0.03\textwidth}
			\text{(c)}
		\end{subfigure}
		\begin{subfigure}[t]{0.45\textwidth}        \centering
			\includegraphics[width=\linewidth,trim={0cm 0cm 0cm 0cm}]{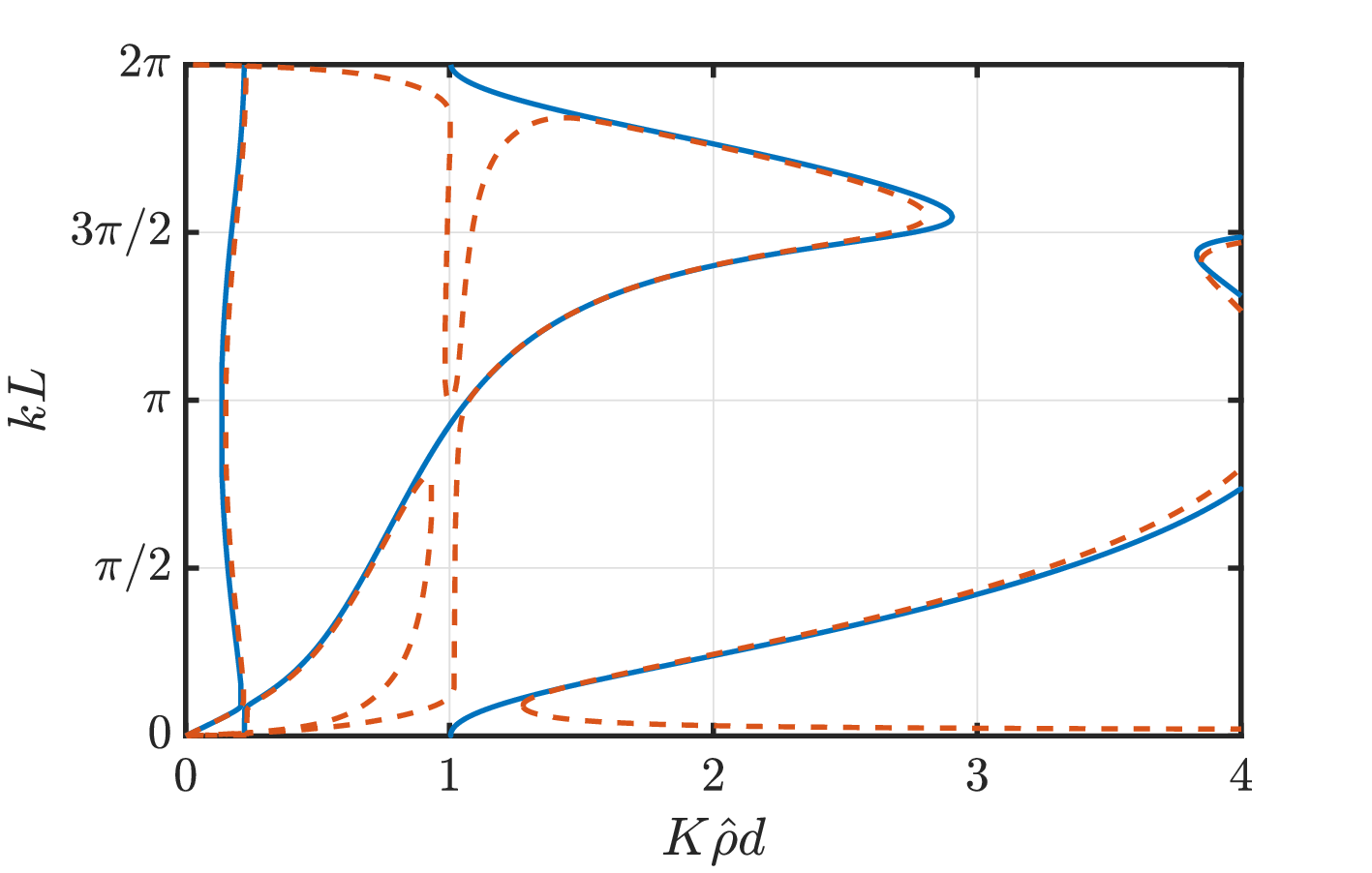}
		\end{subfigure}\hfill
		\begin{subfigure}[t]{0.03\textwidth}
			\text{(d)}
		\end{subfigure}
		\begin{subfigure}[t]{0.45\textwidth}        \centering
			\includegraphics[width=\linewidth,trim={0cm 0cm 0cm 0cm}]{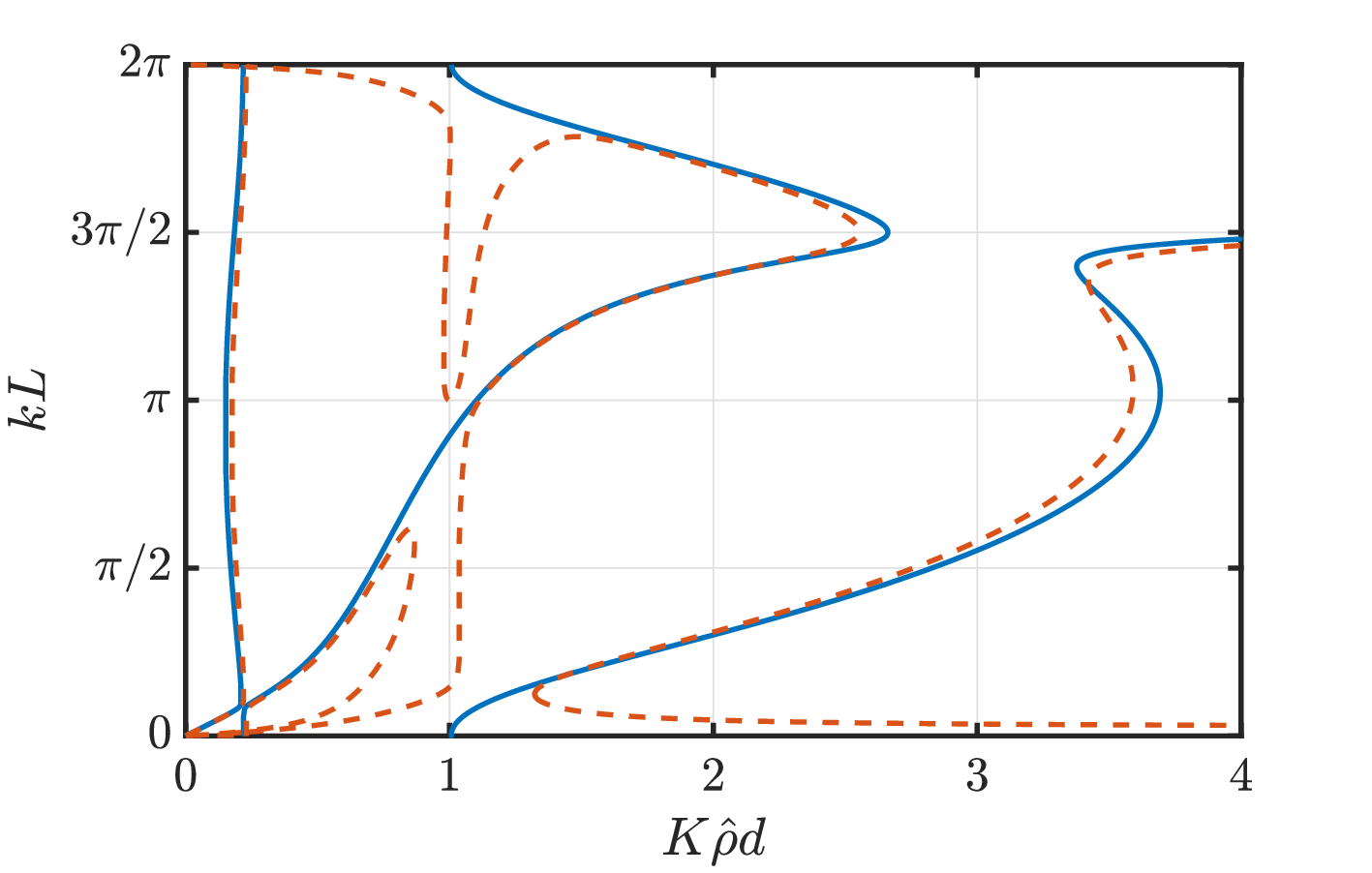}
		\end{subfigure}
		\caption{\label{fig8} Dispersion curves ($kL$ versus $K \rhoh d$) for fully unconstrained motions in the case $\rhoh=0.9$, $L/d = 1 + \epsilon$ (square ice floes). Exact results (blue solid curves) and small-gap approximations (orange, dashed) for gap sizes: (a) $\epsilon = 0.01$ with low-frequency results inset, (b) $\epsilon = 0.02$, (c) $\epsilon = 0.08$, (d) $\epsilon = 0.12$.}
	\end{figure}

We now consider results for the problem originally conceived in this paper
whereby waves propagate through freely-floating rectangular floes.
The previous sets of results which describe relations between frequency
and wavenumber for motions which are constrained to move in either one mode
or two coupled modes have been useful in establishing certain
fundamental features associated with heave, surge and pitch motions.
In particular we can observe that when two modes are coupled, the
resulting dispersion diagrams are, to a large extent, a superposition
of the dispersion curves for the two uncoupled modes. Where two
single mode curves intersect, the coupling creates connecting loops
characteristic of saddle points. These features are most clear
to see when the gaps between floes are small. Increasing the size of the
gaps between floes strengthens the coupling, distorting the dispersion
curves further away from the superposition of the two
component dispersion curves. The exceptions have been in the coupling
of surge and pitch, which has resulted in one of the component
branches being dropped from the combined dispersion diagram; and
heave, when coupled to either surge or pitch, suppresses the
heave-constrained asymptote at resonance in the exact results,
although the asymptotic small-gap results do not suppress this effect.

In Fig.~\ref{fig8} we show the dispersion diagram for unconstrained motion
(freely-floating floes) corresponding to the same array of
square floes and gaps that have been the subject of Figs.~\ref{fig2}--\ref{fig7}.
Again, the solid blue curves represent the numerical solutions based on
the exact method and the orange dashed curves correspond to
approximate solutions that have used the asymptotic results for each of
the component hydrodynamic forces (that is, the definitions (\ref{eqn:3.40}), (\ref{eqn:3.41}), (\ref{eqn:4.1}), (\ref{eqn:4.5}),
(\ref{eqn:4.11}), (\ref{eqn:4.15}) have been used in (\ref{eqn:2.15})).

The combined dispersion diagram in Fig.~\ref{fig8} combines curves from each
component heave, surge and pitch modes as we might expect from our
analysis of the previous sets of results. Thus, only the
low-frequency branch of the pitch dispersion curve is represented in
Fig.~\ref{fig8} and the branch that extends into $K \hat{\rho} d > 1$ is
missing from Fig.~\ref{fig8}, as it was from the combined surge/pitch results
in Fig.~\ref{fig7}; the heave resonance is also suppressed. The approximations
obtained from the asymptotic evaluations of the forces under the small-gap
approximation are generally good apart from close to resonance, $K \hat{\rho} d = 1$.

The intended purpose of this work relates to low frequency wave
propagation, well below $K \hat{\rho} d = 1$, and here we can
see that the small-gap results capture quite accurately the
low frequency dispersion curves. The dominant branch extending
away from $(0,0)$ in the dispersion diagrams in Fig.~\ref{fig8} follows very closely
the curves in Fig.~\ref{fig2} based on heave-constrained motion. What is
surprising in these results is the existence of a low-frequency branch
representing pitch-constrained motions.

\begin{figure}[!htbp]
	\centering
	\begin{subfigure}[t]{0.03\textwidth}
		\text{(a)}
	\end{subfigure}
	\begin{subfigure}[t]{0.45\textwidth}        \centering
		\includegraphics[width=\linewidth,trim={0cm 0cm 0cm 0cm}]{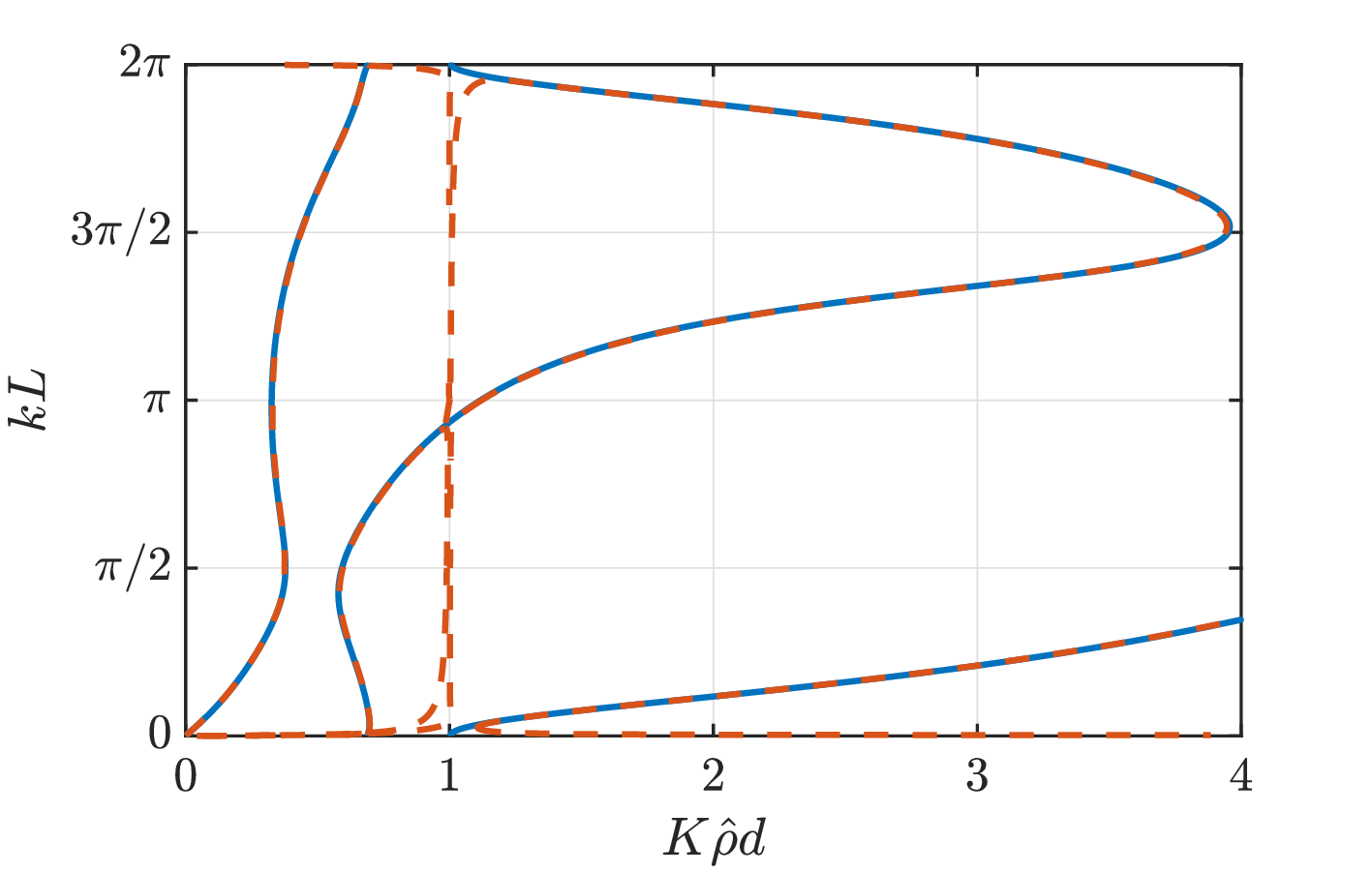}
	\end{subfigure}\hfill
	\begin{subfigure}[t]{0.03\textwidth}
		\text{(b)}
	\end{subfigure}
	\begin{subfigure}[t]{0.45\textwidth}        \centering
		\includegraphics[width=\linewidth,trim={0cm 0cm 0cm 0cm}]{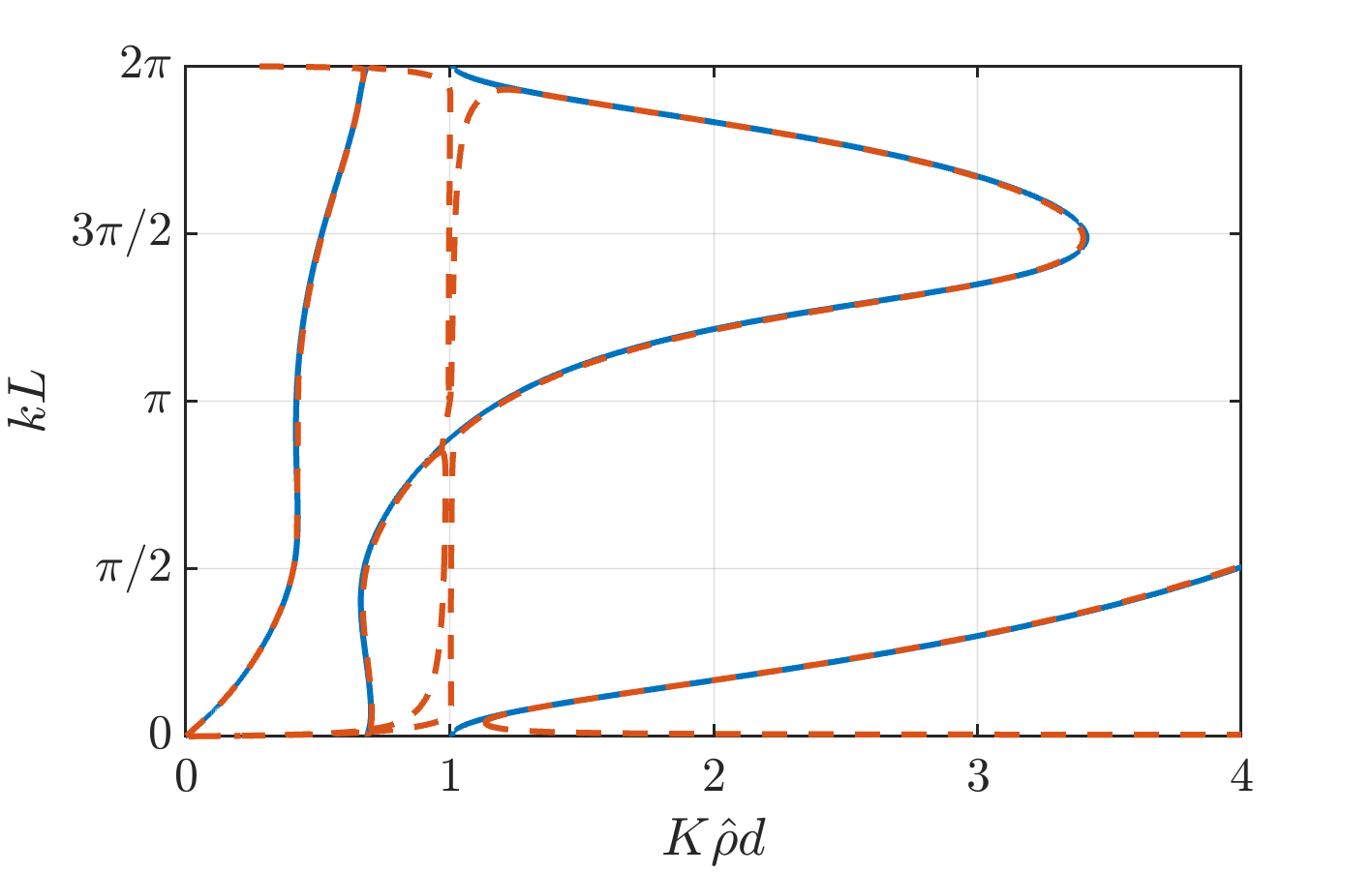}
	\end{subfigure}
	\begin{subfigure}[t]{0.03\textwidth}
		\text{(c)}
	\end{subfigure}
	\begin{subfigure}[t]{0.45\textwidth}        \centering
		\includegraphics[width=\linewidth,trim={0cm 0cm 0cm 0cm}]{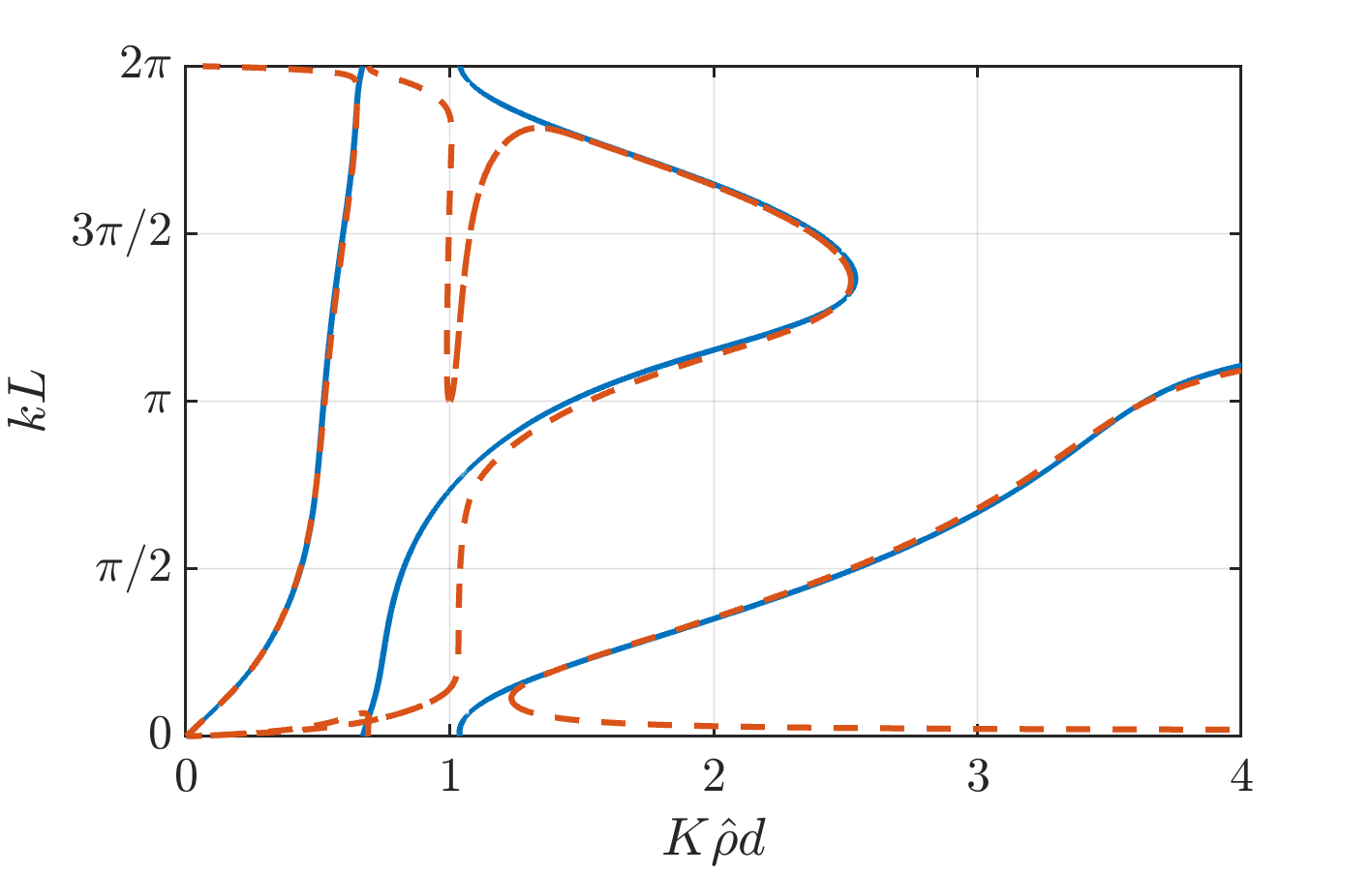}
	\end{subfigure}\hfill
	\begin{subfigure}[t]{0.03\textwidth}
		\text{(d)}
	\end{subfigure}
	\begin{subfigure}[t]{0.45\textwidth}        \centering
		\includegraphics[width=\linewidth,trim={0cm 0cm 0cm 0cm}]{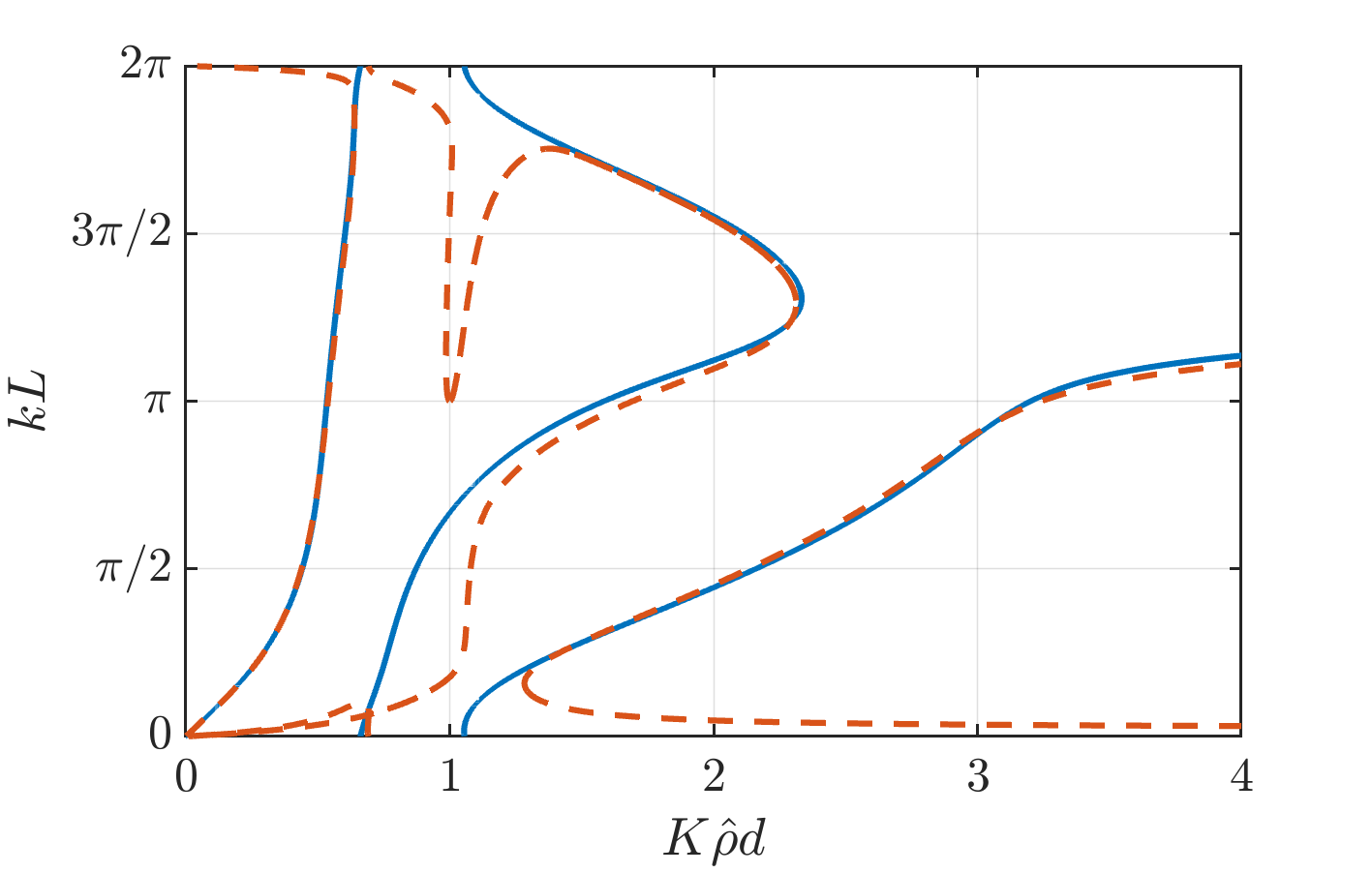}
	\end{subfigure}
	\caption{\label{fig_ar2} Dispersion curves ($kL$ versus $K \rhoh d$) for fully unconstrained motions in the case $\rhoh=0.9$, $L/d = 2 + \epsilon$ (2:1 aspect ratio floes). Exact results (blue solid curves) and small-gap approximations (orange, dashed) for gap sizes: (a) $\epsilon = 0.01$, (b) $\epsilon = 0.02$, (c) $\epsilon = 0.08$, (d) $\epsilon = 0.12$.}
\end{figure}

\begin{figure}[!htbp]
	\centering
	\begin{subfigure}[t]{0.03\textwidth}
		\text{(a)}
	\end{subfigure}
	\begin{subfigure}[t]{0.45\textwidth}        \centering
		\includegraphics[width=\linewidth,trim={0cm 0cm 0cm 0cm}]{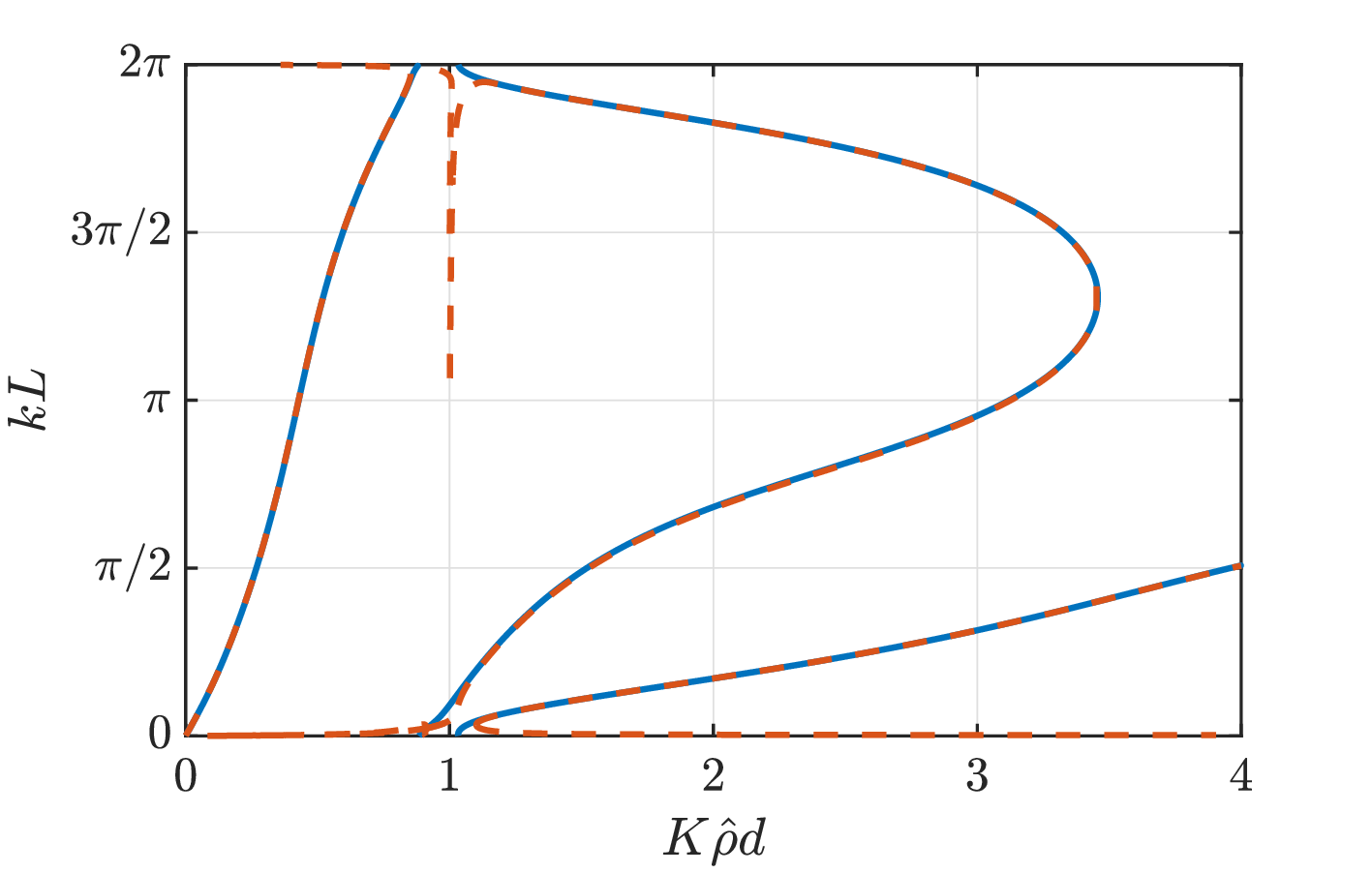}
	\end{subfigure}\hfill
	\begin{subfigure}[t]{0.03\textwidth}
		\text{(b)}
	\end{subfigure}
	\begin{subfigure}[t]{0.45\textwidth}        \centering
		\includegraphics[width=\linewidth,trim={0cm 0cm 0cm 0cm}]{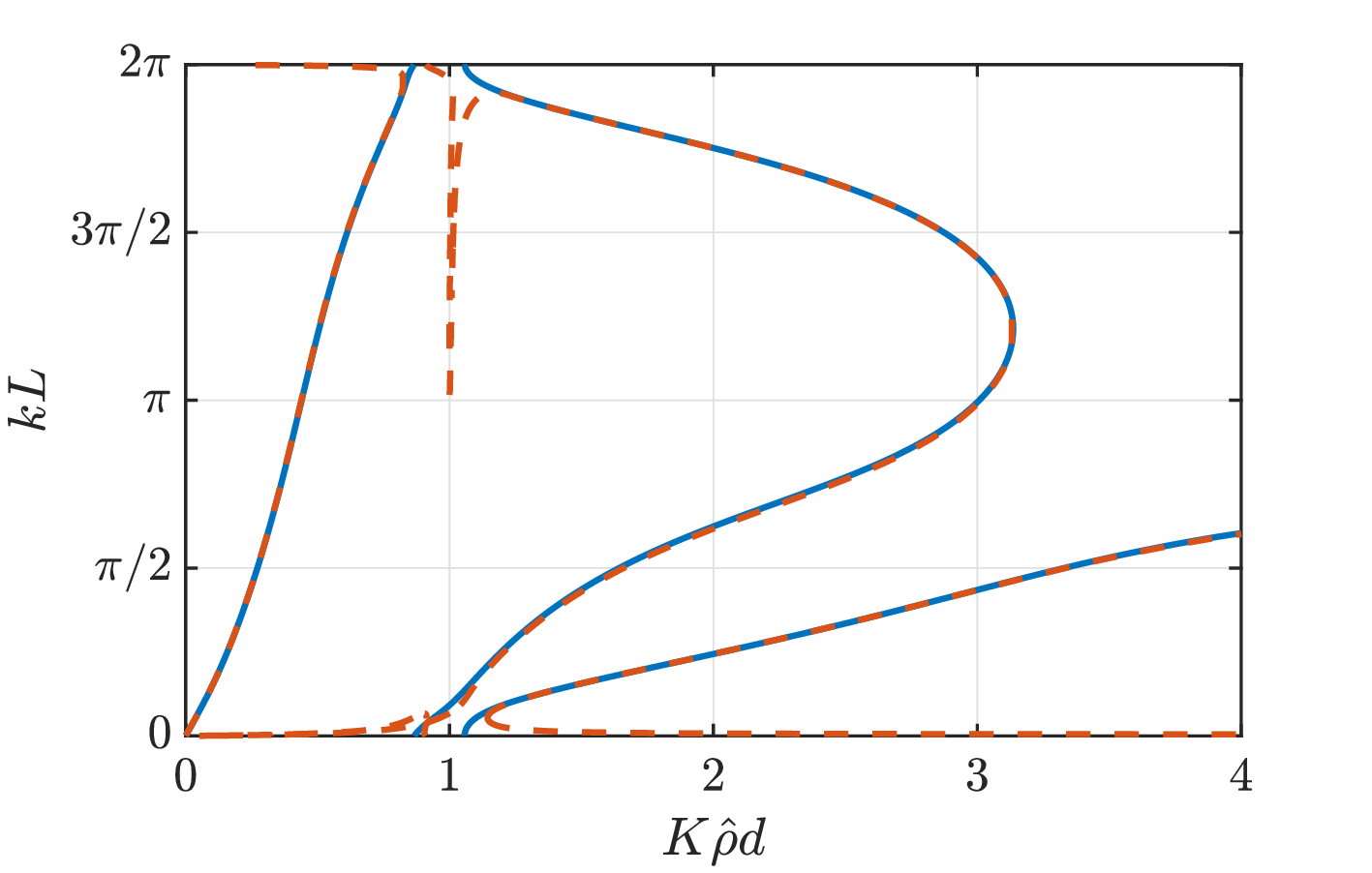}
	\end{subfigure}
	\begin{subfigure}[t]{0.03\textwidth}
		\text{(c)}
	\end{subfigure}
	\begin{subfigure}[t]{0.45\textwidth}        \centering
		\includegraphics[width=\linewidth,trim={0cm 0cm 0cm 0cm}]{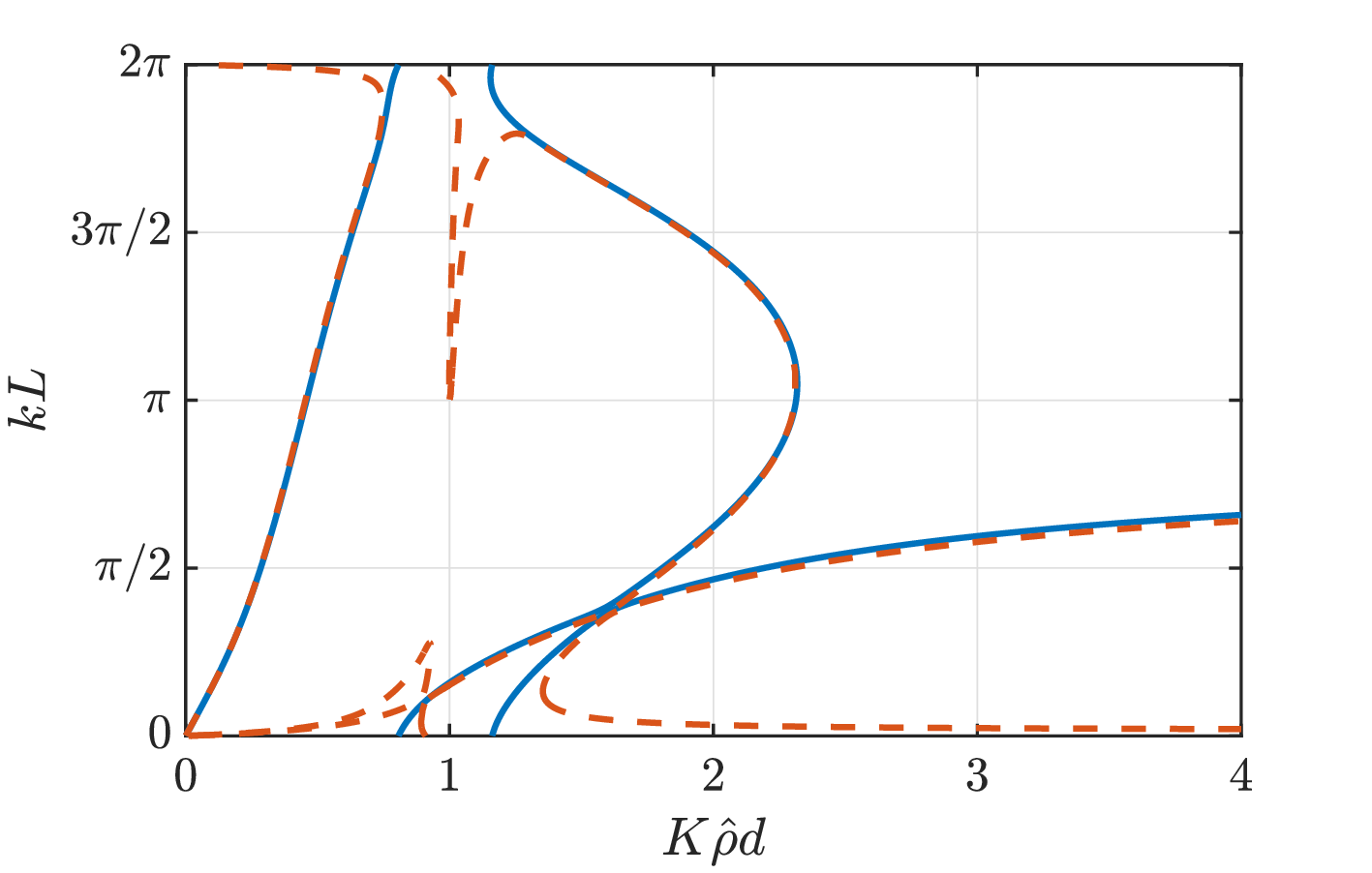}
	\end{subfigure}\hfill
	\begin{subfigure}[t]{0.03\textwidth}
		\text{(d)}
	\end{subfigure}
	\begin{subfigure}[t]{0.45\textwidth}        \centering
		\includegraphics[width=\linewidth,trim={0cm 0cm 0cm 0cm}]{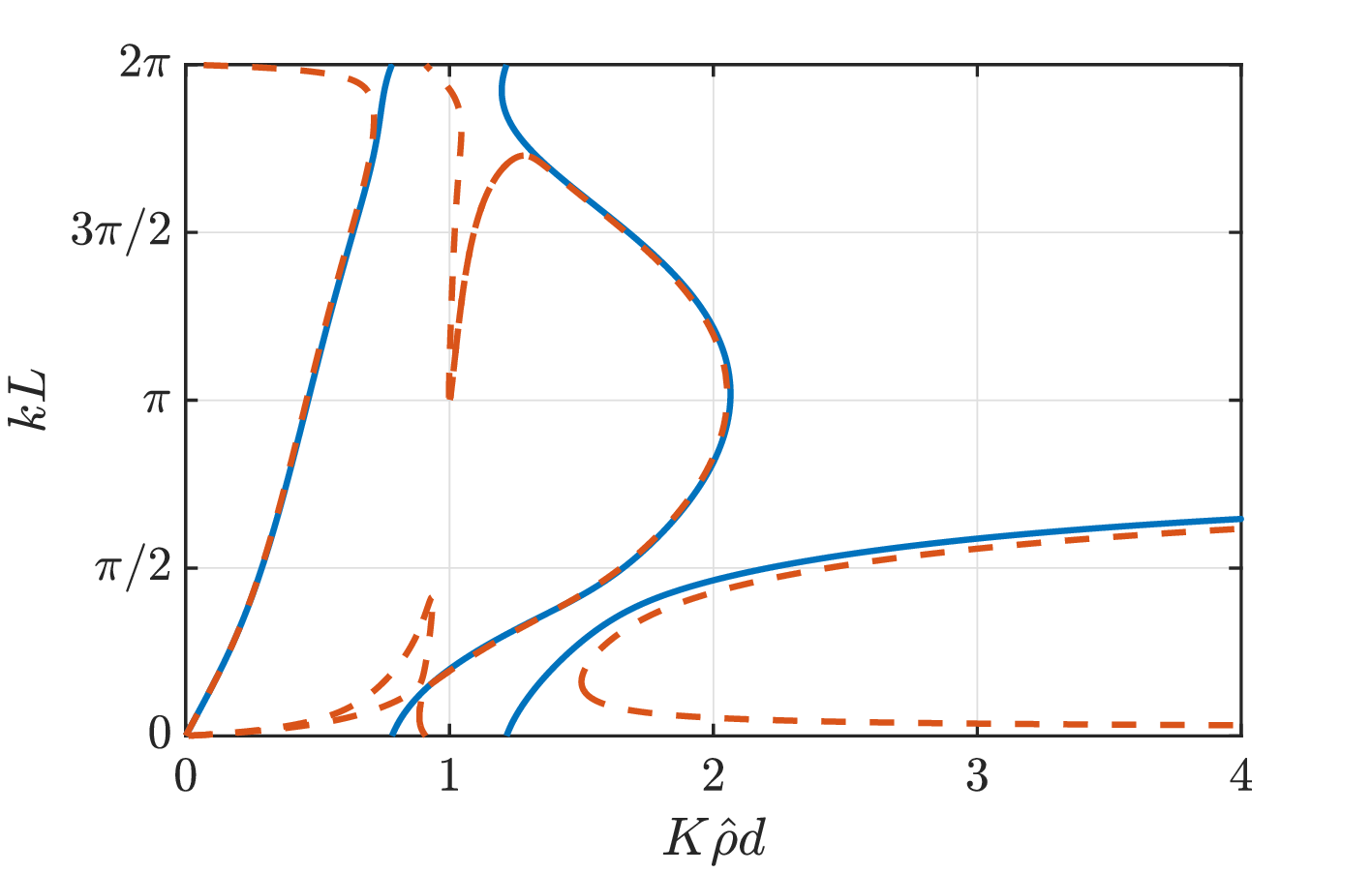}
	\end{subfigure}
	\caption{\label{fig_ar4} Dispersion curves ($kL$ versus $K \rhoh d$) for fully unconstrained motions in the case $\rhoh=0.9$, $L/d = 4 + \epsilon$ (4:1 aspect ratio floes). Exact results (blue solid curves) and small-gap approximations (orange, dashed) for gap sizes: (a) $\epsilon = 0.01$, (b) $\epsilon = 0.02$, (c) $\epsilon = 0.08$, (d) $\epsilon = 0.12$.}
\end{figure}

\begin{figure}[!htbp]
	\centering
	\begin{subfigure}[t]{0.03\textwidth}
		\text{(a)}
	\end{subfigure}
	\begin{subfigure}[t]{0.45\textwidth}        \centering
		\includegraphics[width=\linewidth,trim={0cm 0cm 0cm 0cm}]{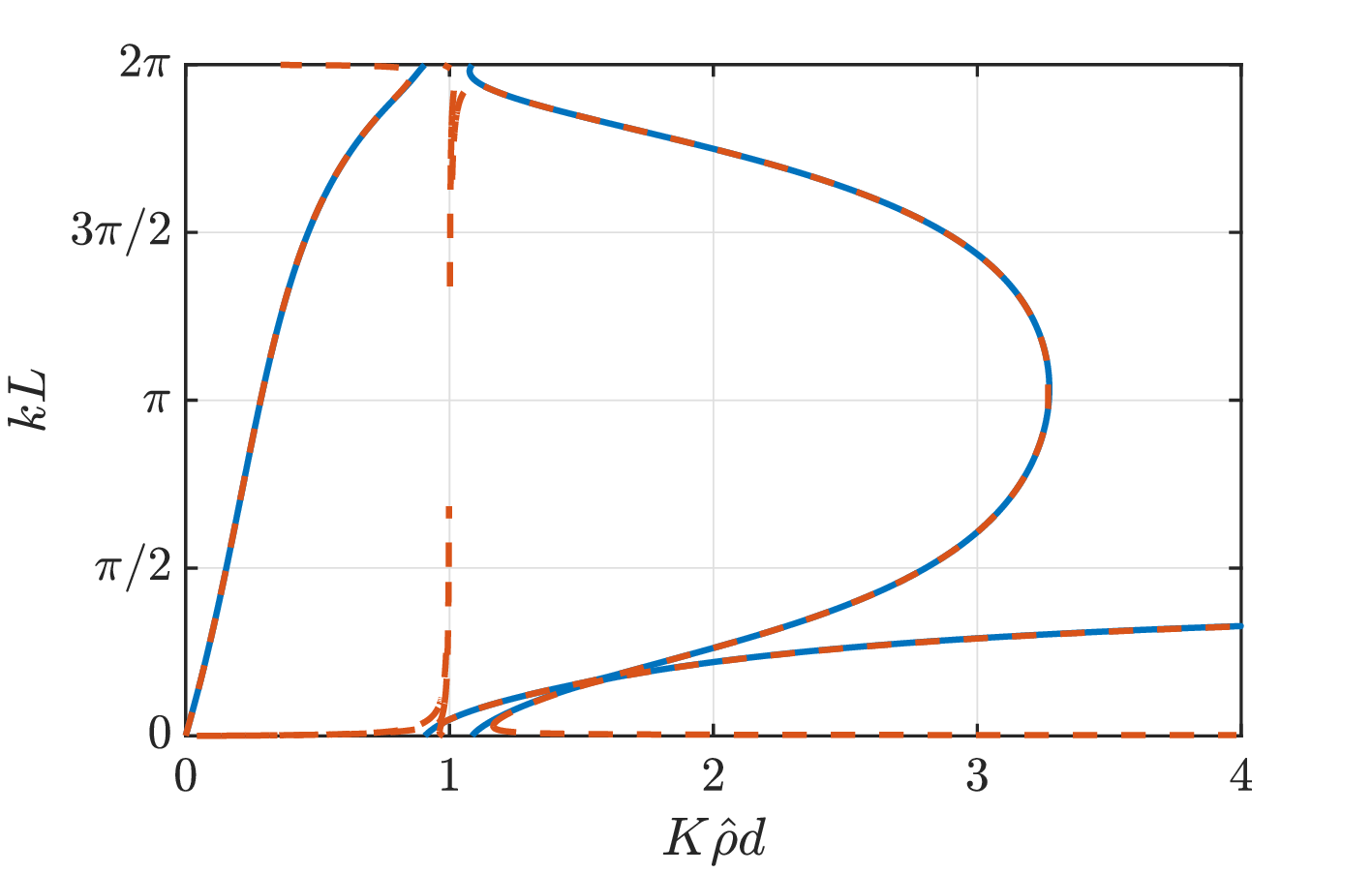}
	\end{subfigure}\hfill
	\begin{subfigure}[t]{0.03\textwidth}
		\text{(b)}
	\end{subfigure}
	\begin{subfigure}[t]{0.45\textwidth}        \centering
		\includegraphics[width=\linewidth,trim={0cm 0cm 0cm 0cm}]{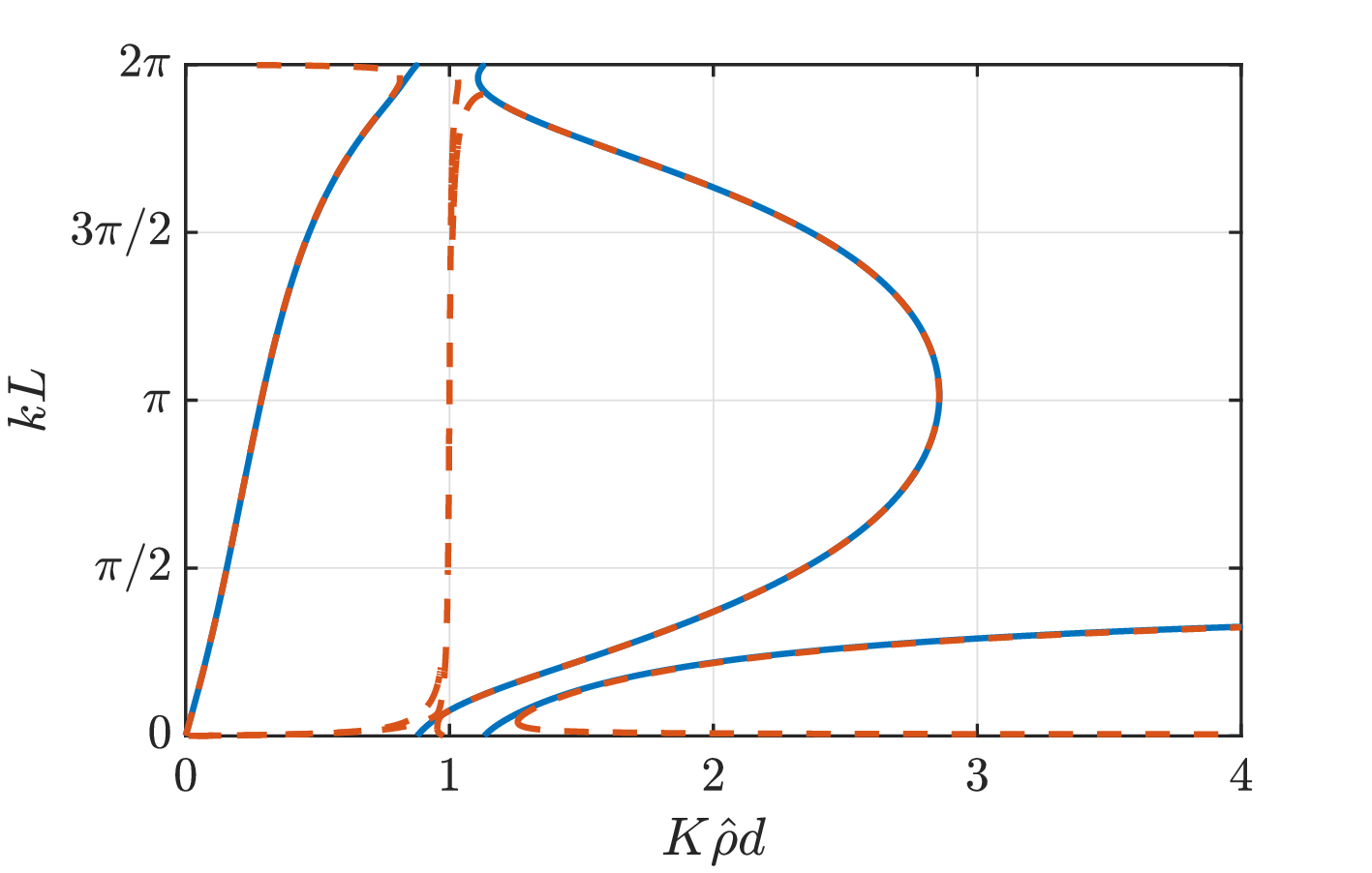}
	\end{subfigure}
	\begin{subfigure}[t]{0.03\textwidth}
		\text{(c)}
	\end{subfigure}
	\begin{subfigure}[t]{0.45\textwidth}        \centering
		\includegraphics[width=\linewidth,trim={0cm 0cm 0cm 0cm}]{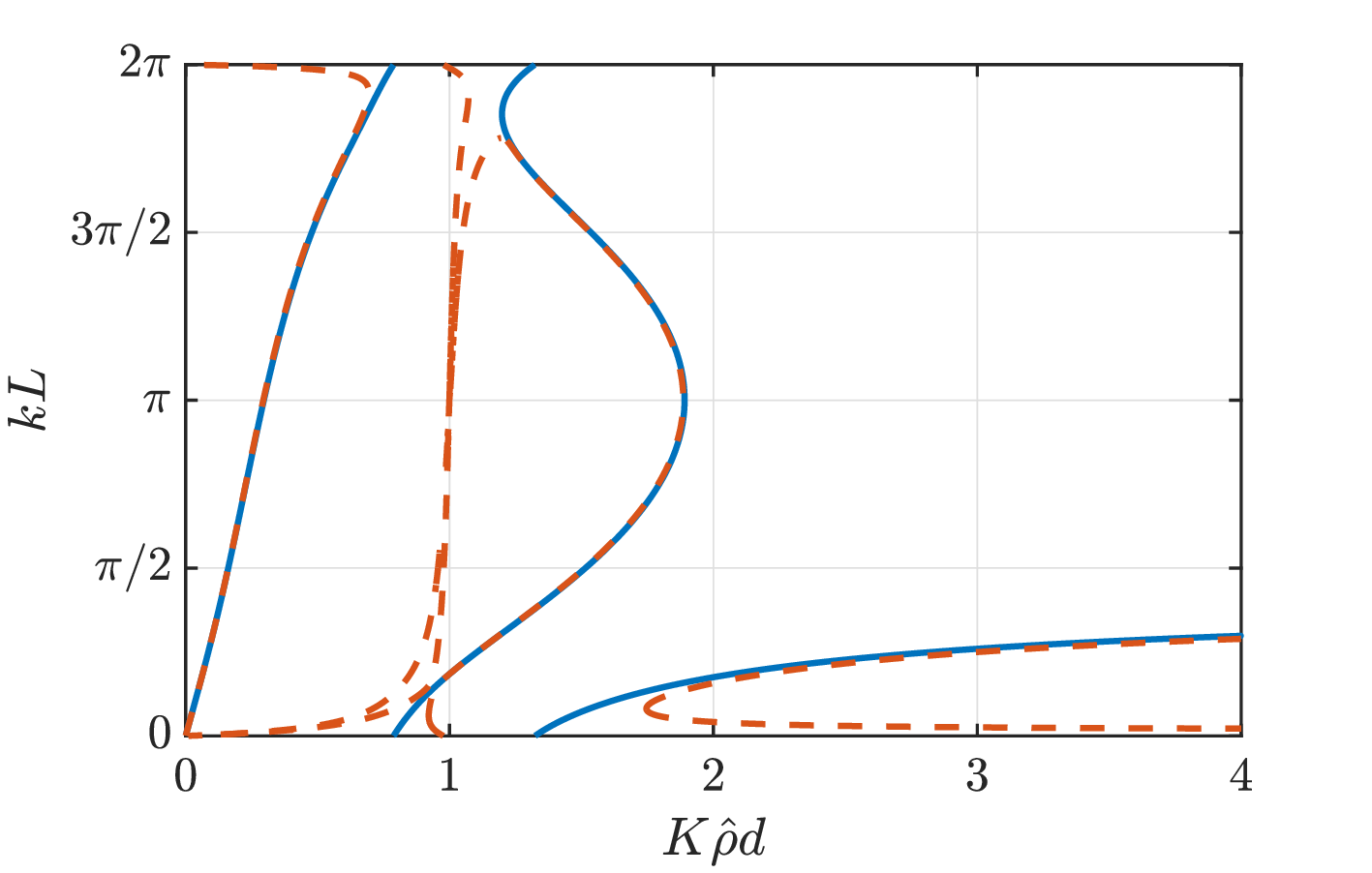}
	\end{subfigure}\hfill
	\begin{subfigure}[t]{0.03\textwidth}
		\text{(d)}
	\end{subfigure}
	\begin{subfigure}[t]{0.45\textwidth}        \centering
		\includegraphics[width=\linewidth,trim={0cm 0cm 0cm 0cm}]{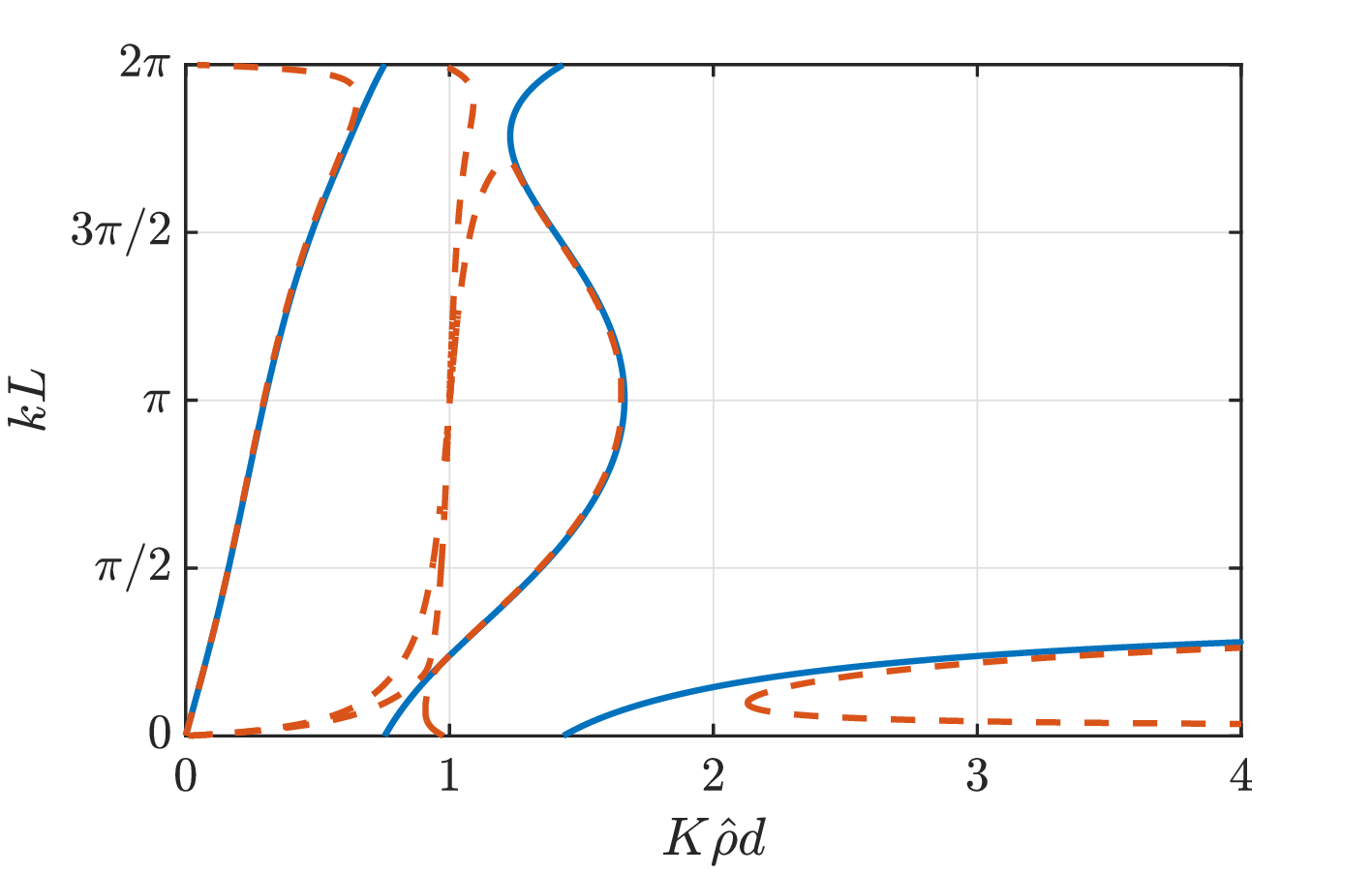}
	\end{subfigure}
	\caption{\label{fig_ar8} Dispersion curves ($kL$ versus $K \rhoh d$) for fully unconstrained motions in the case $\rhoh=0.9$, $L/d = 8 + \epsilon$ (8:1 aspect ratio floes). Exact results (blue solid curves) and small-gap approximations (orange, dashed) for gap sizes: (a) $\epsilon = 0.01$, (b) $\epsilon = 0.02$, (c) $\epsilon = 0.08$, (d) $\epsilon = 0.12$.}
\end{figure}

In Figs.~\ref{fig_ar2},~\ref{fig_ar4} and~\ref{fig_ar8} we consider the effect that changing the aspect ratio of the floes has on the results for unconstrained motions (that is, we make no detailed analysis of the constrained components as we have done for square floes). We can see that as the floes become longer, the low-frequency branch associated with pitch moves towards higher frequencies (though still below $K \hat{\rho} d = 1$). In particular, since $F^{(p,p)}$ in (\ref{eqn:4.10}) is proportional to $(d/L)^3$, pitch effects become increasingly suppressed as the aspect ratio grows. For the 4:1 aspect ratio the pitch-dominated branch is barely visible in the low-frequency regime and the dispersion diagram is dominated by the heave mode. The saddle point-like bifurcation at the intersection of heave and pitch branches that is clearly visible for square floes coalesces to form a single branch as the aspect ratio increases. This branch initially follows the heave-constrained dispersion relation before becoming a pitch-dominant mode.

\begin{figure}[!htbp]
	\centering
	\begin{subfigure}[t]{0.03\textwidth}
		\text{(a)}
	\end{subfigure}
	\begin{subfigure}[t]{0.45\textwidth}        \centering
		\includegraphics[width=\linewidth,trim={0cm 0cm 0cm 0cm}]{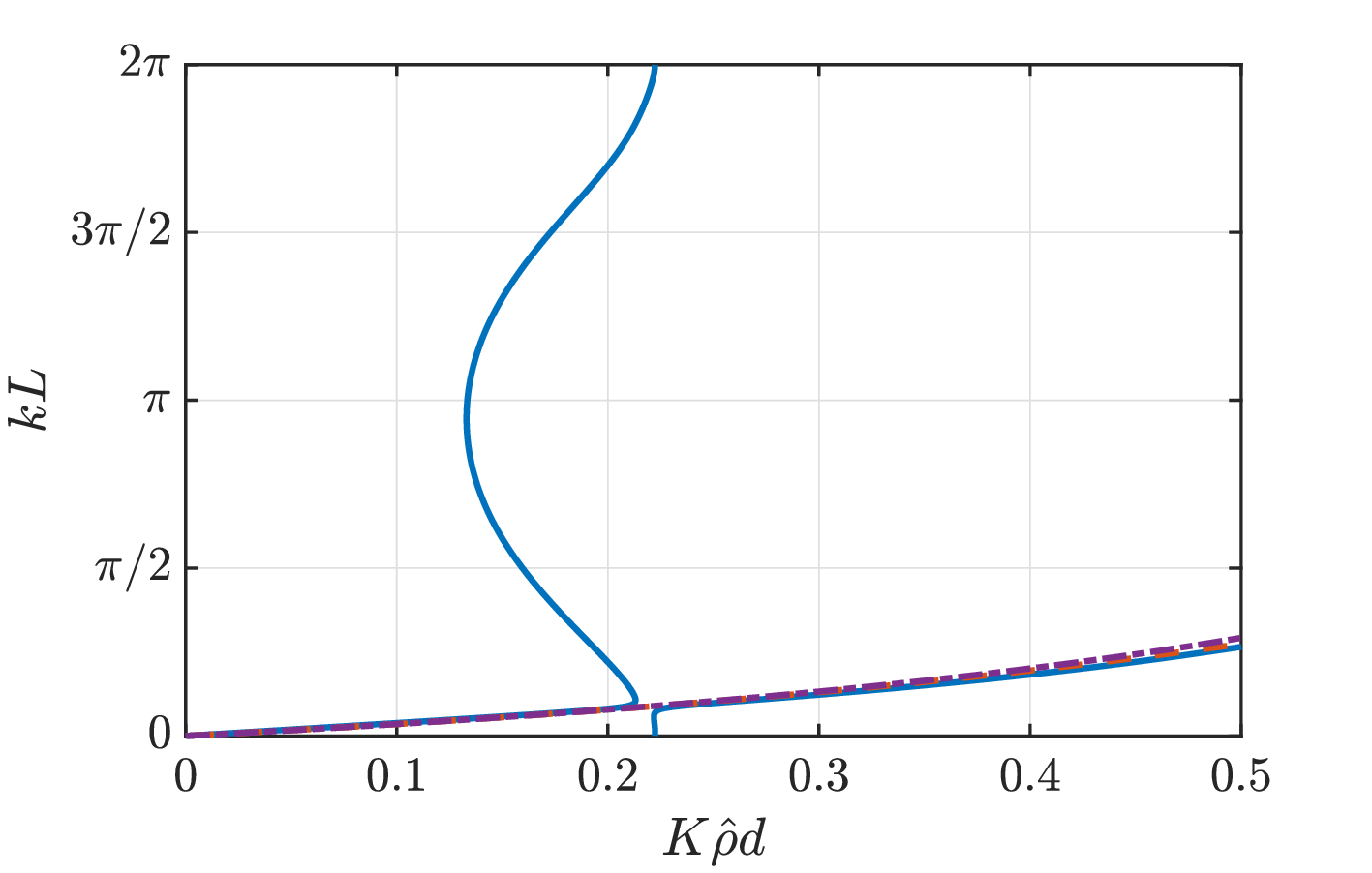}
	\end{subfigure}\hfill
	\begin{subfigure}[t]{0.03\textwidth}
		\text{(b)}
	\end{subfigure}
	\begin{subfigure}[t]{0.45\textwidth}        \centering
		\includegraphics[width=\linewidth,trim={0cm 0cm 0cm 0cm}]{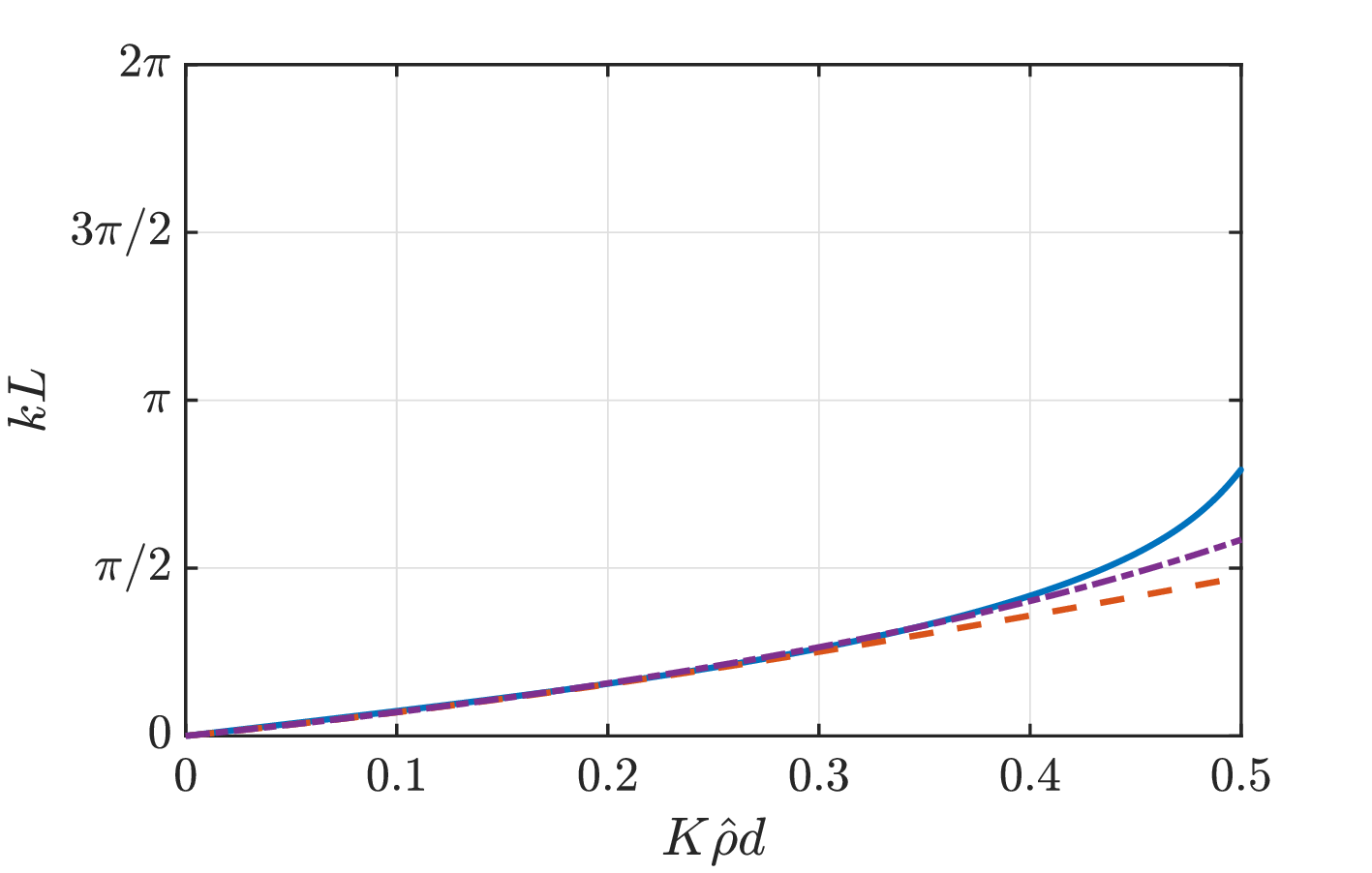}
	\end{subfigure}
	\begin{subfigure}[t]{0.03\textwidth}
		\text{(c)}
	\end{subfigure}
	\begin{subfigure}[t]{0.45\textwidth}        \centering
		\includegraphics[width=\linewidth,trim={0cm 0cm 0cm 0cm}]{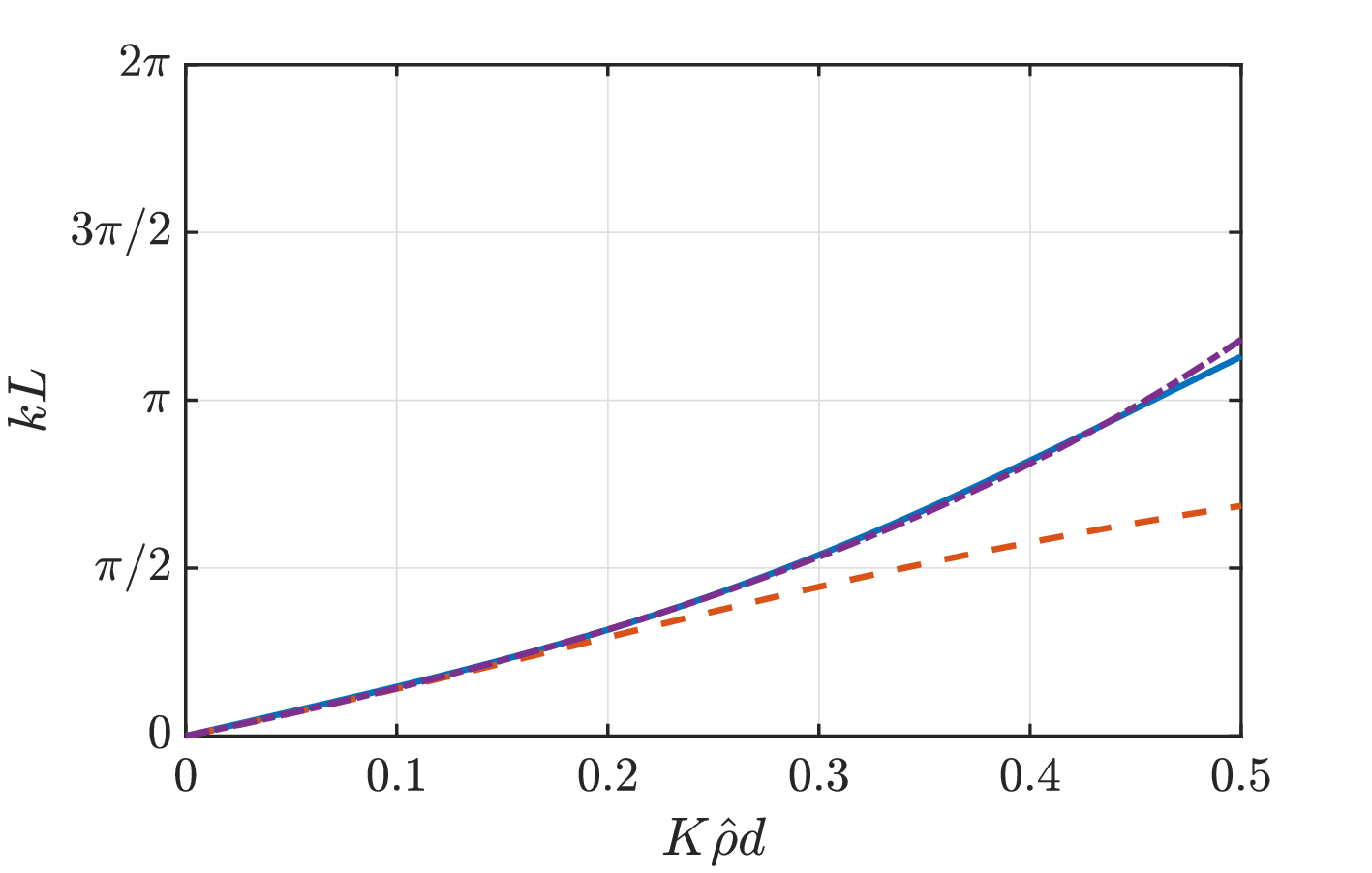}
	\end{subfigure}\hfill
	\begin{subfigure}[t]{0.03\textwidth}
		\text{(d)}
	\end{subfigure}
	\begin{subfigure}[t]{0.45\textwidth}        \centering
		\includegraphics[width=\linewidth,trim={0cm 0cm 0cm 0cm}]{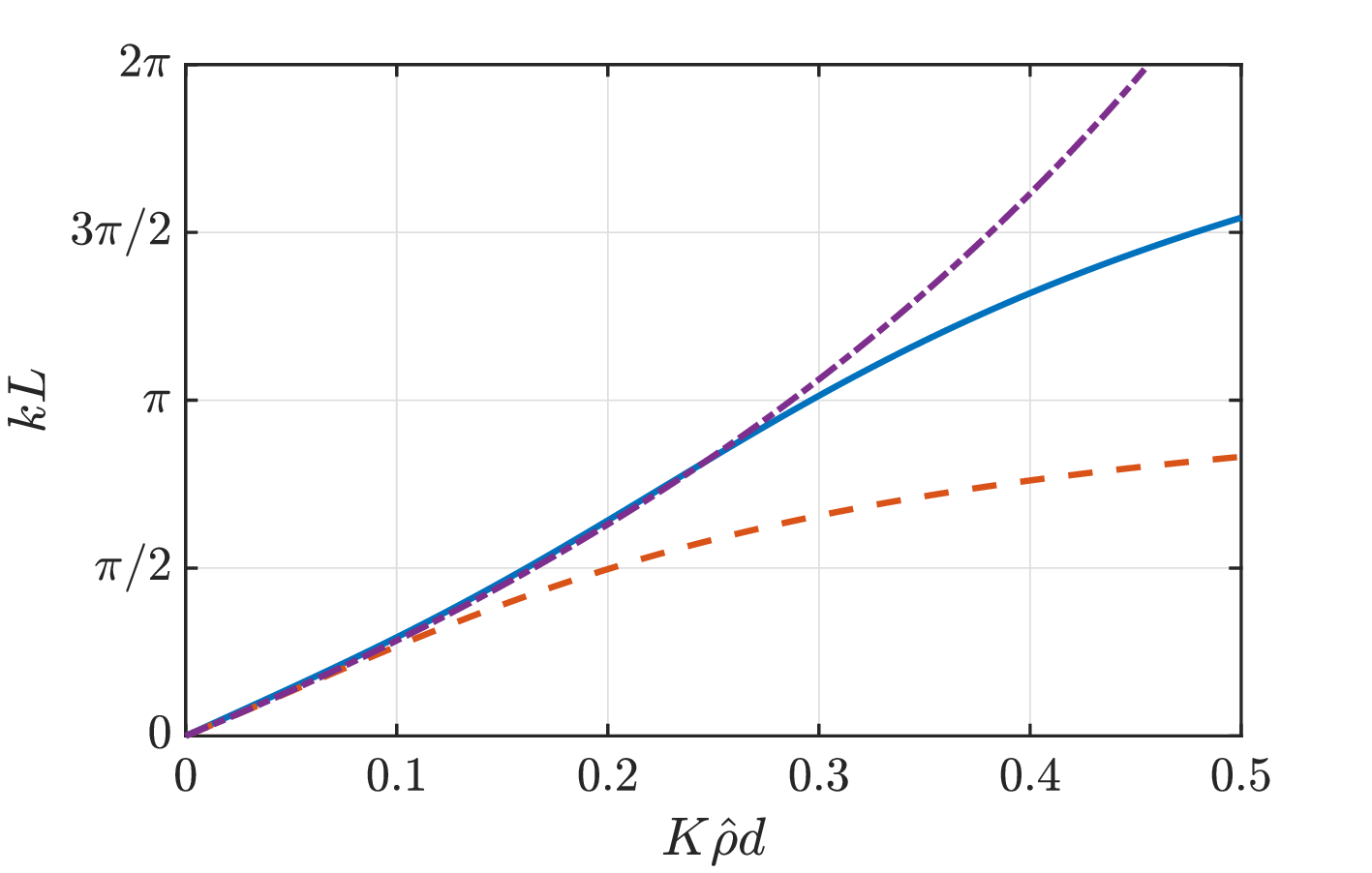}
	\end{subfigure}
	\caption{\label{fig_analytical_eps8} Low-frequency dispersion curves ($kL$ versus $K \rhoh d$ for $0 < K\rhoh d < 0.5$) for fully unconstrained motions in the case $\rhoh=0.9$ and $\epsilon = 0.08$. Exact results (blue solid curves), leading-order heave approximation (orange, dashed) and mass-loading solution of \citet{dafandpor26} (purple, dot-dashed) for aspect ratios: (a) $L/d = 1 + \epsilon$, (b) $L/d = 2 + \epsilon$, (c) $L/d = 4 + \epsilon$, (d) $L/d = 8 + \epsilon$.}
\end{figure}

In Fig.~\ref{fig_analytical_eps8} we focus on the low-frequency regime ($0 < K\rhoh d < 0.5$) for $\epsilon = 0.08$ across the four aspect ratios. Here, the leading-order heave approximation (orange, dashed) is compared alongside the mass-loading solution of \citet{dafandpor26} (purple, dot-dashed). The mass-loading dispersion relation is accurate for all values of $L/d$ considered, which is surprising given that the leading-order heave solution fails to describe it as accurately. In fact, we expect the leading-order asymptotic heave solution to fail as terms proportional to $L/d$ become dominant and terms proportional to $d/L$ get small, resulting in the leading-order behaviour of the asymptotics no longer being concretely described by the leading-order heave solution.

\begin{figure}[!htbp]
\centering
\begin{subfigure}[t]{0.03\textwidth}
\text{(a)}
\end{subfigure}
\begin{subfigure}[t]{0.45\textwidth}        \centering
\includegraphics[width=\linewidth,trim={0cm 0cm 0cm 0cm}]{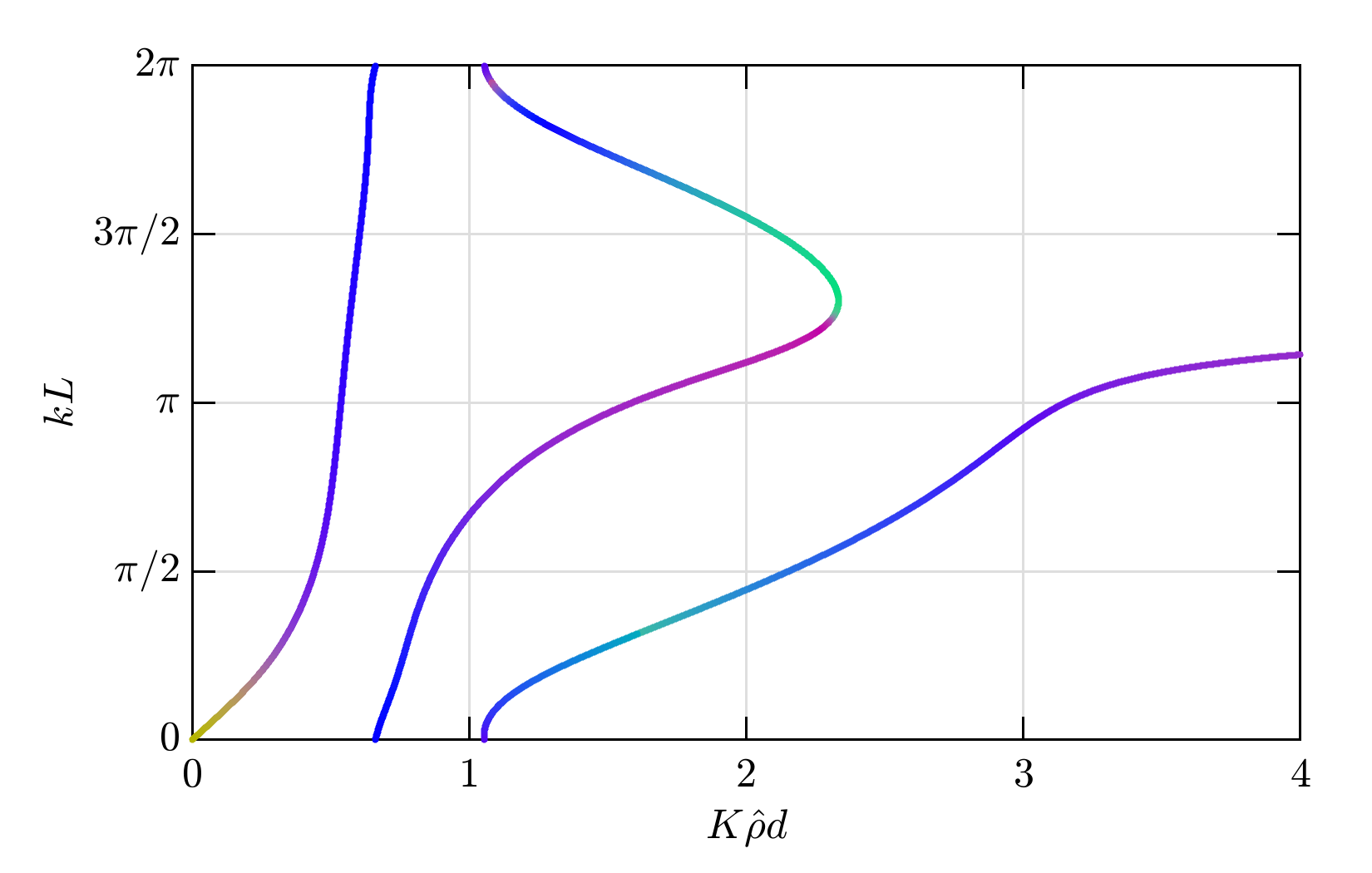}
\end{subfigure}\hfill
\begin{subfigure}[t]{0.03\textwidth}
\text{(b)}
\end{subfigure}
\begin{subfigure}[t]{0.45\textwidth}        \centering
\includegraphics[width=\linewidth,trim={0cm 0cm 0cm 0cm}]{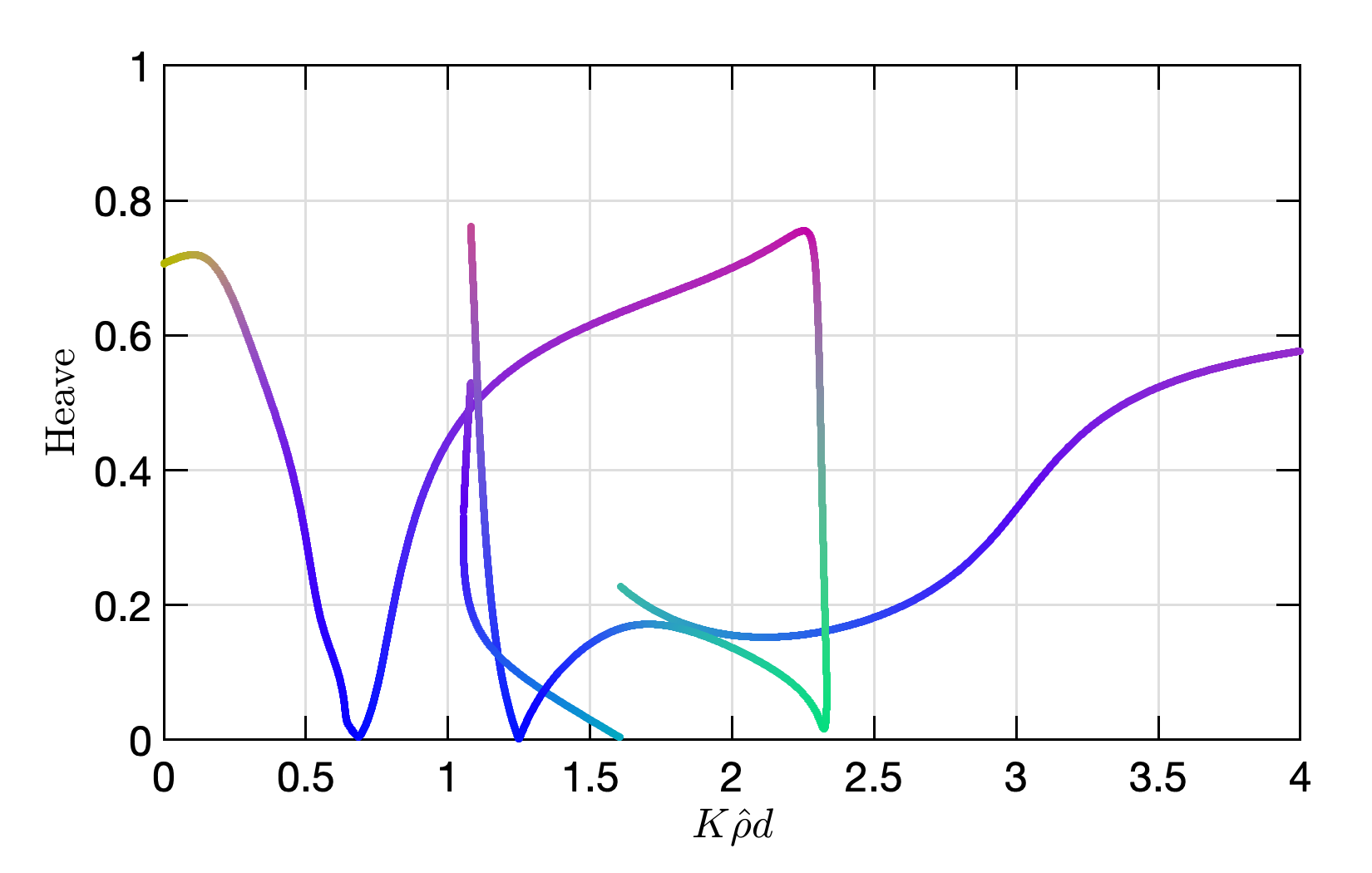}
\end{subfigure}
\begin{subfigure}[t]{0.03\textwidth}
\text{(c)}
\end{subfigure}
\begin{subfigure}[t]{0.45\textwidth}        \centering
\includegraphics[width=\linewidth,trim={0cm 0cm 0cm 0cm}]{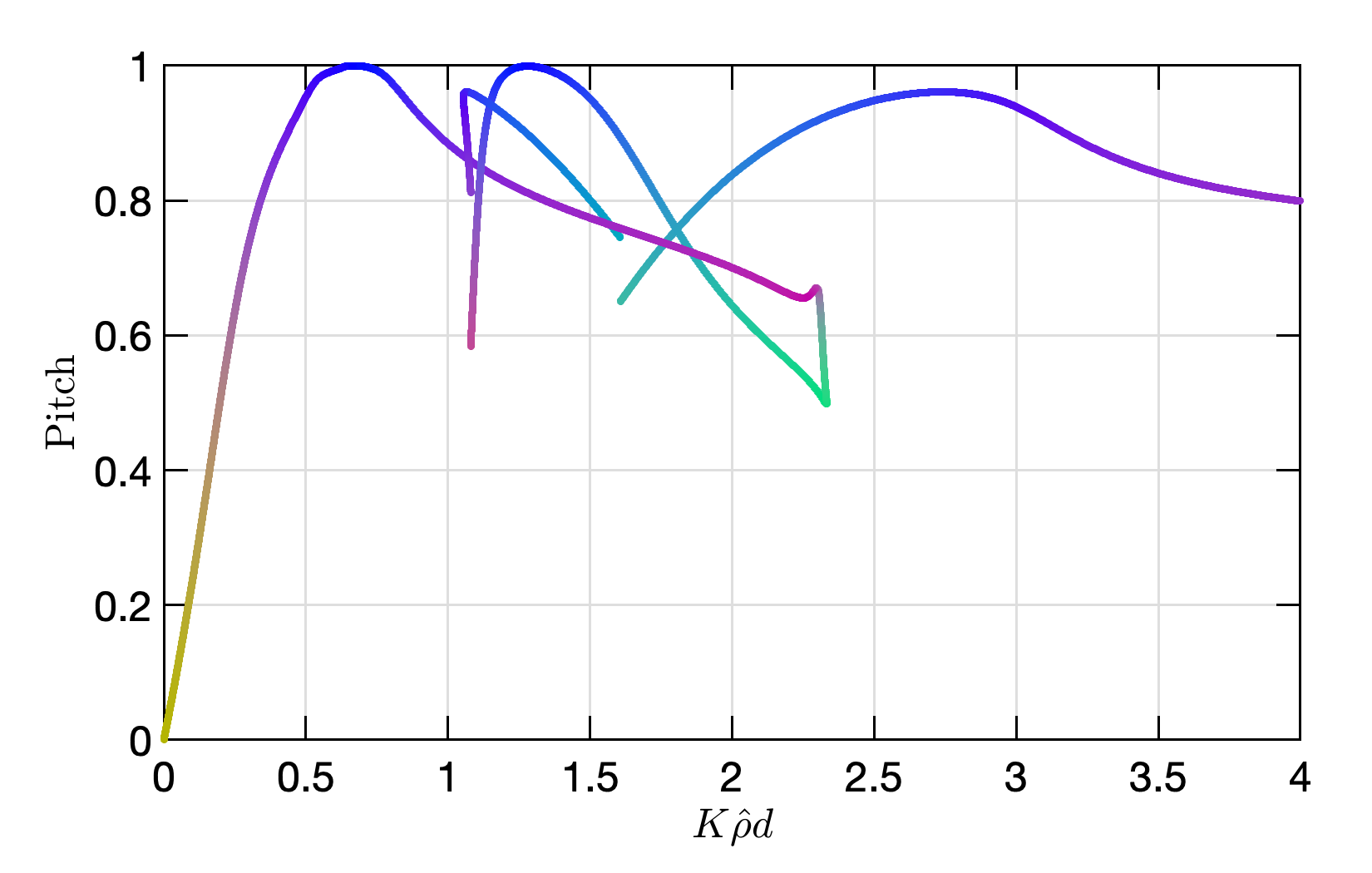}
\end{subfigure}\hfill
\begin{subfigure}[t]{0.03\textwidth}
\text{(d)}
\end{subfigure}
\begin{subfigure}[t]{0.45\textwidth}        \centering
\includegraphics[width=\linewidth,trim={0cm 0cm 0cm 0cm}]{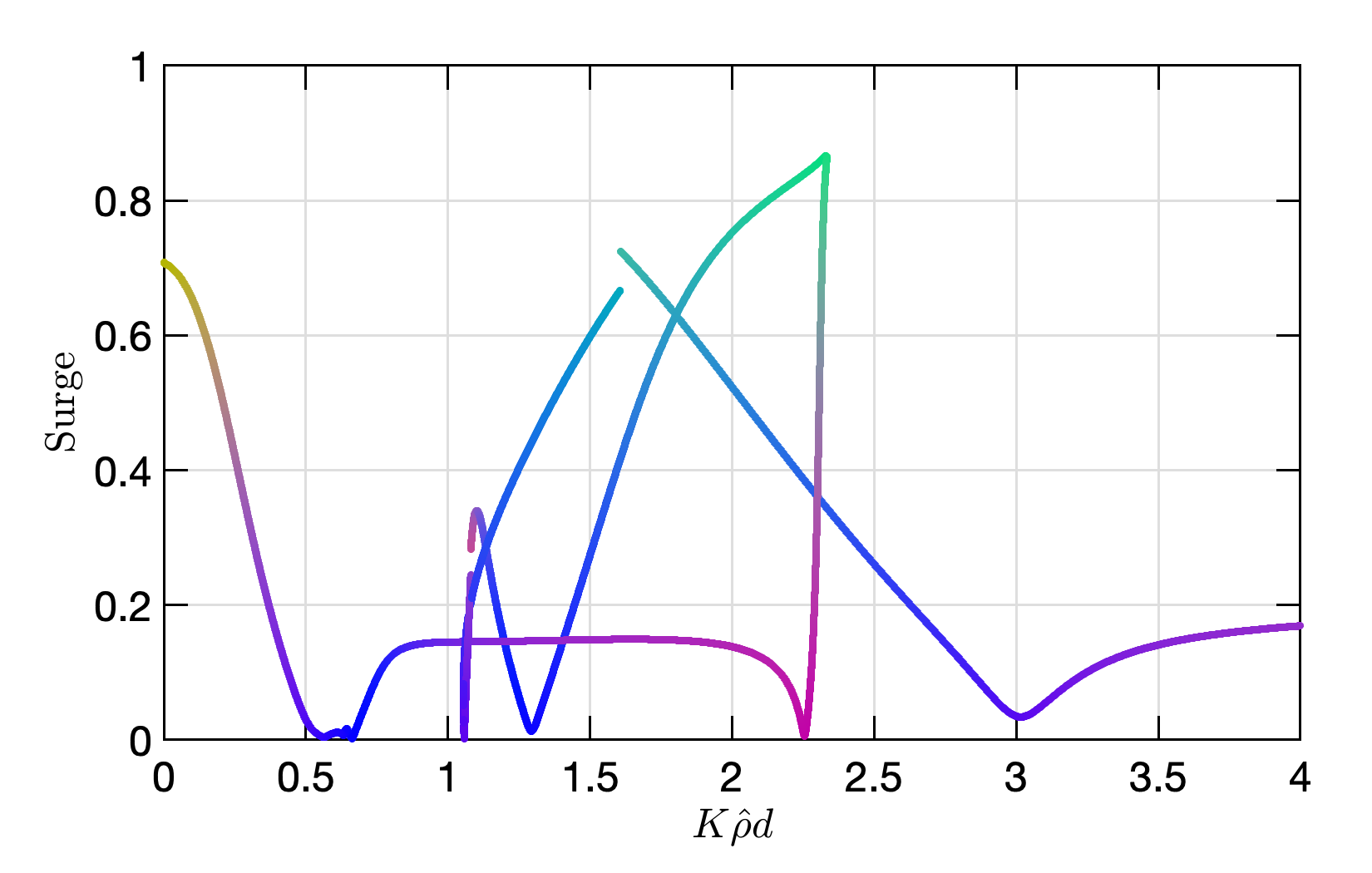}
\end{subfigure}
\caption{Colour coded plots of the heave-surge-pitch model for $\epsilon = 0.12$, $L/d = 2 + \epsilon$, $\rhoh = 0.9$. (a) Wavenumber plot, (b) absolute heave value, (c) absolute surge value, (d) absolute pitch value.\label{fig11}}
\end{figure}

\begin{figure}[!htbp]
\centering
\begin{subfigure}[t]{0.03\textwidth}
    \text{(a)}
\end{subfigure}
\begin{subfigure}[t]{0.45\textwidth}\centering
	\includegraphics[width=\linewidth,trim={4cm 4cm 4cm 4cm},clip]{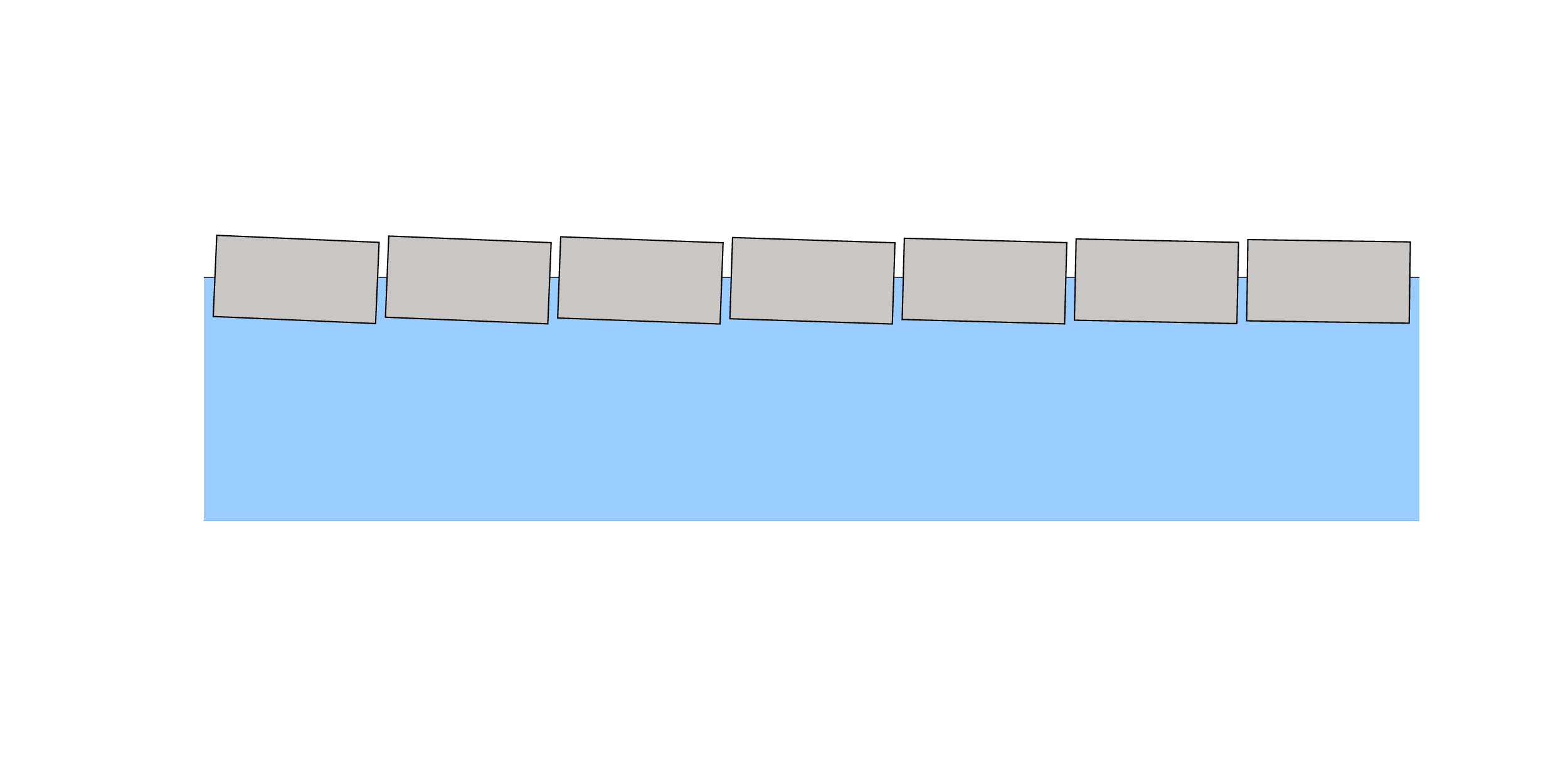}
\end{subfigure}
\begin{subfigure}[t]{0.03\textwidth}
    \text{(b)}
\end{subfigure}
\begin{subfigure}[t]{0.45\textwidth}\centering
	\includegraphics[width=\linewidth,trim={4cm 4cm 4cm 4cm},clip]{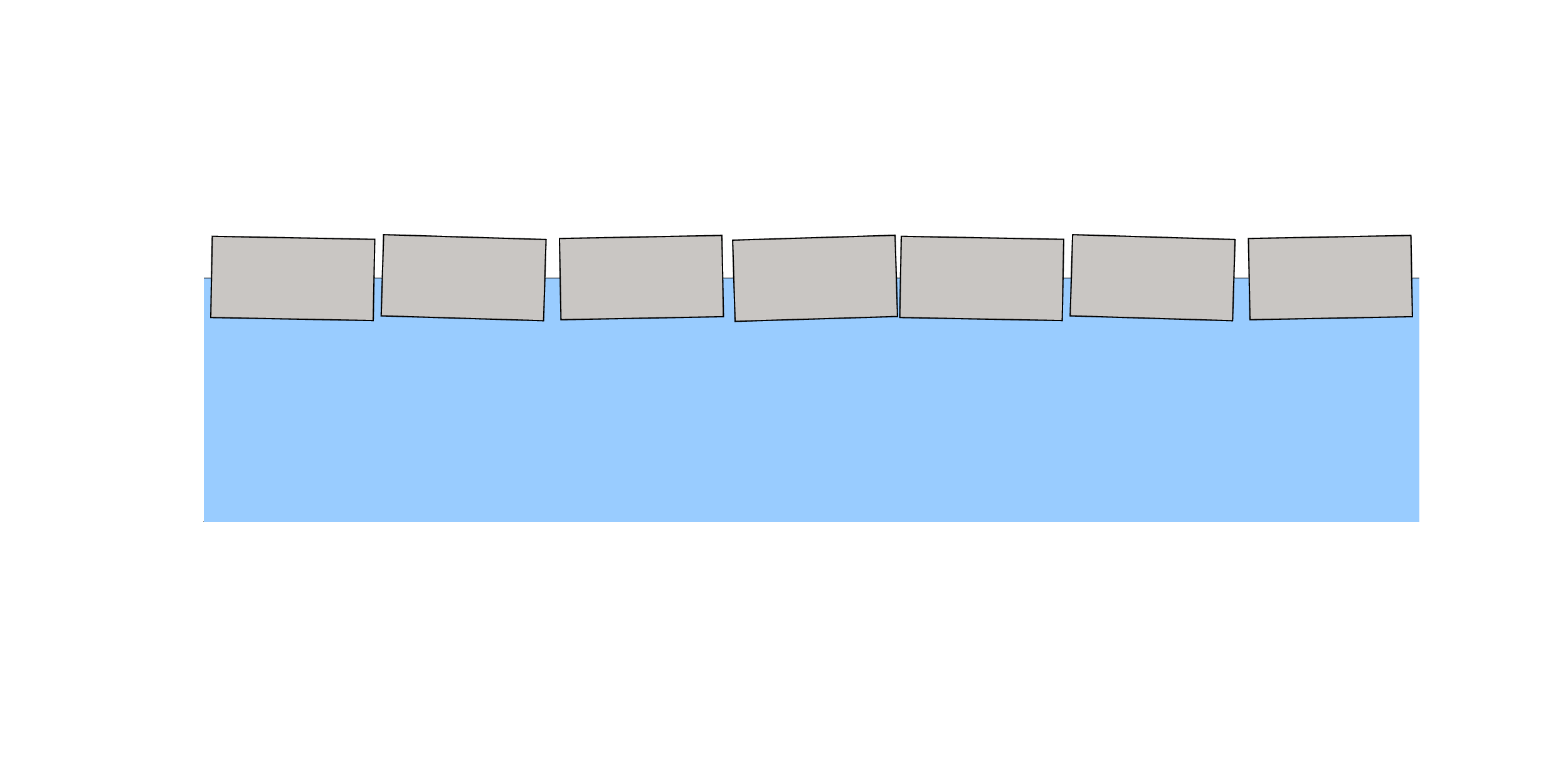}
\end{subfigure}
\\
\begin{subfigure}[t]{0.03\textwidth}
    \text{(c)}
\end{subfigure}
\begin{subfigure}[t]{0.45\textwidth}\centering
	\includegraphics[width=\linewidth,trim={4cm 4cm 4cm 4cm},clip]{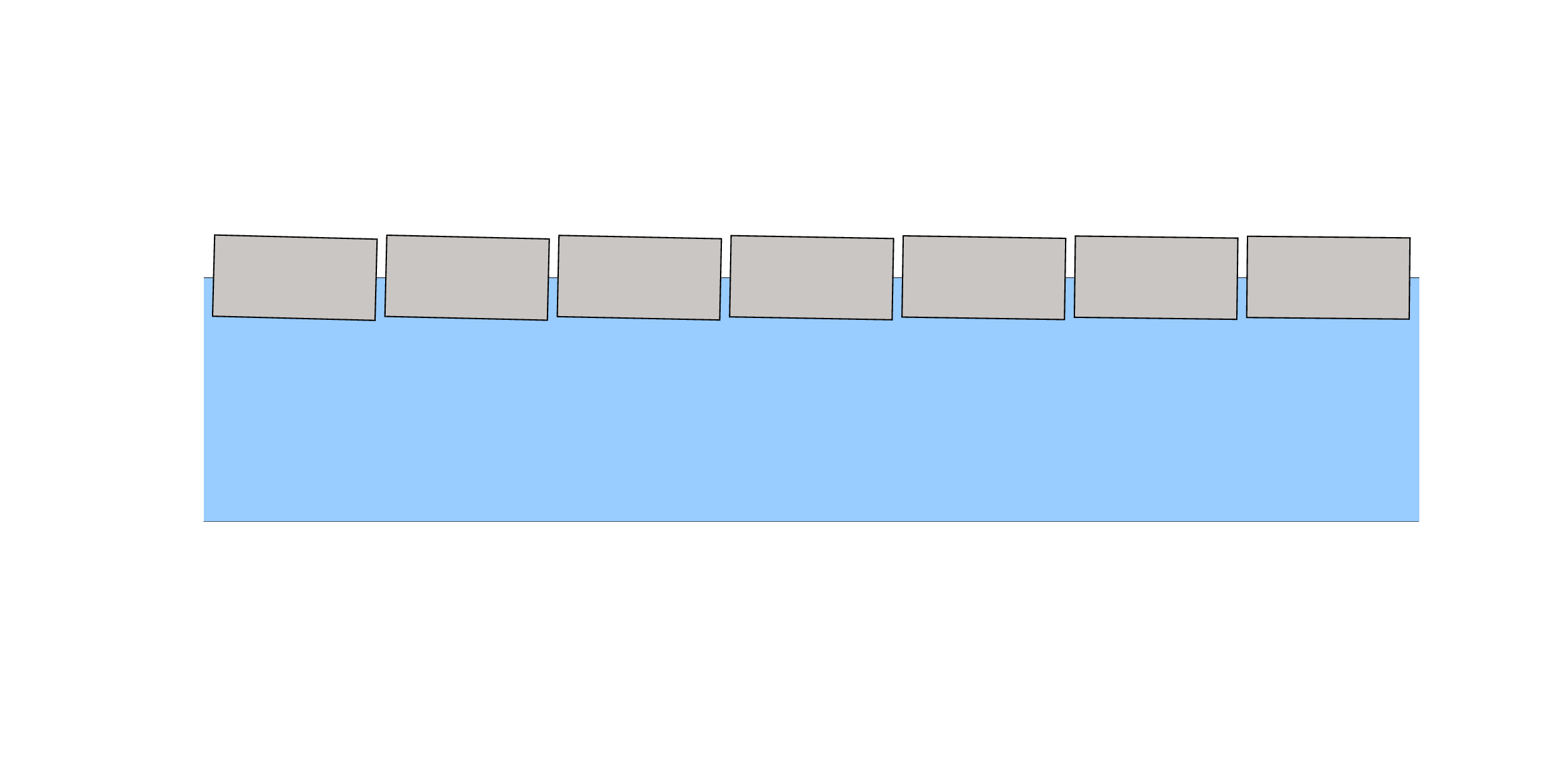}
\end{subfigure}
\begin{subfigure}[t]{0.03\textwidth}
    \text{(d)}
\end{subfigure}
\begin{subfigure}[t]{0.45\textwidth}\centering
	\includegraphics[width=\linewidth,trim={4cm 4cm 4cm 4cm},clip]{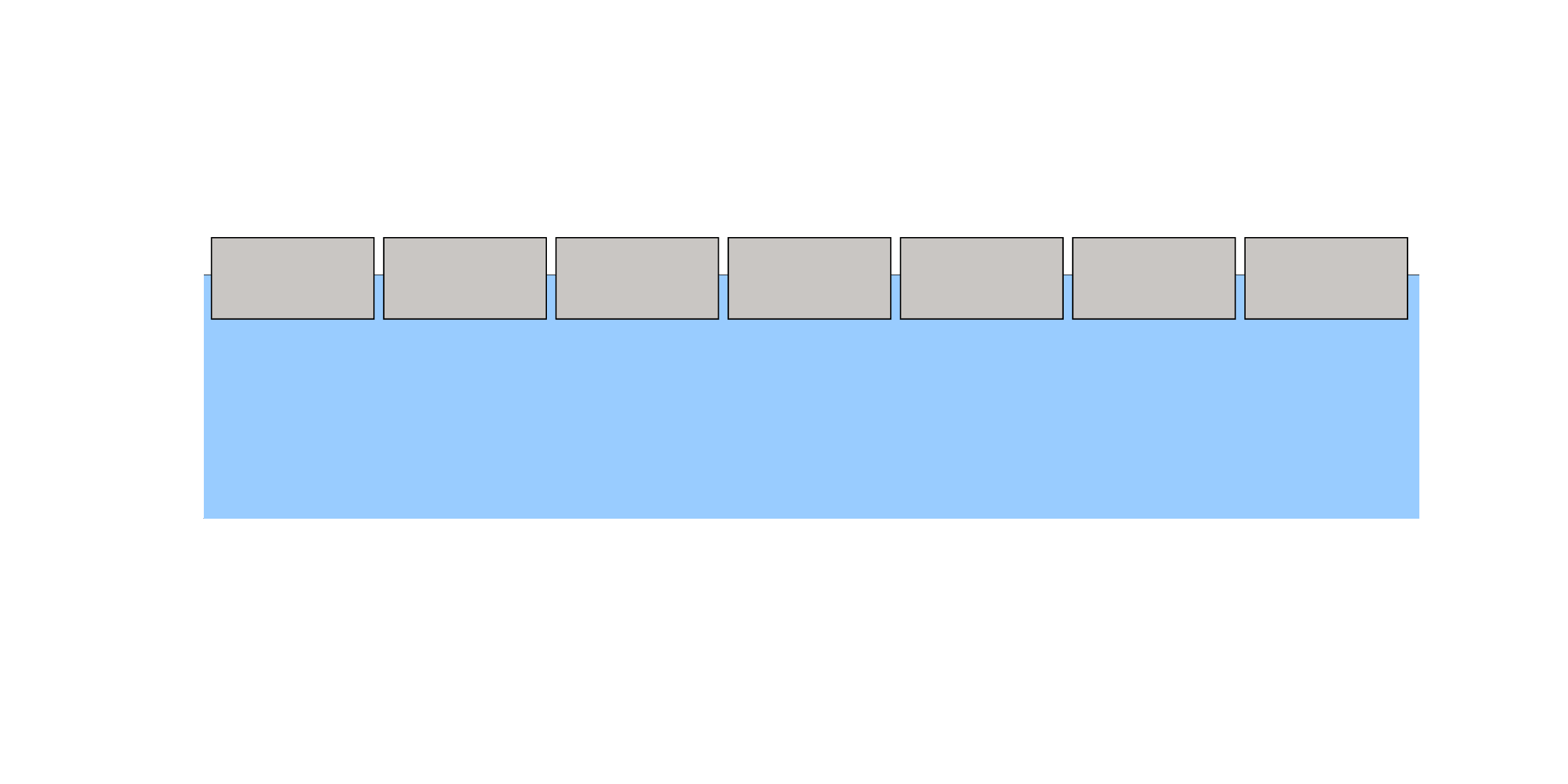}
\end{subfigure}
\\
\begin{subfigure}[t]{0.03\textwidth}
    \text{(e)}
\end{subfigure}
\begin{subfigure}[t]{0.45\textwidth}\centering
	\includegraphics[width=\linewidth,trim={4cm 4cm 4cm 4cm},clip]{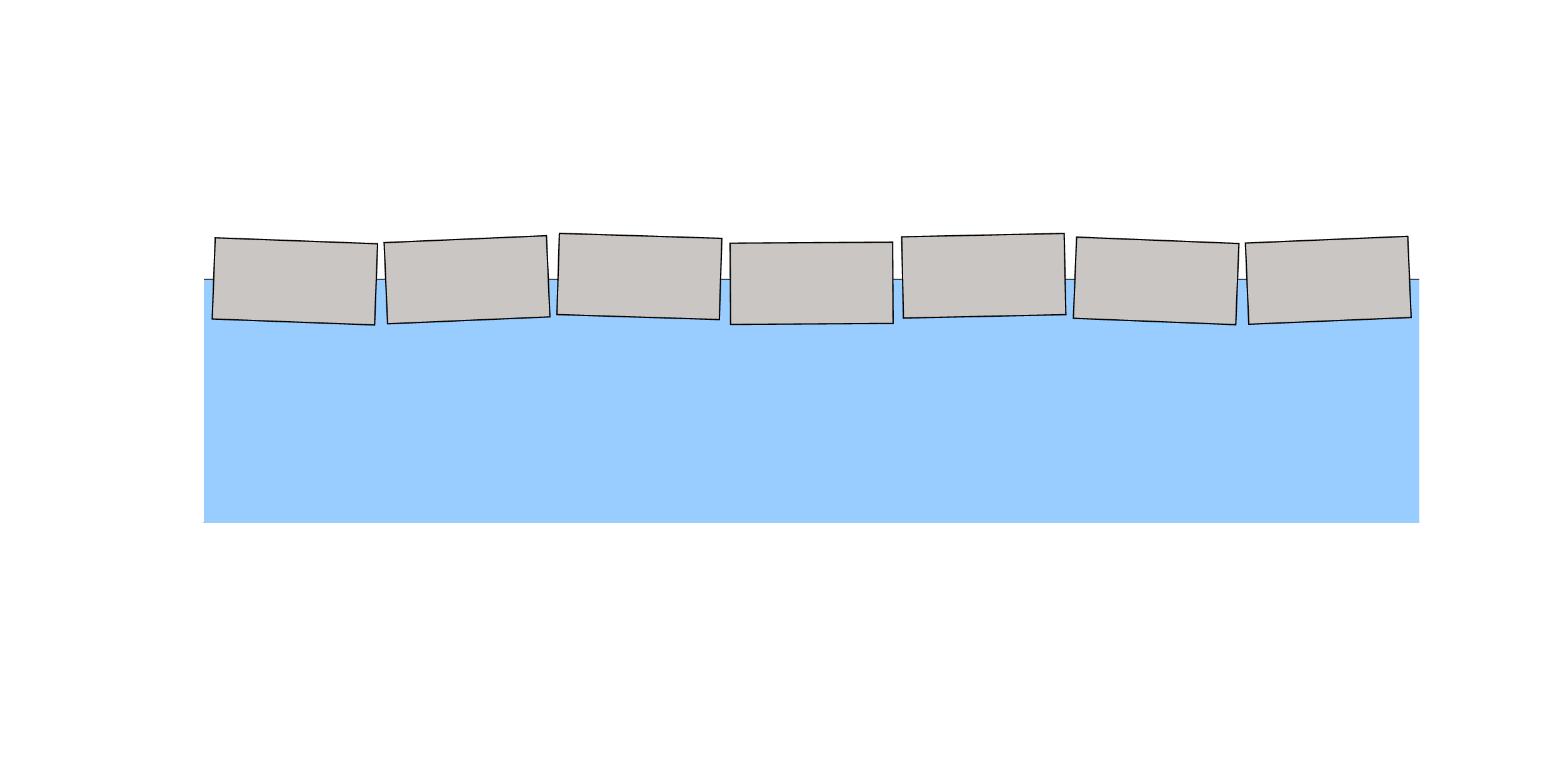}
\end{subfigure}
\begin{subfigure}[t]{0.03\textwidth}
    \text{(f)}
\end{subfigure}
\begin{subfigure}[t]{0.45\textwidth}\centering
	\includegraphics[width=\linewidth,trim={4cm 4cm 4cm 4cm},clip]{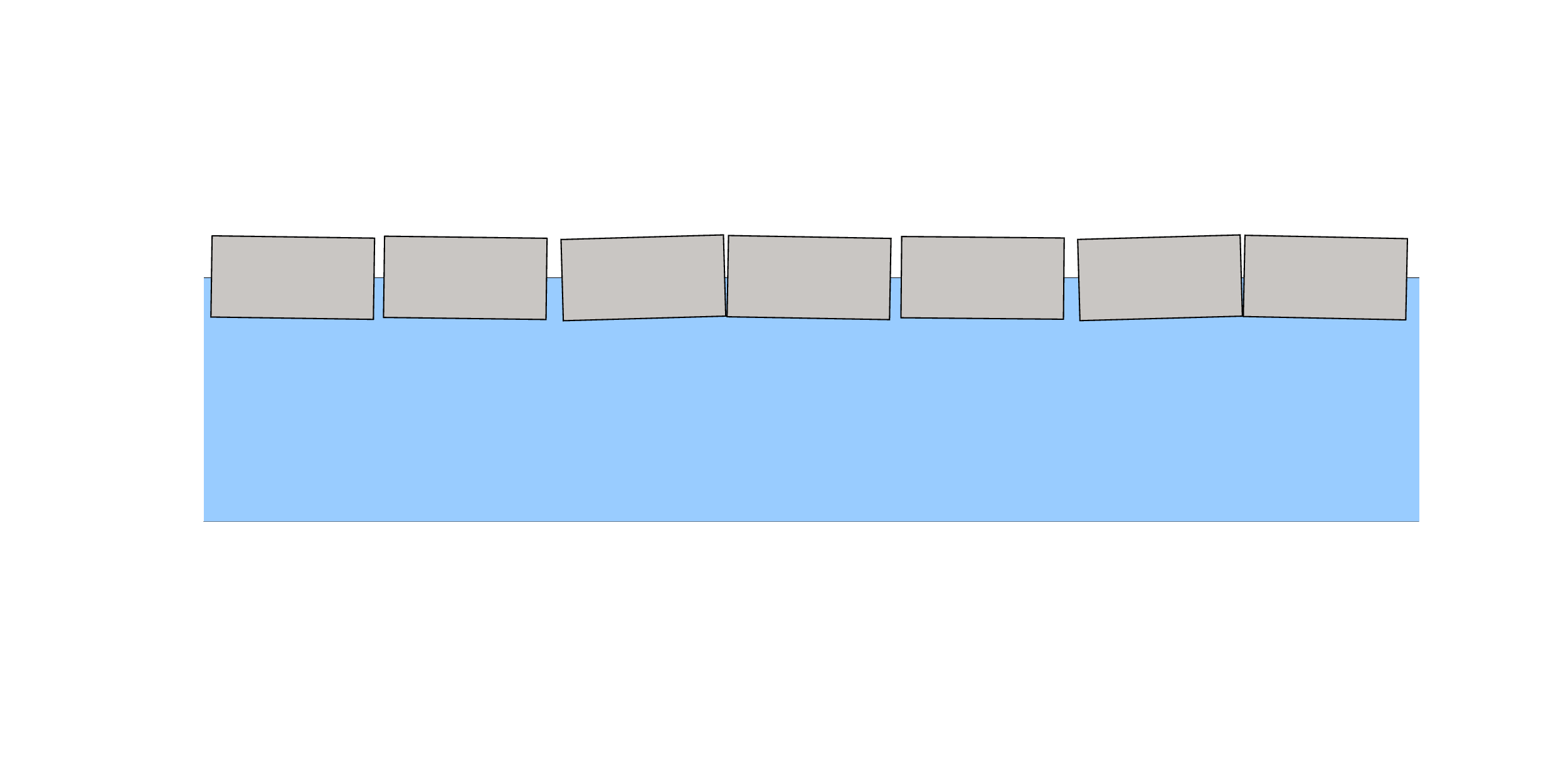}
\end{subfigure}
\\
\begin{subfigure}[t]{0.03\textwidth}
    \text{(g)}
\end{subfigure}
\begin{subfigure}[t]{0.45\textwidth}\centering
	\includegraphics[width=\linewidth,trim={4cm 4cm 4cm 4cm},clip]{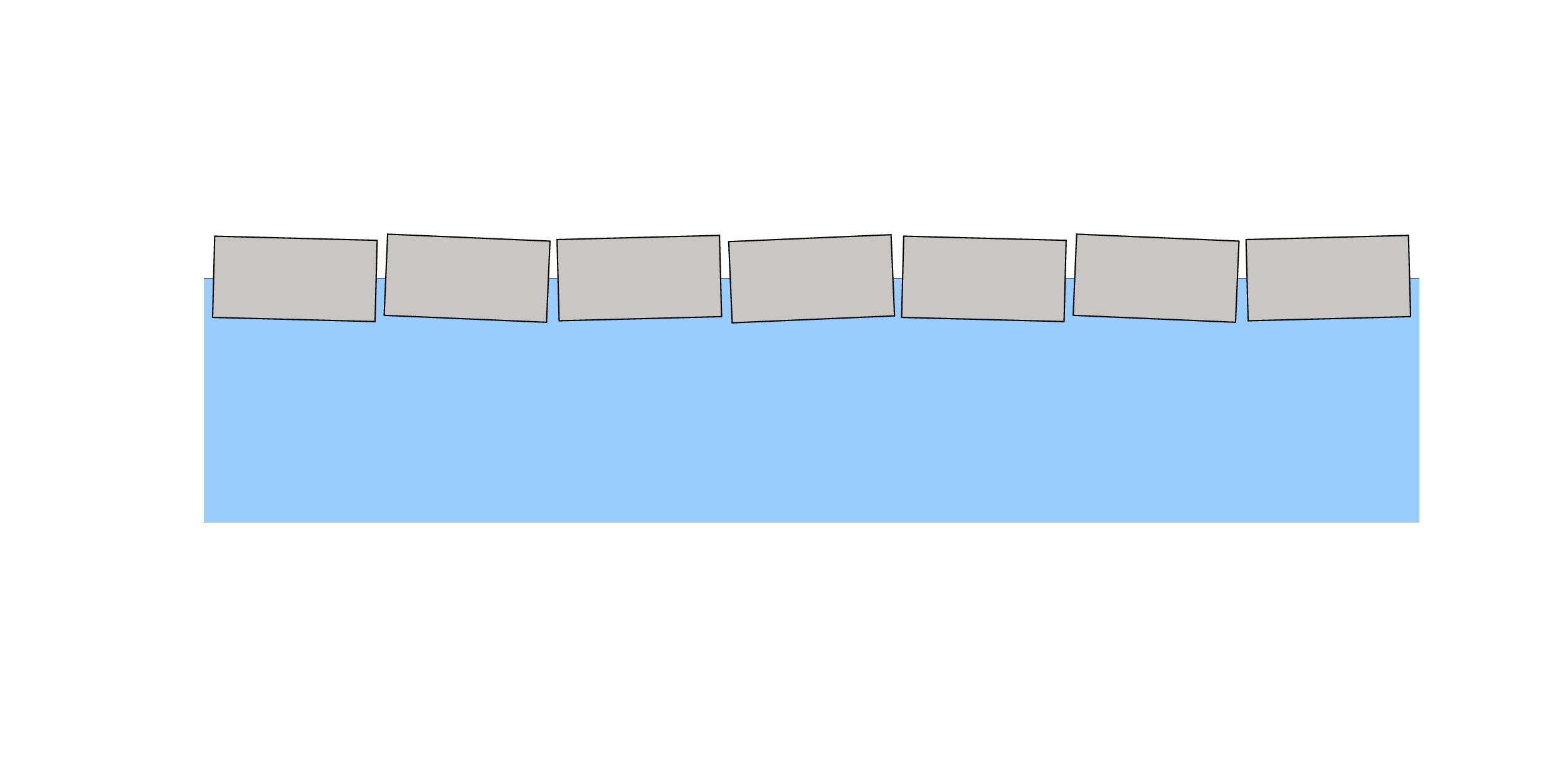}
\end{subfigure}
\begin{subfigure}[t]{0.03\textwidth}
    \text{(h)}
\end{subfigure}
\begin{subfigure}[t]{0.45\textwidth}\centering
	\includegraphics[width=\linewidth,trim={0cm 0cm 0cm 0cm}]{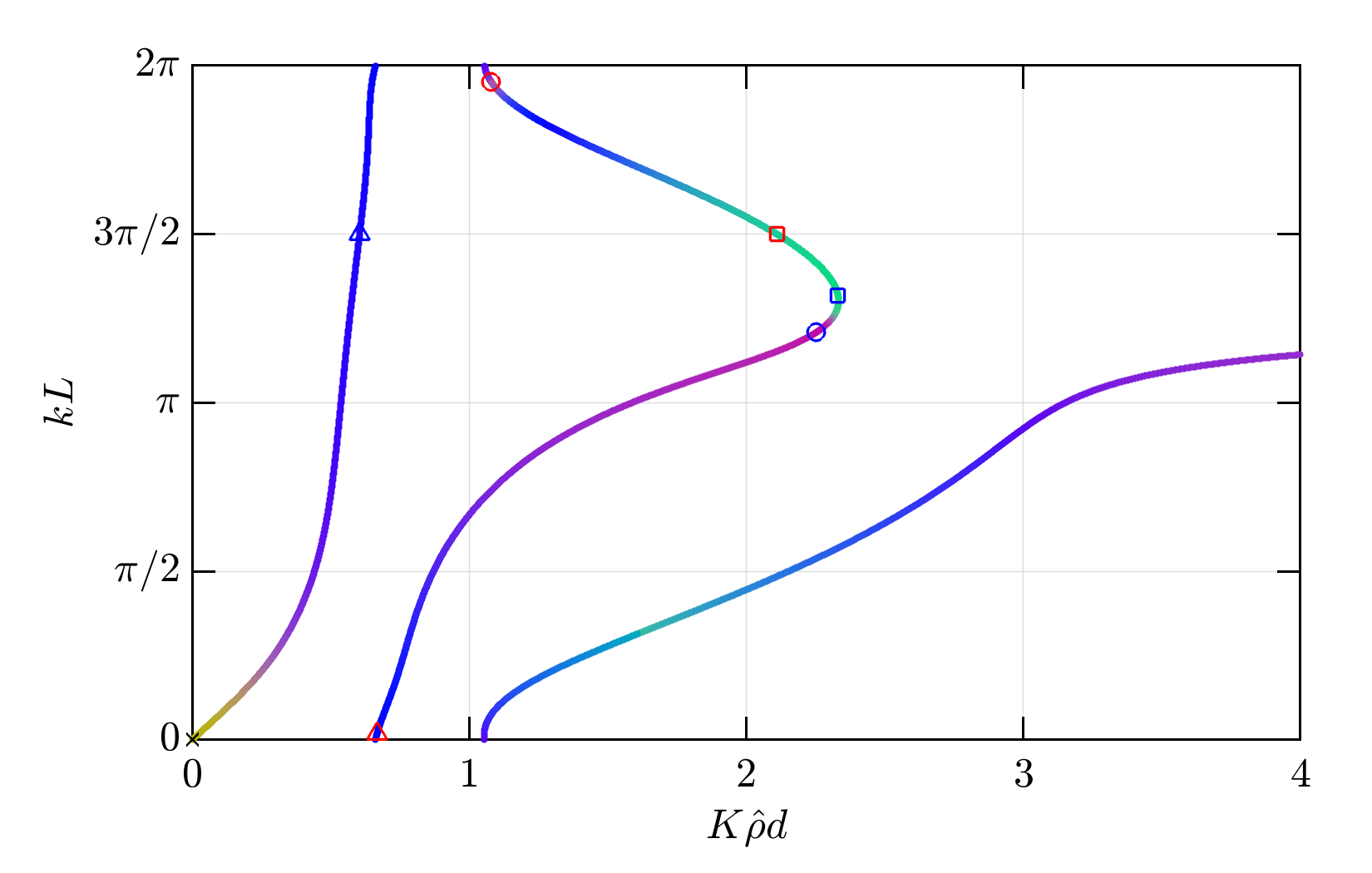}
\end{subfigure}
\caption{Representative snapshots at time $t=2\pi/3$ from the seven animations together with the scatter plot that locates each snapshot in $(K\rhoh d,\,kL)$ space. (a), (b), (c): the most heave-, surge- and pitch-dominant in-phase modes. (d): the point closest to the origin. (e), (f), (g): the most heave-, surge- and pitch-dominant out-of-phase modes. (h) the scatter plot showing each point's spectral coordinates where blue represents in-phase, red represents out-of-phase and circles, squares and triangles represent heave-, surge- and pitch-dominant modes respectively. The black cross denotes the point chosen near the origin.\label{fig12}}
\end{figure}

Here we find Fig.~\ref{fig11} useful. In Fig.~\ref{fig11}(a) we have
reproduced the dispersion diagram of Fig.~\ref{fig_ar2} corresponding to $\epsilon = 0.12$ using R-G-B colour
coding to represent the relative amplitude of (respectively) the
heave-surge-pitch motions. The normalised amplitudes of each mode are
displayed on individual plots in Figs.~\ref{fig11}(b,c,d) with the same
colour coding. Thus, we can determine that the low-frequency
dispersion curve which approximately follows the heave-constrained curve for
small $K \hat{\rho} d$ is not simply heave motion, but a roughly
equal combination of heave and surge motion. Furthermore, these
two motions are a quarter of a period out of phase with one another
and thus the motion along this low-frequency branch is actually
circular floe motion, even though the dispersion relation is
approximately that satisfied by constrained heave motion.

Further along this branch we see that pitch becomes increasingly dominant.
At a particular frequency corresponding to $kL = 2 \pi$ the motion is
entirely pitch; each floe is now pitching in phase with its neighbours at a
frequency some way below the resonant frequency of the fluid in the narrow
channel.
We can also see that for an interval of frequency, just above
$K \hat{\rho} d = 1$ corresponding to fluid resonance and extending to
roughly $K \hat{\rho} d = 2.3$,
three possible propagating wave modes are now possible.
Each one has its own frequency-dependent blend of heave, surge and
pitch components.

The solution landscape is evidently remarkably rich and complicated
and hard to understand physically. To help in developing some
understanding, in Fig.~\ref{fig12}(a--g) we have plotted
sample snapshots in time of the floe motion corresponding to
various points in the corresponding dispersion diagram.
Much more useful is the online tool (see \citet{dafydd_fff_2026}) that we have
developed which provides time-varying simulations of fluid/floe
motion and the user can pick
any point they choose within the dispersion diagram, not just the
seven used to produce Fig.~\ref{fig12}.

\section{Conclusions}\label{sec:conclusions}

The central question of this paper was to determine whether the heave-only, zero-gap model used by \citet{dafandpor24,dafandpor26} provides a reasonable approximation to wave propagation through an array of freely-floating floes separated by small gaps. We address this question here.

In the low-frequency regime, which is the range of primary interest for the application to wave propagation through broken ice, the heave-constrained zero-gap dispersion relation does indeed provide a good approximation to the dispersion relation for freely-floating floes with small gaps. For all cases considered in this paper, the dominant low-frequency branch extending from $(kL, K\rhoh d) = (0,0)$ in the fully unconstrained results (Fig.~\ref{fig8}) follows very closely the heave-constrained mass-loading curve of \citet{dafandpor26}. The close agreement persists for floes with larger aspect ratios (Figs.~\ref{fig_ar2},~\ref{fig_ar4},~\ref{fig_ar8}).

However, the situation is more nuanced than the dispersion relation alone suggests. An examination of the modal amplitudes (Fig.~\ref{fig11}) reveals that along this dominant low-frequency branch, the floe motion is not simply heave: it is a roughly equal combination of heave and surge with a quarter-period phase shift between the two components, producing a circular floe motion. Thus, even though the dispersion relation is well approximated by the heave-only model, the actual floe kinematics are fundamentally different.

A further complication arises from the pitch mode. For floes of small aspect ratio (we have shown results for square floes, or $d/(L-\ell) = 1$), a low-frequency pitch-dominated branch of solutions co-exists alongside the heave-dominated curve (see Figs.~\ref{fig5},~\ref{fig8}). Since $F^{(p,p)}$ in (\ref{eqn:4.10}) is proportional to $(d/L)^3$, pitch effects are suppressed as the aspect ratio grows. For floes with aspect ratio 4:1, the pitch branch is shifted to higher frequencies and is far less prominent. However, for the small aspect-ratio floes assumed in \citet{dafandpor24,dafandpor26}, the pitch-dominated branch lies firmly within the low-frequency regime and represents an additional mode of wave propagation not captured by the heave-only model. It is unclear at this stage how multi-modal wave propagation would be incorporated into the ensemble-averaged multiple-scattering framework of \citet{dafandpor24,dafandpor26}, but it could potentially alter their theoretical prediction of wave attenuation. It should be noted that the leading order heave solution of Fig.~\ref{fig2} is shown to be increasingly poor at modeling the low-frequency branch of in unconstrained system in Fig.~\ref{fig12} as aspect ratios increase due to $d/L$ becoming $O(\epsilon)$ or smaller and changing the balance of (\ref{eqn:2.15}), but the mass-loading model remains surprisingly accurate throughout.

The small-gap asymptotic formulae developed in this paper provide computationally efficient approximations to the dispersion relation. Away from the narrow-channel resonance ($K\rhoh d \not\approx 1$) and for $kL$ not close to $0$ or $2\pi$, the leading-order expressions reproduce the numerical results well and reduce to the heave-constrained result at leading order when the motions are coupled. Close to resonance, the asymptotic expansion breaks down and the full numerical solution is required to capture the strong modal interactions.

Possible extensions of this work include the introduction of finite-depth effects, which will alter the positions and widths of the channel resonances, and the extension to three-dimensional floes to assess the role of roll and other out-of-plane motions on the modal structure.

\bibliography{references}

@article{allaire92,
  author  = {Allaire, A.},
  title   = {Homogenization and two-scale convergence},
  journal = {SIAM J. Math. Anal.},
  volume  = {23},
  pages   = {1482--1518},
  year    = {1992},
}

@article{carandmci13,
  author  = {Carter, B. G. and McIver, P.},
  title   = {Water-wave propagation through an infinite array of floating structures},
  journal = {J. Eng. Math.},
  volume  = {81},
  pages   = {9--45},
  year    = {2013},
}

@article{chou98,
  author  = {Chou, T.},
  title   = {Band structure of surface flexural-gravity waves along periodic interfaces},
  journal = {J. Fluid Mech.},
  volume  = {369},
  pages   = {333--350},
  year    = {1998},
}

@unpublished{dafandpor26,
  author = {Dafydd, L. and Porter, R.},
  title  = {On the attenuation of waves through broken ice of randomly-varying thickness on water of finite depth},
  year   = {2026},
  note   = {Submitted for publication},
}

@article{dafandpor24,
  author  = {Dafydd, L. and Porter, R.},
  title   = {Attenuation of long waves through regions of irregular floating ice and bathymetry},
  journal = {J. Fluid Mech.},
  volume  = {996},
  pages   = {A43},
  year    = {2024},
}

@manual{dafydd_fff_2026,
  author = {Dafydd, Lloyd},
  title = {Wave-floe Response Visualiser},
  year = {2026},
  url = {https://lloyddafydd.github.io/fffWebapp/}
}

@article{davandhea84,
  author  = {Davies, A. G. and Heathershaw, A. D.},
  title   = {Surface-wave propagation over sinusoidally varying topography},
  journal = {J. Fluid Mech.},
  volume  = {144},
  pages   = {419--443},
  year    = {1984},
}

@article{diaandvid14,
  author  = {Dias, G. S. and Videman, J. H.},
  title   = {Trapped modes along a periodic array of freely floating obstacles},
  journal = {Math. Meth. Appl. Sci.},
  volume  = {38},
  pages   = {4038--4051},
  year    = {2014},
}

@article{dobetal15,
  author  = {Doble, M. J. and De Carolis, G. and Meylan, M. H. and Bidlot, J.-R. and Wadhams, P.},
  title   = {Relating wave attenuation to pancake ice thickness, using field measurements and model results},
  journal = {Geophys. Res. Lett.},
  volume  = {42},
  pages   = {4473--4481},
  year    = {2015},
}

@article{evaandfer95,
  author  = {Evans, D. V. and Fernyhough, M.},
  title   = {Edge waves along periodic coastlines. Part 2},
  journal = {J. Fluid Mech.},
  volume  = {297},
  pages   = {307--325},
  year    = {1995},
}

@book{graandryz81,
  author    = {Gradshteyn, I. S. and Ryzhik, I. M.},
  title     = {Table of integrals, series, and products},
  publisher = {Academic Press},
  address   = {Waltham, MA},
  year      = {1981},
}

@article{hosetal20,
  author  = {Ho\v{s}ekov\'a, L. and Malila, M. P. and Rogers, W. E. and Roach, L. A. and Eidam, E. and Rainville, L.},
  title   = {Attenuation of ocean surface waves in pancake and frazil sea ice along the coast of the Chukchi Sea},
  journal = {J. Geophys. Res.: Oceans},
  volume  = {125},
  pages   = {e2020JC016746},
  year    = {2020},
}

@article{huaandpor25,
  author  = {Huang, J. and Porter, R.},
  title   = {Water wave propagation through arrays of closely spaced surface-piercing vertical barriers},
  journal = {J. Fluid Mech.},
  volume  = {960},
  pages   = {A20},
  year    = {2023},
}

@article{lietal17,
  author  = {Li, J. and Kohout, A. L. and Doble, M. J. and Wadhams, P. and Guan, C. and Shen, H. H.},
  title   = {Rollover of apparent wave attenuation in ice covered seas},
  journal = {J. Geophys. Res.: Oceans},
  volume  = {122},
  pages   = {8557--8566},
  year    = {2018},
}

@article{linton11,
  author  = {Linton, C. M.},
  title   = {Water waves over arrays of horizontal cylinders: band gaps and Bragg resonance},
  journal = {J. Fluid Mech.},
  volume  = {670},
  pages   = {504--526},
  year    = {2011},
}

@book{newman77,
  author    = {Newman, J. N.},
  title     = {Marine Hydrodynamics},
  publisher = {MIT Press},
  address   = {Cambridge, MA},
  year      = {1977},
}

@book{meietal05,
  author    = {Mei, C. C. and Stiassnie, M. and Yu, D. K. P.},
  title     = {Theory and application of ocean surface waves: Part I},
  publisher = {Wiley Interscience},
  address   = {New York},
  year      = {2005},
}

@article{meyetal14,
  author  = {Meylan, M. H. and Bennetts, L. G. and Kohout, A. L.},
  title   = {In situ measurements and analysis of ocean waves in the Antarctic marginal ice zone},
  journal = {Geophys. Res. Lett.},
  volume  = {41},
  number  = {14},
  pages   = {5046--5051},
  year    = {2014},
}

@article{meyetal15,
  author  = {Meylan, M. H. and Yiew, L. J. and Bennetts, L. G. and French, B. J. and Thomas, G. A.},
  title   = {Surge motion of an ice floe in waves: comparison of a theoretical and an experimental model},
  journal = {Ann. Glaciol.},
  volume  = {56},
  number  = {69},
  pages   = {155--159},
  year    = {2015},
}

@article{meyetal18,
  author  = {Meylan, M. H. and Bennetts, L. G. and Mosig, J. E. M. and Rogers, W. E. and Doble, M. J. and Peter, M. A.},
  title   = {Dispersion relations, power laws, and energy loss for waves in the marginal ice zone},
  journal = {J. Geophys. Res.: Oceans},
  volume  = {123},
  pages   = {3322--3335},
  year    = {2018},
}

@article{mosetal17,
  author  = {Mosig, J. E. M. and Montiel, F. and Squire, V. A.},
  title   = {Water wave scattering from a mass loading ice floe of random length using generalised polynomial chaos},
  journal = {Wave Motion},
  volume  = {70},
  pages   = {222--239},
  year    = {2017},
}

@article{pitandben26,
  author  = {Pitt, J. P. A. and Bennetts, L. G.},
  title   = {Model study of ocean wave propagation through broken sea ice covers with variable ice concentration},
  journal = {Wave Motion},
  volume  = {141},
  pages   = {103655},
  year    = {2026},
}

@article{rogetal16,
  author  = {Rogers, W. E. and Thomson, J. and Shen, H. H. and Doble, M. J. and Wadhams, P. and Cheng, S.},
  title   = {Dissipation of wind waves by pancake and frazil ice in the autumn Beaufort Sea},
  journal = {J. Geophys. Res.: Oceans},
  volume  = {121},
  number  = {11},
  pages   = {7991--8007},
  year    = {2016},
}

@article{sheandack91,
  author  = {Shen, H. H. and Ackley, S. F.},
  title   = {A one-dimensional model for wave-induced ice-floe collisions},
  journal = {Ann. Glaciol.},
  volume  = {15},
  pages   = {87--95},
  year    = {1991},
}

@article{squire18,
  author  = {Squire, V. A.},
  title   = {A fresh look at how ocean waves and sea ice interact},
  journal = {Phil. Trans. Roy. Soc. A.},
  volume  = {376},
  pages   = {20170342},
  year    = {2018},
}

@article{squire20,
  author  = {Squire, V. A.},
  title   = {Ocean wave interactions with sea ice: A reappraisal},
  journal = {Ann. Rev. Fluid Mech.},
  volume  = {52},
  number  = {1},
  pages   = {37--60},
  year    = {2025},
}

@article{thoetal21,
  author  = {Thomson, J. and Ho\v{s}ekov\'a, L. and Meylan, M. H. and Kohout, A. L. and Kumar, N.},
  title   = {Spurious rollover of wave attenuation rates in sea ice caused by noise in field measurements},
  journal = {J. Geophys. Res.: Oceans},
  volume  = {126},
  number  = {3},
  pages   = {e2020JC016606},
  year    = {2021},
}

@article{wadetal88,
  author  = {Wadhams, P. and Squire, V. A. and Goodman, D. J. and Cowan, A. M. and Moore, S. C.},
  title   = {The attenuation rates of ocean waves in the Marginal Ice Zone},
  journal = {J. Geophys. Res.},
  volume  = {93},
  number  = {C6},
  pages   = {6799--6818},
  year    = {1988},
}

@article{weiandkel50,
  author  = {Weitz, M. and Keller, J. B.},
  title   = {Reflection of water waves from floating ice in water of finite depth},
  journal = {Comm. Pure \& Appl. Math.},
  volume  = {3},
  number  = {3},
  pages   = {305--318},
  year    = {1950},
}
\end{document}


\maketitle

\setcounter{equation}{0}
\renewcommand{\theequation}{S.\arabic{section}.\arabic{equation}}
\newcounter{appen}

\begin{abstract}
Using the definitions and notation introduced in the 
journal submission, we outline details of the calculations 
carried out for surge and pitch motions and the associated
hydrodynamic forces that depend on these motions.
\end{abstract}

\section{Solution in surge}

A general expression for $\phi^{(s)}(x,y)$ in the fluid region $V^+ :=\left\{0<x<\ell,-\rhoh d<z<0\right\}$ satisfying both \eqref{eqn:2.1} and \eqref{eqn:2.12}
is given by
\begin{equation}
	\phi^{(s)}(x,z) = \varphi^{(s)}(x,z) + \sum_{m=0}^{\infty}a_m Z_m(z) \cos{\alpha_m x}
 \label{s1}
\end{equation}
where $\alpha_m = m\pi/\ell$ and
\begin{equation}
 Z_0(z) = 1+Kz,\quad Z_m(z) = \cosh{\alpha_m z} + (K/\alpha_m)\sinh{\alpha_m z}  \label{s2}
\end{equation}
are chosen to satisfy \eqref{eqn:2.2}. In (\ref{s1}), we have defined the
particular solution satisfying \eqref{eqn:2.1}, \eqref{eqn:2.2} and \eqref{eqn:2.11}
and the boundary condition $\varphi^{(s)}_z(x,-\rhoh d) = 0$
\begin{eqnarray}
 \varphi^{(s)}(x,z) 
 &=&
 \frac{1}{2\ell}\left(x^2 - (\ell-x)^2\exp^{-\im kL}\right) - \frac{1}{2\ell}\left(1-\exp^{-\im kL}\right)\left(z^2 + 2\rhoh d z\right)
 \nonumber \\ 
 & &- \left(\frac{\rhoh d}{K \ell}+\frac{\ell}{6}\right)\left(1-\exp^{-\im kL}\right) + \sum_{m=1}^{\infty}\frac{2K\left((-1)^m - \exp^{-\im kL}\right)\cos\left({\alpha_m x}\right)\cosh\left[\alpha_m\left(z+\rhoh d\right)\right]}{\ell {\alpha_m}^2 \left(\alpha_m\sinh\left(\alpha_m\rhoh d\right) - K\cosh\left(\alpha_m\rhoh d\right)\right)}.
 \label{s3}
\end{eqnarray}

Over $0<x<\ell$ we let $W^{(s)}(x) = \phi_z^{(s)}(x,-\rhoh d)$ 
to represent the unknown vertical component of the fluid velocity 
across the interface between the two rectangular fluid sub-domains $V^+$ and $V^-$, above and below the level 
$z = -\rhoh d$. 

Using the orthogonality of the functions $\cos \alpha_m x$ (see \eqref{eqn:3.3})
then gives
\begin{equation}
 \left(K\cosh\left(\alpha_m \rhoh d\right) - \alpha_m\sinh\left(\alpha_m \rhoh d\right)\right)a_m = \frac{2}{\ell}\int_{0}^{\ell}W^{(s)}(x)\cos{\alpha_m x}\;\wrt x
 \label{s4}
\end{equation}
for $m=0,1,\ldots$.

In $z<-\rhoh d$, the general solution satisfying \eqref{eqn:2.1}, \eqref{eqn:2.4} and \eqref{eqn:2.11}
can be written
\begin{equation}
	\phi^{(s)}(x,z) = \sum_{m=-\infty}^{\infty} b_m\exp^{\im \beta_m x}\exp^{\beta_m(z+\rhoh d)}
 \label{s5}
\end{equation}
where $\beta_m = k + 2\pi m/L$. Using \eqref{eqn:3.7} now gives us
\begin{equation}
	b_m = \frac{1}{\beta_m L}\int_{0}^{\ell}W^{(s)}(x)\exp^{-\im \beta_m x}\;\wrt x
 \label{s6}
\end{equation}
for all $m$. To complete the solution we match $\phi^{(s)}$ from
(\ref{s1}) and (\ref{s5}) across $z=-\rhoh d$, $0<x<\ell$ and 
substitute expressions for $a_m$ and $b_m$ from (\ref{s3}), (\ref{s6}),
a process which results in the integral equation
\begin{equation}
	(\mathcal{T} W^{(s)})(x) \equiv \frac{1}{\ell}\int_{0}^{\ell}W^{(s)}(x')T(x,x')\;\wrt x' = G^{(s)}(x),\quad 0<x<\ell 
 \label{s7}
\end{equation}
for the unknown function $W^{(s)}(x)$ where $T(x,x')$ is defined by \eqref{eqn:3.18} 
and we have defined
\begin{eqnarray}
 G^{(s)}(x)  = (1/\ell)\varphi^{(s)}(x,-\rhoh d)
 & = &
 \frac{x^2}{2\ell^2} - \frac{(\ell-x)^2}{2\ell^2} \exp^{-\ci k L}
 +\left( \frac{d^2}{\ell^2} \left( \frac{\rhoh^2}{2} - \frac{\rhoh}{Kd} \right)
 - \frac{1}{6} \right) (1-\exp^{-\ci k L})
 \nonumber \\
 & & - \frac{2}{\ell^2} \sum_{m=1}^\infty \frac{( (-1)^m - \exp^{-\ci k L}) \cos \alpha_m x }
 {\alpha_m^2 (\cosh (\alpha_m \hat{\rho} d) - 
 (\alpha_m/K) \sinh (\alpha_m \hat{\rho} d)) }
 \label{s8}
\end{eqnarray}
which has been determined from (\ref{s3}).

In order to use the solution $W^{(s)}(x)$ of (\ref{s7}) to calculate the 
surge-induced surge force $F^{(s,s)}$ directly, we use Green's identity 
with the functions $\phi^{(s)}(x,z)$ and $\overline{\varphi^{(s)}}(x,z)$ 
in the region $V^+$ and this gives us
\begin{eqnarray}
	0 &=& \int_{\partial V^+} \frac{\partial \phi^{(s)}}{\partial n} \overline{\varphi^{(s)}} -  \frac{\partial \overline{\varphi^{(s)}}}{\partial n} {\phi^{(s)}}\;\wrt s
 \nonumber \\
	&=& -\int_{0}^{\ell} \left(\frac{\partial \phi^{(s)}}{\partial z} \overline{\varphi^{(s)}} -  \frac{\partial \overline{\varphi^{(s)}}}{\partial z} {\phi^{(s)}}\right)\Bigg|_{z=-\rhoh d}\;\wrt x + \int_{0}^{\ell} \left(\frac{\partial \phi^{(s)}}{\partial z} \overline{\varphi^{(s)}} -  \frac{\partial \overline{\varphi^{(s)}}}{\partial z} {\phi^{(s)}}\right)\Bigg|_{z=0}\;\wrt x
 \nonumber \\
	& & + \int_{-\rhoh d}^{0} \left(\frac{\partial \phi^{(s)}}{\partial x} \overline{\varphi^{(s)}} -  \frac{\partial \overline{\varphi^{(s)}}}{\partial x} {\phi^{(s)}}\right)\Bigg|_{x=\ell}\;\wrt z - \int_{-\rhoh d}^{0} \left(\frac{\partial \phi^{(s)}}{\partial x} \overline{\varphi^{(s)}} -  \frac{\partial \overline{\varphi^{(s)}}}{\partial x} {\phi^{(s)}}\right)\Bigg|_{x=0}\;\wrt z
 \nonumber \\
	 &=&  -\int_{0}^{\ell} W^{(s)}(x) \overline{\varphi^{(s)}}(x,-\rhoh d)\;\wrt x + \int_{-\rhoh d}^{0}  \overline{\varphi^{(s)}}(\ell,z) -  {\phi^{(s)}}(\ell,z)\;\wrt z - \int_{-\rhoh d}^{0} \overline{\varphi^{(s)}}(0,z)\exp^{-\im kL} -   {\phi^{(s)}}(0,z)\exp^{\im kL}\;\wrt z
 \label{s9}
\end{eqnarray}
using the conditions satisfied by the two functions 
on the four boundaries and rearranging gives
\begin{equation}
 \int_{-\rhoh d}^{0}\phi^{(s)}(\ell,z) - \exp^{\im kL}\phi^{(s)}(0,z)\;\wrt z =  \int_{-\rhoh d}^{0} \overline{\varphi^{(s)}}(\ell,z) - \overline{\varphi^{(s)}}(0,z)\exp^{- \im kL} \;\wrt z -\int_{0}^{\ell}W^{(s)}(x)\overline{\varphi^{(s)}}(x,-\rhoh d)\;\wrt x.
 \label{s10}
\end{equation}
From the definition \eqref{eqn:2.13} we have
\begin{eqnarray}
	F^{(s,s)} &=&
 \frac{1}{d(L-\ell)}\int_{-\rhoh d}^{0} \phi^{(s)}(\ell,z) - \phi^{(s)}(0,z)\exp^{\im kL}\;\wrt z
 \nonumber \\
	&=& -\frac{\ell^2}{d(L-\ell)}  A^{(s,s)} +   C^{(s,s)} \label{s11}
\end{eqnarray}
after substituting from (\ref{s10}) and using the definition (\ref{s7}).
For convenience we have defined
\begin{equation}
 A^{(s,s)} = \frac{1}{\ell}
 \int_{0}^{\ell}W^{(s)}(x)\overline{G^{(s)}}(x)\;\wrt x
 \label{s12}
\end{equation}
and
\begin{equation}
 C^{(s,s)} = \frac{1}{d(L-\ell)} \int_{-\rhoh d}^{0} \overline{\varphi^{(s)}}(\ell,z) - \overline{\varphi^{(s)}}(0,z)\exp^{- \im kL} \;\wrt x
 \label{s13}
\end{equation}
in the above. Using the definition (\ref{s3}), we can calculate
\begin{equation}
 C^{(s,s)} 
 = 
 \frac{\ell \hat{\rho}}{L-\ell} - \frac{4 \hat{\rho} d^2 \sin^2 (kL/2))}{\ell(L-\ell)} \left( 
 \frac{\ell^2}{6d^2} + \frac{\hat{\rho}}{Kd}
 - \frac{\hat{\rho}^2}{3} \right)
 + \frac{4}{\ell(L-\ell) d} \sum_{m=1}^\infty \frac{(1-(-1)^m \cos kL)}{\alpha_m^3
 ((\alpha_m/K) - \coth (\alpha_m \hat{\rho} d))}.
 \label{s14}
\end{equation}

\subsection{Numerical approximation}

Following the example of heave in the main paper, we write
\begin{equation}
	W^{(s)}(x)\approx\sum_{n=0}^{N} A_n^{(s)} u_n(x)
 \label{s15}
\end{equation}
for unknowns ${A_n}^{(s)}$ and with $u_n(x)$ defined by \eqref{eqn:3.30}. 
After substituting into (\ref{s7}), multiplying by $\overline{u_m(x)}$ 
and integrating over $0<x<\ell$ one obtains
\begin{equation}
	\sum_{n=0}^N A_n^{(s)} T_{mn} = E_{m}^{(s)}
 \label{s16}
\end{equation}
where $T_{mn}$ is given by \eqref{eqn:3.28} whilst
\begin{equation}
	E_m^{(s)} = \int_{0}^{\ell} G^{(s)}(x)
 \overline{u(x)}\;\wrt x.
 \label{s17}
\end{equation}
From the definition (\ref{s3}) we find
\begin{equation}
	E_m^{(s)} = \mathcal{K}_m  + \left(-\frac{\rhoh^2 d^2}{2\ell^2} + \frac{\rhoh d}{K\ell^2} + \frac{1}{6}\right)\left(1-\exp^{-\im k L}\right)\overline{\mathcal{I}_{m0}} +\sum_{n=1}^\infty \frac{2K\left((-1)^n - \exp^{-\im k L}\right)\overline{\mathcal{I}_{mn}}}{{\alpha_n}^2\ell^2\left(\alpha_n\sinh(\alpha_n\rhoh d)-K\cosh(\alpha_n\rhoh d)\right)},
 \label{s18}
\end{equation}
where $\mathcal{I}_{mn}$ is defined by \eqref{eqn:3.31}, \eqref{eqn:3.32} and we have had to 
make a new calculation
$\mathcal{K}_{m} = \left\langle g,u_m\right\rangle$
where $g(x) = (x^2 - (\ell -x)^2\exp^{-\im k L})/(2\ell^2)$ and find,
using some standard integral results for products of 
polynomials and Gegenbauer polynomials (REF), that
\begin{equation}
	\mathcal{K}_m = \frac{\im^m 2^{1/6}\Gamma(8/3)}{\sqrt{\pi}\Gamma\left(m+10/3\right)\Gamma(3-m)}((-1)^m-\exp^{-\im k L})
 \label{s19}
\end{equation}
noting that, for $m\geq3$, we have $\mathcal{K}_m = 0$. 

Finally, it follows from using (\ref{s15}) in (\ref{s12}) that
\begin{equation}
  A^{(s,s)} \approx \sum_{n=0}^{N} A_n^{(s)} \overline{E_n^{(s)}}.
 \label{s20}
\end{equation}

\subsection{Asymptotic results for small gaps}

Letting $\ell/d = \epsilon \ll 1$ and assuming $|K \hat{\rho} d -1| \gg \epsilon$ (away from resonance) we have as before, from \eqref{eqn:3.40},
\begin{equation}
 T(x,x') \approx \epsilon^{-1} \tilde{T}, \qquad \mbox{where }
 \tilde{T} = \frac{K\rhoh d-1}{Kd} + \frac12\epsilon \cot\left(\frac{kL}{2}\right).
 \label{s21}
\end{equation}
We also have from (\ref{s8}) the leading order behaviour
\begin{equation}
  G^{(s)}(x) \approx \frac{1}{\epsilon^2} \tilde{G}^{(s)},
 \qquad \mbox{where }
 \tilde{G}^{(s)} =
 \left(\frac{\rhoh^2}{2}-\frac{\rhoh}{Kd}\right)\left(1-\exp^{-\im kL}\right).
 \label{s22}
\end{equation}
Using (\ref{s21}) and (\ref{s22}) in (\ref{s7}) gives
\begin{equation}
 \frac{1}{\ell} \int_{0}^{\ell}W^{(s)}(x') \;\wrt x' \approx 
 \frac{1}{\epsilon} \frac{\tilde{G}^{(s)}}{\tilde{T}}
 \label{s23}
\end{equation}
which when used in (\ref{s13}) gives
\begin{equation}
 A^{(s,s)} \approx \frac{1}{\epsilon^3} \frac{\left|\tilde{G}^{(s)}\right|^2}
 {\tilde{T}}. 
 \label{s24}
\end{equation}
Also, the leading order behaviour in (\ref{s14}) gives
\begin{equation}
 C^{(s,s)} \approx \frac{1}{\epsilon}
 \frac{4\rhoh^2}{KL}  \left(\frac{1}{3}K\rhoh d-{1}\right)
 \sin^2\left(\frac{kL}{2}\right).
 \label{s25}
\end{equation}
Using (\ref{s24}) and (\ref{s25}) with (\ref{s21}) and (\ref{s22}) 
in (\ref{s11}) and simplifying shows us that
\begin{equation}
    F^{(s,s)} \approx \frac{1}{\epsilon}\frac{4\rhoh^2d}{KL}\sin^2\left(\frac{kL}{2}\right)\left(\frac13K\rhoh d-1 - \frac{\left( K\rhoh d -2\right)^2}{4 (K\rhoh d-1) + 2 \epsilon Kd \cot(kL/2)}\right)\label{s26}
\end{equation}
is the leading order behaviour of the surge-induced surge force for
$\epsilon \ll 1$ away from resonance.

\section{Combined heave-surge forces}

We are now in a position to consider the
heave-induced surge force and the surge-induced heave force (theory
dictates that are complex conjugates).

We see from the definition \eqref{eqn:2.13} that
\begin{equation}
	F^{(h,s)} = \frac{1}{d(L-\ell)}\int_{-\rhoh d}^{0}\phi^{(h)}(\ell,z)\; \wrt z - \frac{\exp^{\im k L}}{d(L-\ell)} \int_{-\rhoh d}^{0} \phi^{(h)}(0,z)\;\wrt z
 \label{s27}
\end{equation}
and
\begin{equation}
	F^{(s,h)} = \frac{1}{d(L-\ell)}\int_{\ell}^{L}\phi^{(s)}(x,-\rhoh d)\; \wrt x.
 \label{s28}
\end{equation}
Applying Green's identity to $\phi^{(h)}(x,z)$ and 
$\overline{\varphi^{(s)}}(x,z)$ in the region $V^+$ we get, using all
of the equations and boundary conditions defining those functions
\begin{equation}
	\int_{-\rhoh d}^{0}\phi^{(h)}(\ell,z)\; \wrt z - \exp^{\im k L} \int_{-\rhoh d}^{0} \phi^{(h)}(0,z)\;\wrt z = -\ell \int_{0}^{\ell} W^{(h)}(x)\overline{G^{(s)}(x)}\;\wrt x.
 \label{s29}
\end{equation}
Instead, applying Green's identity to $\phi^{(s)}(x,z)$ and 
$\overline{\varphi^{(h)}}(x,z)$ in the region $V^-$ and again
using all
of the equations and boundary conditions defining those functions
\begin{equation}
	\int_{\ell}^{L}\phi^{(s)}(x,-\rhoh d)\; \wrt x  = -\ell \int_{0}^{\ell} W^{(s)}(x)\overline{G^{(h)}}(x)\;\wrt x.
 \label{s30}
\end{equation}
Combining (\ref{s29}), (\ref{s30}) with (\ref{s27}), (\ref{s28})
results in
\begin{equation}
	F^{(h,s)} = -\frac{\ell^2}{d(L-\ell)} A^{(h,s)}, \qquad
 \mbox{where } A^{(h,s)} = \frac{1}{\ell}
 \int_{0}^{\ell} W^{(h)}(x)\overline{G^{(s)}(x)}\;\wrt x
 \label{s31}
\end{equation}
and
\begin{equation}
	F^{(s,h)} = -\frac{\ell^2}{d(L-\ell)} A^{(s,h)},
 \qquad \mbox{where }
  A^{(s,h)} = \int_{0}^{\ell} W^{(s)}(x)\overline{G^{(h)}(x)}\;\wrt x.
 \label{s32}
\end{equation}

\subsection{Numerical approximation}

The numerical approximations to these quantities above are 
obtained from
\begin{equation}
	A^{(h,s)} \approx \sum_{n=0}^{N}A_n^{(h)}\overline{E_n^{(s)}}
 \label{s33}
\end{equation}
and
\begin{equation}
	A^{(s,h)} \approx \sum_{n=0}^{N}A_n^{(s)}\overline{E_n^{(h)}}.
 \label{s33b}
\end{equation}
Use of the integral equations,
$({\mathcal T} W^{(s,h)})(x) = G^{(s,h)}(x)$, satisfied by $W^{(s,h)}(x)$
gives
\begin{equation}
 A^{(h,s)} = \langle W^{(h)}, G^{(s)} \rangle 
 =
 \langle W^{(h)}, \mathcal{T} W^{(s)} \rangle
 =
 \langle \mathcal{T} W^{(h)},  W^{(s)} \rangle
 =
 \langle G^{(h)},  W^{(s)} \rangle
 = \overline{A^{(s,h)}}.
 \label{s84}
\end{equation}
This means $F^{(h,s)} = \overline{F^{(s,h)}}$ confirming the 
identity \eqref{eqn:2.17} in the main paper and thus we only need consider
$F^{(h,s)}$.

\subsection{Asymptotic analysis for small gaps}

Using \eqref{eqn:3.37}, (\ref{s21}) and (\ref{s22}) we can immediately write
\begin{equation}
 A^{(h,s)} \approx \frac{1}{\epsilon^2} \frac{\tilde{G}^{(h)} \overline{\tilde{G}^{(s)}}}{\tilde{T}}, \qquad
 \mbox{and} 
 \qquad
 A^{(s,h)} \approx \frac{1}{\epsilon^2} \frac{\tilde{G}^{(s)} \overline{\tilde{G}^{(h)}}}{\tilde{T}}
 \label{s34}
\end{equation}
and so, by (\ref{s31}), (\ref{s32}) the leading order approximation of 
the surge/heave forces for small gaps are defined by the expression
\begin{equation}
    F^{(h,s)} = i\hat{\rho} \frac{\frac{1}{2}K\hat{\rho} d - 1}{K\hat{\rho} d - 1 + \frac{1}{2}\epsilon Kd\cot\left(\frac{kL}{2}\right)}.\label{s35}
\end{equation}

\section{Solution in Pitch}

This is the most complicated problem to solve because the forcing applies
on all sides of the submerged rectangular floe and so particular solutions
are required in both $V^+$ and $V^-$.

A general expression for the potential $\phi^{(p)}(x,z)$ in the fluid 
region $V^+ :=\left\{0<x<\ell,-\rhoh d<z<0\right\}$ satisfying 
\eqref{eqn:2.1}, \eqref{eqn:2.2} and \eqref{eqn:2.12} can be written as
\begin{equation}
	\phi^{(p)}(x,z) = \varphi^{(+p)}(x,z) + \sum_{m=0}^{\infty}a_m Z_m(z) \cos{\alpha_m x}
 \label{s36}
\end{equation}
where $\alpha_m = m\pi/\ell$ and
\begin{equation}
	Z_0(z) = 1+Kz,\quad Z_m(z) = \cosh{\alpha_m z} + (K/\alpha_m)\sinh{\alpha_m z}
 \label{s37}
\end{equation}
are defined as before. The function $\varphi^{(+p)}(x,z)$ satisfies
The function $\varphi^{(+p)}(x,z)$ satisfies
\eqref{eqn:2.1}, \eqref{eqn:2.2} and the inhomogeneous condition \eqref{eqn:2.11}
as well as the imposed condition
$\psi_z^{(+p)}(x,-\rhoh d) = 0$. After considerable effort, we
find it can be written
\begin{eqnarray}
	\varphi^{(+p)}(x,z) & = & 
 \frac{1}{2\ell}\left(x^2 - (\ell-x)^2\exp^{-\im kL}\right)\frac{(z-z^c)}{L-\ell} - \frac{1}{6\ell}\left(1-\exp^{-\im kL}\right)\frac{((z-z^c)^3 - 3\left(z^c + \rhoh d\right)^2 z + \ell^2 (z-z^c))}{L-\ell}
 \nonumber \\ 
 & &+ \frac{(3\rhoh d(2z^c + \rhoh d)-K{z^c}^3)}{6K (L-\ell)\ell}\left(1-\exp^{-\im kL}\right)
 \nonumber \\
 & & - \frac{2}{\ell(L-\ell)}
 \sum_{m=1}^{\infty} \frac{((-1)^m -\exp^{-\im k L})}
 {\alpha_m^3\cosh(\alpha_m\rhoh d)}
 \cos\left({\alpha_m x}\right)\left(\sinh\left[\alpha_mz\right] + 
 C_m\cosh\left[\alpha_m(z+\rhoh d)\right]\right)
 \label{s38}
\end{eqnarray}
where
\begin{equation}
	C_m = \frac{\alpha_m(\text{sech}(\alpha_m\rhoh d))-1-Kz^c)}{K - \alpha_m\tanh(\alpha_m\rhoh d)}.
 \label{s39}
\end{equation}
We are reminded that $z^c = (\frac12 -\rhoh) d$ is the $z$ coordinate
of the centre of mass.
In deriving this expression we have, in turn, satisfied the 
inhomogenous boundary conditions \eqref{eqn:2.11} on $x=0$, $x=\ell$,
and then Laplace's equation, followed by the condition on
$z=-\rhoh d$ and finally the condition on $z=0$.

After taking the $z$-derivative of (\ref{s36}), evaluated on
$z=-\rhoh d$, multiplying by $\cos{\alpha_m x}$ and integrating over 
$0<x<\ell$ we find, for $m=0$,
\begin{equation}
	a_0 = \frac{1}{K\ell}\int_{0}^{\ell}W^{(p)}(x)\;\wrt x
 \label{s40}
\end{equation}
where we have let $W^{(p)}(x)=\phi_z^{(p)}(x,-\rhoh d)$ for $0<x<\ell$.
For $m\geq 1$, 
\begin{equation}
	\left(K\cosh\left(\alpha_m \rhoh d\right) - \alpha_m\sinh\left(\alpha_m \rhoh d\right)\right)a_m = \frac{2}{\ell}\int_{0}^{\ell}W^{(p)}(x)\cos{\alpha_m x}\;\wrt x.
 \label{s41}
\end{equation}
In $z<-\rhoh d$, we require a general solution to satisfy \eqref{eqn:2.1}, \eqref{eqn:2.4} and \eqref{eqn:2.11} and thus write
\begin{equation}
	\phi^{(p)}(x,z) = \varphi^{(-p)}(x,z) + \sum_{m=-\infty}^{\infty} b_m\exp^{\im \beta_m x}\exp^{\beta_m(z+\rhoh d)}
 \label{s42}
\end{equation}
where $\beta_m = k + 2\pi m/L$ as before. Here, we have defined
$\varphi^{(-p)}(x,z)$ to be the harmonic function satisfying \eqref{eqn:2.4} and decaying 
into the depth in addition to the imposed condition
\begin{equation}
	\varphi^{(-p)}_z(x,-\rhoh d) = \begin{cases}
		-(x-x^c)/(L-\ell), \quad \ell<x<L\\
		0, \quad 0<x<\ell
	\end{cases}
 \label{s43}
\end{equation}
which accounts for the forcing on the underside of the floe in much
the same way as $\varphi^{(+p)}(x,z)$ accounted for the forcing on the
sides of the floe in $V^+$.

The particular solution can be determined by writing the separation
series satisfying \eqref{eqn:2.1} and \eqref{eqn:2.4}
\begin{equation}
 \varphi^{(-p)}(x,z) = \ell \sum_{m=-\infty}^{\infty} \frac{g_m^{(p)} \exp^{\im \beta_m x}}{\beta_m L}\exp^{\beta_m(z+\rhoh d)}
 \label{s44}
\end{equation}
where (\ref{s43}) tells us that
\begin{equation}
	\ell g_m^{(p)} = -\frac{1}{(L-\ell)}\int_{\ell}^{L}(x-x^c)\exp^{-\im\beta_m x}\;\wrt x = \frac{\exp^{-\im\beta_m \ell}-\exp^{-\im\beta_m L}}{{\beta_m}^2(L-\ell)} + \frac12 \frac{\exp^{-\im\beta_m\ell} + \exp^{-\im\beta_m L}}{\im\beta_m}
 \label{s45}
\end{equation}
after using the orthogonality of the functions $\exp^{\im \beta_m x}$
on $0 < x < L$ expressed by \eqref{eqn:3.10}.
As a result of taking the $z$-derivative of (\ref{s42}) on $z = -\rhoh d$, 
we find
\begin{equation}
 b_m = \frac{1}{\beta_m L}\int_{0}^{\ell}W^{(p)}(x)\exp^{-\im \beta_m x}\;\wrt x
 \label{s46}
\end{equation}
for all $m$. Next, applying continuity of $\phi^{(p)}$ across the 
common interface $z=-\rhoh d$, $0<x<\ell$ between the two regions 
and substituting for $a_m$ and $b_m$ from (\ref{s41}), (\ref{s46})
results in the integral equation
\begin{equation}
(\mathcal{T} W^{(p)})(x) \equiv 
 \frac{1}{\ell}\int_{0}^{\ell}W^{(p)}(x')T(x,x')\;\wrt x' = G^{(p)}(x),\quad 0<x<\ell \label{s47}
\end{equation}
where $T(x,x')$ is defined by \eqref{eqn:3.18} and we have defined the function
\begin{equation}
	G^{(p)}(x) = (1/\ell)\left(\varphi^{(+p)}(x,-\rhoh d)- \varphi^{(-p)}(x,-\rhoh d)\right).
 \label{s48}
\end{equation}
Using Green's identity with the functions two harmonic functions
$\phi^{(p)}(x,z)$ and $\overline{\varphi^{(-p)}}(x,z)$ in the region $V^-$ 
and using all of the conditions of these functions on the boundary of
$V^-$ gives us
\begin{eqnarray}
 0 &= &
 \int_{V^-} \frac{\partial \phi^{(p)}}{\partial n} \overline{\varphi^{(-p)}} -  \frac{\partial \overline{\varphi^{(-p)}}}{\partial n} {\phi^{(p)}}\;\wrt s
 \nonumber \\
 & & = \int_{0}^{\ell} \left(\frac{\partial \phi^{(p)}}{\partial z} \overline{\varphi^{(-p)}} -  \frac{\partial \overline{\varphi^{(-p)}}}{\partial z} {\phi^{(p)}}\right)\Bigg|_{z=-\rhoh d}\;\wrt x + \int_{\ell}^{L} \left(\frac{\partial \phi^{(p)}}{\partial z} \overline{\varphi^{(-p)}} -  \frac{\partial \overline{\varphi^{(-p)}}}{\partial z} {\phi^{(p)}}\right)\Bigg|_{z=-\rhoh d}\;\wrt x
 \nonumber \\
 & &= \int_{0}^{\ell}W^{(h)}(x)\overline{\varphi^{(-p)}}(x,-\rhoh d)\;\wrt x - \int_{\ell}^{L} \frac{x-x^c}{L-\ell}\left(\overline{\varphi^{(-p)}}(x,-\rhoh d) - \phi(x,-\rhoh d)\right)\;\wrt x
 \label{s49}
\end{eqnarray}
from which we can clearly see
\begin{equation}
 \int_{\ell}^{L}\frac{x-x^c}{L-\ell}\phi^{(p)}(x,-\rhoh d)\;\wrt x =- \int_{0}^{\ell}W^{(p)}(x)\overline{\varphi^{(-p)}}(x,-\rhoh d)\;\wrt x + \int_{\ell}^{L}\frac{x-x^c}{L-\ell} \overline{\varphi^{(-p)}}(x,-\rhoh d) \;\wrt x.
 \label{s50}
\end{equation}
Separately we use Green's identity with $\phi^{(p)}(x,z)$ and 
$\overline{\varphi^{(+p)}}(x,z)$ in $V^+$ which results, after 
some algebra, in the relation
\begin{eqnarray}
	\int_{-\rhoh d}^{0}\frac{z-z^c}{L-\ell}\left(\phi^{(p)}(\ell,z) - \exp^{\im kL}\phi^{(p)}(0,z)\right)\;\wrt z
 & = &
  -\int_{0}^{\ell}W^{(p)}(x)\overline{\varphi^{(+p)}}(x,-\rhoh d)\;\wrt x
 \nonumber \\
  & & +
  \int_{-\rhoh d}^{0} \frac{z-z^c}{L-\ell}\left(\overline{\varphi^{(+p)}}(\ell,z) - \overline{\varphi^{(+p)}}(0,z)\exp^{- \im kL}\right) \;\wrt z .
 \label{s51}
\end{eqnarray}
Now, from the general formula \eqref{eqn:2.13} we can deduce that the pitch-induced
pitch force is expressed in terms of the potential $\phi^{(p)}$ as
\begin{equation}
 F^{(p,p)} = -\frac{1}{d(L-\ell)^2} \int_{\ell}^{L}\phi^{(p)}(x,-\rhoh d)(x-x^c)\;\wrt x + \frac{1}{d(L-\ell)^2}\int_{-\rhoh d}^{0} (z-z^c)\left(\phi^{(p)}(\ell,z)-\phi^{(p)}(0,z)\exp^{\im k L}\right)\;\wrt z.
 \label{s52}
\end{equation}
Using (\ref{s50}) and (\ref{s51}) in (\ref{s52}) along with the definition 
(\ref{s48}) gives
\begin{eqnarray}
	F^{(p,p)} 
 & = &
 -\frac{\ell}{d(L-\ell)}\int_{0}^{\ell}W^{(p)}(x)\overline{G^{(p)}(x)}\;\wrt x 
 - \frac{1}{d(L-\ell)^2}\int_{\ell}^{L}(x-x^c)\overline{\varphi^{(-p)}}(x,-\rhoh d)\;\wrt x
 \nonumber \\ 
 & &
 +  \frac{1}{d(L-\ell)^2}\int_{-\rhoh d}^{0}(z-z^c)\left(\overline{\varphi^{(+p)}}(\ell,z) - \overline{\varphi^{(+p)}}(0,z)\exp^{- \im kL}\right)\;\wrt z.
 \label{s53}
\end{eqnarray}
Thus, we have written $F^{(p,p)}$ in terms of the solution, $W^{(p)}(x)$,
of (\ref{s47}) plus integrals of known functions.
It helps to write
\begin{equation}
 F^{(p,p)} = - \frac{\ell^2}{d(L-\ell)} A^{(p,p)} + C^{(p,p)}
 \label{s54}
\end{equation}
where
\begin{equation}
 A^{(p,p)} = \frac{1}{\ell}
 \int_{0}^{\ell}W^{(p)}(x)\overline{G^{(p)}(x)}\;\wrt x 
 \label{s55}
\end{equation}
and
\begin{equation}
 C^{(p,p)} = C_1^{(p,p)} + C_2^{(p,p)}
 \label{s56}
\end{equation}
with: (i)
\begin{equation}
 C_1^{(p,p)} = - \frac{1}{d(L-\ell)^2}\int_{\ell}^{L}(x-x^c)\overline{\varphi^{(-p)}}(x,-\rhoh d)\;\wrt x
 = \frac{\ell^2}{d(L-\ell)} \sum_{m=-\infty}^\infty \frac{|g_m^{(p)}|^2}{\beta_m L}
 = 0
 \label{s57}
\end{equation}
a result which is proved in the Appendix; and (ii)
\begin{equation}
 C_2^{(p,p)} =  \frac{1}{d(L-\ell)^2}\int_{-\rhoh d}^{0}(z-z^c)\left(\overline{\varphi^{(+p)}}(\ell,z) - \overline{\varphi^{(+p)}}(0,z)\exp^{- \im kL}\right)\;\wrt z.
 \label{s58}
\end{equation}
A considerable amount of algebra allows us to determine
\begin{eqnarray}
 C^{(p,p)} & = & 
 \frac{\ell d^2 \rhoh}{(L-\ell)^3} \left(
 \frac{\rhoh^2}{3}
 -\frac{\rhoh}{2}
 +\frac{1}{4}
 \right)
 + \frac{d^4 \sin^2 (kL/2)}{\ell (L-\ell)^3} \left(
 \frac{\rhoh^5}{5}
 -\frac{\rhoh^4}{2}
 +\frac{\rhoh^3}{3}
 -\frac{\rhoh^2 (1-\rhoh)^2}{Kd} 
 - \frac{\ell^2}{d^2} \left(
 \frac{2\rhoh^3}{9}
 -\frac{\rhoh^2}{3}
 +\frac{\rhoh}{6}
 \right) \right)
 \nonumber \\
 & & - \frac{4}{d \ell (L-\ell)^3}
 \sum_{m=1}^\infty \frac{(1-(-1)^m \cos(kL))}{\alpha_m^5 \cosh (\alpha_m \rhoh d)}
 \left( \alpha_n \rhoh d -(1+ {\textstyle \frac12} C_m \alpha_m d)
 (1-\exp^{-\alpha_m \rhoh d}) \right).
 \label{s59}
\end{eqnarray}

\subsection{Numerical approximation}

We follow the same methods as before, and start by writing
\begin{equation}
	W^{(p)}(x) \approx \sum_{n=0}^{N} A_n^{(p)}u_n(x)
 \label{s60}
\end{equation}
which, when used in (\ref{s46}) under Galerkin's method, 
results in the expansion coefficients $A_n^{(p)}$ being 
determined by the linear system of equations
\begin{equation}
 	\sum_{n=0}^{N} A_n^{(p)}T_{mn} = E_m^{(p)},
 \label{s61}
\end{equation}
where 
\begin{equation}
 E_m^{(p)} = \frac{1}{\ell} \int_0^\ell G^{(p)}(x) \overline{u_m(x)} \; \wrt x.
 \label{s62}
\end{equation}
Using the definitions of $G^{(p)}(x)$ and $u_m(x)$ from (\ref{s48}) and
(\eqref{eqn:3.35}) we find, again after considerable algebra,
\begin{eqnarray}
 E_m^{(p)} 
 & = &
 -\sum_{n=-\infty}^{\infty}\frac{g_n^{(p)}}{\beta_nL}{\overline{\mathcal{J}_{mn}}} - \frac{d}{2(L-\ell)}{\mathcal{K}_m} 
 + \frac{d^3(1-\exp^{-\im k L})}{6 \ell^2 (L-\ell)} \left( \frac{3 \rhoh(1-\rhoh)}{Kd} - \frac{\ell^2}{2 d^2} + \rhoh^3 - \frac{3 \rhoh^2}{2} \right)
 \overline{\mathcal{I}_{m0}}
 \nonumber \\
 & &
 - \frac{2}{\ell^2(L-\ell)}
 \sum_{n=1}^{\infty} \frac{((-1)^n - \exp^{-\im kL})}{\alpha_n^3 \cosh (\alpha_n \rhoh d)} \left(C_n - \sinh (\alpha_n \rhoh d)\right){\mathcal{I}_{mn}}.
 \label{s63}
\end{eqnarray}
In the above $\mathcal{K}_m$ is defined by (\ref{s19}) and 
$\mathcal{J}_{mn}$ and $\mathcal{I}_{mn}$ are defined in the main paper.

Finally, from (\ref{s61}) in (\ref{s63}) we have
\begin{equation}
 A^{(p,p)} \approx \sum_{n=0}^{N} A_n^{(p)} \overline{E_m^{(p)}}.
 \label{s64}
\end{equation}

\subsection{Asymptotic results for small gaps}

First we note from the definition (\ref{s48}) that we can write
\begin{equation}
 G^{(p)} \approx \frac{1}{\epsilon^2} \tilde{G}^{(p)}
 \label{s65}
\end{equation}
to leading order where
\begin{equation}
 \tilde{G}^{(p)} = \frac{\rhoh d}{2 L} (1-\exp^{-\im kL})
 \left( \frac{\rhoh^2}{3} - \frac{\rhoh}{2} + \frac{(1-\rhoh)}{Kd} \right).
 \label{s66}
\end{equation}
Using $T(x,x') \approx \epsilon^{-1} \tilde{T}$ where $\tilde{T}$ is
given by \eqref{eqn:3.35}, the solution of (\ref{s47}) is approximated by
\begin{equation}
 \frac{1}{\ell} \int_{0}^{\ell} W^{(p)}(x')\;\wrt x' \approx \frac{1}{\epsilon}
 \frac{\tilde{G}^{(p)}}{\tilde{T}}.
 \label{s67}
\end{equation}
It follows from (\ref{s55}) that
\begin{equation}
 A^{(p,p)} \approx \frac{1}{\epsilon^3}
 \frac{|\tilde{G}^{(p)}|^2}{\tilde{T}}.
 \label{s68}
\end{equation}
Simultaneously, we can show that the leading order behaviour of 
$C^{(p,p)}$ defined in (\ref{s59}), is
\begin{equation}
 C^{(p,p)} \approx \frac{1}{\epsilon}
 \frac{d^3}{L^3} \rhoh^2 \sin^2(kL/2) \left( 
 \frac{\rhoh^3}{5}
 -\frac{\rhoh^2}{2}
 +\frac{\rhoh}{3}
 -\frac{(1-\rhoh)^2}{Kd}
 \right).
 \label{s69}
\end{equation}
Bringing both results together in (\ref{s54}) gives
\begin{equation}
    F^{(p,p)} = \frac{1}{\epsilon} \hat{\rho}^2 \left(\frac{d}{L}\right)^3 \left(\frac{\hat{\rho}^3}{5} - \frac{\hat{\rho}^2}{2} + \frac{\hat{\rho}}{3} - \frac{(1-\hat{\rho})^2}{Kd} - \frac{\left(\frac{\hat{\rho}^2}{3} - \frac{\hat{\rho}}{2} + \frac{1-\hat{\rho}}{Kd}\right)^2}{\frac{K\hat{\rho} d - 1}{Kd} + \frac{1}{2}\epsilon\cot\left(\frac{kL}{2}\right)}\right) \sin^2\left(\frac{kL}{2}\right)\label{s70}
\end{equation}
as the leading order approximation to the pitch-induced pitch force
away from resonance.

\section{Combined heave-pitch forces}

From \eqref{eqn:2.13} we express the pitch-induced heave force as
\begin{equation}
 F^{(p,h)} =
 \frac{1}{d(L-\ell)}\int_{\ell}^{L} \phi^{(p)} (x,-\rhoh d)\;\wrt x
 \label{s71}
\end{equation}
after using the conditions satisfied by the $\phi^{(h)}$ on $z=-\rhoh d$,
$x=0,\ell$. In order to express this in terms of solutions to the integral
equations formulated for each problem, we use
$\phi^{(p)}$ and $\varphi^{(h)}$ over $V^-$ in Green's identity to get
\begin{eqnarray}
	0 &=& \int_{\partial V^-} \frac{\partial \phi^{(p)}}{\partial n} \overline{\varphi^{(h)}} -  \frac{\partial \overline{\varphi^{(h)}}}{\partial n} {\phi^{(p)}}\;\wrt s
 \nonumber \\
	&=&
 \int_{0}^{\ell} \left(\frac{\partial \phi^{(p)}}{\partial z} \overline{\varphi^{(h)}} -  \frac{\partial \overline{\varphi^{(h)}}}{\partial z} {\phi^{(p)}}\right)\Bigg|_{z=-\rhoh d}\;\wrt x + \int_{\ell}^{L} \left(\frac{\partial \phi^{(p)}}{\partial z} \overline{\varphi^{(h)}} -  \frac{\partial \overline{\varphi^{(h)}}}{\partial z} {\phi^{(p)}}\right)\Bigg|_{z=-\rhoh d}\;\wrt x
 \nonumber \\
	&=&
 \int_{0}^{\ell}W^{(p)}(x)\overline{\varphi^{(h)}}(x,-\rhoh d)\;\wrt x - \int_{\ell}^{L} \frac{x-x^c}{L-\ell}\overline{\varphi^{(h)}}(x,-\rhoh d) + \phi^{(p)}(x,-\rhoh d)\;\wrt x.
 \label{s72}
\end{eqnarray}
Therefore, we can write
\begin{equation}
	F^{(p,h)} = -\frac{\ell^2}{d(L-\ell)} A^{(p,h)} + C^{(p,h)}
 \label{s73}
\end{equation}
where
\begin{equation}
 A^{(p,h)} = \frac{1}{\ell} \int_{0}^{\ell}W^{(p)}(x)
 \overline{G^{(h)}(x)}\;\wrt x
 \label{s74}
\end{equation}
with $G^{(h)}$ defined by \eqref{eqn:3.14} and
\begin{equation}
 C^{(p,h)} = \frac{\ell}{d(L-\ell)}\int_{\ell}^{L} \frac{x-x^c}{L-\ell}\overline{G^{(h)}(x)}\; \wrt x.
 \label{s75}
\end{equation}
Using the definition \eqref{eqn:3.14} with (\ref{s45}) allows us to write
\begin{equation}
 C^{(p,h)} = \frac{\ell^2}{d(L-\ell)}\sum_{m=-\infty}^{\infty}
 \frac{\overline{{g_m}^{(h)}}{g_m}^{(p)}}{\beta_m L}.
 \label{s76}
\end{equation}
In Appendix A we show that
\begin{equation}
 C^{(p,h)} = -\im \frac{(L-\ell)}{12d}.
 \label{s77}
\end{equation}

We now consider the reciprocal heave-induced pitch force.
From \eqref{eqn:2.13} we express the heave-induced pitch force as
\begin{equation}
 F^{(h,p)} 
	= -\frac{1}{d(L-\ell)^2}\int_{\ell}^{L}(x-x^c)\phi^{(h)}(x,-\rhoh d)\;\wrt x + \frac{1}{d(L-\ell)^2}\int_{-\rhoh d}^{0} (z-z^c)\left(\phi^{(h)}(\ell,z) - \exp^{\im k L}\phi^{(h)}(0,z)\right)\; \wrt z
 \label{s78}
\end{equation}
after using the conditions satisfied by the $\phi^{(p)}$ on $z=-\rhoh d$,
$x=0,\ell$.
Using $\phi^{(h)}$ and $\varphi^{(+p)}$ in Green's identity over the
region $V^+$ we find
\begin{eqnarray}
	0 
	&= &
 \int_{0}^{\ell} \left(\frac{\partial \phi^{(h)}}{\partial z} \overline{\varphi^{(+p)}} -  \frac{\partial \overline{\varphi^{(+p)}}}{\partial z} {\phi^{(h)}}\right)\Bigg|_{z=-\rhoh d}\;\wrt x - \int_{0}^{\ell} \left(\frac{\partial \phi^{(h)}}{\partial z} \overline{\varphi^{(+p)}} -  \frac{\partial \overline{\varphi^{(+p)}}}{\partial z} {\phi^{(h)}}\right)\Bigg|_{z=0}\;\wrt x
 \nonumber \\
	& &+\int_{-\rhoh d}^{0} \left(\frac{\partial \phi^{(h)}}{\partial x} \overline{\varphi^{(+p)}} -  \frac{\partial \overline{\varphi^{(+p)}}}{\partial x} {\phi^{(h)}}\right)\Bigg|_{x=\ell}\;\wrt z - \int_{-\rhoh d}^{0} \left(\frac{\partial \phi^{(h)}}{\partial x} \overline{\varphi^{(+p)}} -  \frac{\partial \overline{\varphi^{(+p)}}}{\partial x} {\phi^{(h)}}\right)\Bigg|_{x=0}\;\wrt z
 \nonumber \\
	& = & -\int_{0}^{\ell}W^{(h)}(x)\overline{\varphi^{(+p)}}(x,-\rhoh d)\;\wrt x - \int_{-\rhoh d}^{0} \frac{(z-z^c)}{(L-\ell)}\left({\phi^{(h)}}(\ell,z) - \exp^{\im k L}\phi^{(h)}(0,z)\right)\;\wrt z.
 \label{s79}
\end{eqnarray}
Also, using $\phi^{(h)}$ and $\varphi^{(-p)}$ in Green's identity in the 
domain $V^-$ gives
\begin{eqnarray}
	0 
 &=&
		\int_{0}^{L} \left(\frac{\partial \phi^{(h)}}{\partial z} \overline{\varphi^{(p)}} -  \frac{\partial \overline{\varphi^{(p)}}}{\partial z} {\phi^{(h)}}\right)\Bigg|_{z=-\rhoh d}\;\wrt x
 \nonumber \\
	&= &
 \int_{0}^{\ell}W^{(h)}(x)\overline{\varphi^{(p)}}(x,-\rhoh d)\;\wrt x  + \int_{\ell}^{L} \overline{\varphi^{(-p)}}(x,-\rhoh d) + \frac{(x-x^c)}{(L-\ell)}\phi^{(h)}(x,-\rhoh d)\;\wrt x.
 \label{s80}
\end{eqnarray}
Using both (\ref{s79}) and (\ref{s80}) in (\ref{s78}) along with (\ref{s48})
results in
\begin{equation}
	F^{(h,p)} = -\frac{\ell^2}{d(L-\ell)} A^{(h,p)} + C^{(h,p)}
 \label{s81}
\end{equation}
where we have defined
\begin{equation}
 A^{(h,p)} = \frac{1}{\ell} \int_{0}^{\ell} W^{(h)}(x) \overline{G^{(p)} (x)} \;\wrt x 
 \label{s82}
\end{equation}
and
\begin{equation}
 C^{(h,p)} = \frac{1}{d(L-\ell)}\int_{\ell}^{L}
 \overline{\varphi^{(-p)}}(x,-\rhoh d)\;\wrt x =
 \frac{\ell^2}{d(L-\ell)}\sum_{m=-\infty}^{\infty}\frac{\overline{g_m^{(p)}}g_m^{(h)}}{\beta_m L}
 \label{s83}
\end{equation}
which follows from use of \eqref{eqn:3.8} and (\ref{s44}).
Clearly $C^{(h,p)} = \overline{C^{(p,h)}}$ whilst
$A^{(h,p)} = \overline{A^{(p,h)}}$ using the same reasoning 
outlined in the surge/heave section.

\subsection{Numerical approximation}

We simply have
\begin{equation}
	\tilde{A}^{(h,p)} \approx \sum_{n=0}^{N}A_n^{(h)}\overline{E_n^{(p)}}
 \label{s85}
\end{equation}
and
\begin{equation}
	\tilde{A}^{(p,h)} \approx \sum_{n=0}^{N}A_n^{(p)}\overline{E_n^{(h)}}
 \label{s86}
\end{equation}
where $A_n^{(h,p)}$ are the coefficients determined from \eqref{eqn:3.23} and
(\ref{s61}) and $E_n^{(h,p)}$ are defined by \eqref{eqn:3.29} and
(\ref{s62}) respectively.

\subsection{Asymptotic results for small gaps}

Using the leading order terms in the integral equations, we find
\begin{equation}
 A^{(h,p)} \approx \frac{1}{\epsilon^2} \frac{ \overline{\tilde{G}^{(p)}}
 \tilde{G}^{(h)}}{\tilde{T}}.
 \label{s87}
\end{equation}
Using \eqref{eqn:3.35}, \eqref{eqn:3.37} and (\ref{s66}) and simplifying results whilst using $C^{(h,p)} = \im L/12 d$ from (\ref{s77}), we have from (\ref{s81}),
\begin{equation}
    F^{(h,p)} = i\left(\frac{L}{12d} + \frac{\hat{\rho}d}{2L} \frac{\frac{\hat{\rho}^2}{3} - \frac{\hat{\rho}}{2} + \frac{1-\hat{\rho}}{Kd}}{\frac{K\hat{\rho} d - 1}{Kd} + \frac{\epsilon}{2}\cot\left(\frac{kL}{2}\right)}\right)\label{s90}
\end{equation}
is the leading order heave-induced pitch force for small gaps.
$F^{(p,h)}$ is the conjugate of (\ref{s90})

\section{Combined surge-pitch forces}

Following the methodology used for the previous problems we are
able to show that
\begin{equation}
 F^{(s,p)} = -\frac{\ell^2}{d(L-\ell)}
 A^{(s,p)} + C^{(s,p)}
 \label{s91}
\end{equation}
where
\begin{equation}
 A^{(s,p)} = \frac{1}{\ell} \int_{0}^{\ell}W^{(s)}(x)\overline{G^{(p)}}(x)\;\wrt x\end{equation}
 \label{s92}
and
\begin{equation}
 C^{(s,p)}
 = \frac{1}{d(L-\ell)}\int_{-\rhoh d}^{0}\overline{\varphi^{(+p)}}(\ell,z) - \exp^{-\im kL}\overline{\varphi^{(+p)}}(0,z)\;\wrt z,
 \label{s93}
\end{equation}
Likewise, for the reciprocal force, we have
\begin{equation}
 F^{(s,p)} = -\frac{\ell^2}{d(L-\ell)} A^{(s,p)} + C^{(s,p)}
 \label{s94}
\end{equation}
where
\begin{equation}
 A^{(s,p)} = \frac{1}{\ell} \int_{0}^{\ell}W^{(p)}(x)\overline{G^{(s)}}(x)\;\wrt x\end{equation}
 \label{s95}
and
\begin{equation}
 C^{(p,s)}
 = \frac{1}{d(L-\ell)^2}\int_{-\rhoh d}^{0}(z-z^c)\left(\overline{\varphi^{(s)}}(\ell,z) - \exp^{-\im kL}\overline{\varphi^{(s)}}(0,z)\right)\;\wrt z.
 \label{s96}
\end{equation}
Here we deduce from (\ref{s96}), after lengthy algebra,
\begin{eqnarray}
 C^{(s,p)} & = &
 -\frac{\ell d \rhoh(1-\rhoh)}{2(L-\ell)^2}
 + \frac{2 \rhoh d^3 \sin^2(kL/2)}{\ell (L-\ell)^2}
 \left( \frac{\rhoh^3}{4} - \frac{\rhoh^2}{3} + (1-\rhoh)
 \left( \frac{\rhoh}{Kd} + \frac{\ell^2}{6 d^2} \right) \right)
 \nonumber \\
 & & 
 + \frac{4}{d \ell (L-\ell)^2} \sum_{m=1}^\infty
 \frac{ (1-(-1)^m \cos kL)(1-\cosh (\alpha_m \rhoh d) + (\frac12 - \rhoh)
 \alpha_m d \sinh (\alpha_m \rhoh d))}{\alpha_m^4 ((\alpha_m/K) \sinh (\alpha_m \rhoh d)
 - \cosh (\alpha_m \rhoh d))}.
 \label{s97}
\end{eqnarray}
We remark that a calculation of $C^{(s,p)}$ reveals it to be 
identical to $C^{(p,s)}$, which is expected by reciprocity.

\subsection{Numerical approximation}

We have
\begin{equation}
	A^{(s,p)} \approx \sum_{n=0}^{N}A_n^{(s)}\overline{E_n^{(p)}}
 \label{s98}
\end{equation}
and
\begin{equation}
	A^{(p,s)} \approx \sum_{n=0}^{N}A_n^{(p)}\overline{E_n^{(s)}}
 \label{s99}
\end{equation}
and, of course, $A^{(p,s)} = \overline{A^{(s,p)}}$, for reasons already 
detailed.

\subsection{Asymptotic results for small gaps}

First we note from (\ref{s97}) that
\begin{equation}
 C^{(p,s)} \approx \frac{1}{\epsilon} \frac{2 \rhoh^2 d^2 \sin^2(kL/2)}{L^2}
 \left( \frac{\rhoh^2}{4} - \frac{\rhoh}{3} + \frac{(1-\rhoh)}{Kd} \right)
 \label{s100}
\end{equation}
The leading order approximation to the solution of the integral
equation is used to find
\begin{equation}
 A^{(p,s)} \approx \frac{1}{\epsilon^3} \frac{\tilde{G}^{(p)} 
 \overline{\tilde{G}^{(s)}}}{\tilde{T}}.
 \label{s101}
\end{equation}
Using (\ref{s22}), (\ref{s66}) and \eqref{eqn:3.35} and simplifying, we can combine our result with (\ref{s100}) in (\ref{s94}) to get
\begin{equation}
    F^{(s,p)} = \frac{1}{\epsilon} \frac{\hat{\rho}^2 d}{KL^2} \left(2Kd\left(\frac{\hat{\rho}^2}{4} - \frac{\hat{\rho}}{3} + \frac{1-\hat{\rho}}{Kd}\right) - \frac{K\hat{\rho} d - 2}{\frac{K\hat{\rho} d - 1}{Kd} + \frac{1}{2}\epsilon\cot\left(\frac{kL}{2}\right)} \left(\frac{\hat{\rho}^2}{3} - \frac{\hat{\rho}}{2} + \frac{1-\hat{\rho}}{Kd}\right)\right)\sin^2\left(\frac{kL}{2}\right).\label{s103}
\end{equation}

\appendix
\renewcommand{\theequation}{S.\Alph{section}.\arabic{equation}}
\refstepcounter{section}\setcounter{equation}{0}
\section*{Appendix \Alph{section}: Evaluation of infinite series}
\addcontentsline{toc}{section}{Appendix \Alph{section}: Evaluation of infinite series}

In this Appendix we establish three main non-trivial results using the following
definitions given in the body of the paper and the supplementary
material: $\beta_m = k + 2 m \pi/L$,
\begin{equation}
 g_m^{(h)} = \ci (\exp^{-\ci \beta_m L} - \exp^{-\ci \beta_m \ell})/(\beta_m \ell)
 \label{sa.1}
\end{equation}
and
\begin{equation}
 g_m^{(p)} = \frac{\ci}{2 \beta_m \ell}(\exp^{-\ci \beta_m L} + \exp^{-\ci \beta_m \ell})
 + \frac{1}{\beta_m^2 \ell a} (\exp^{-\ci \beta_m L} - \exp^{-\ci \beta_m \ell}).
 \label{sa.2}
\end{equation}
where we have used the abbreviation $a = L-\ell$. Then
\begin{equation}
 \mbox{(i)~} C^{(h,h)} = \frac{\ell^2}{d a} \sum_{m=-\infty}^\infty \frac{\left|g_m^{(h)}\right|^2}{\beta_m L} = \frac{a \cot(kL/2)}{2 d}
 \label{sa.3}
\end{equation}
is the key result from which it will be shown that 
\begin{equation}
 \mbox{(ii)~}
 C^{(p,p)} = 
 \sum_{m=-\infty}^\infty \frac{\left|g_m^{(p)}\right|^2}{\beta_m L} = 0
 \label{sa.4}
\end{equation}
and
\begin{equation}
 \mbox{(iii)~}
 C^{(p,h)} = \frac{\ell^2}{d a}
 \sum_{m=-\infty}^\infty \frac{g_m^{(p)} \overline{g_m^{(h)}} }{\beta_m L} = \ci \frac{a^2}{12\ell^2}.
 \label{sa.5}
\end{equation}

First, from (\ref{sa.3}) we note that
\begin{equation}
	C^{(h,h)} = \frac{L^2}{d a} J_1(a), 
 \qquad \mbox{where }
 J_1(a) = \sum_{m=-\infty}^{\infty} \frac{\sin^2\left(\beta_m a/2\right)}{(\beta_m L)^3}.
 \label{sa.6}
\end{equation}
Then clearly $J(0)=0$ and
\begin{equation}
	J_1'(a) = \frac{2}{L}\sum_{m=-\infty}^{\infty} \frac{\sin\left(\beta_m a\right)}{(\beta_m L)^2},
 \label{sa.7}
\end{equation}
so that $J_1'(0) = 0$ and then finally
\begin{equation}
	J_1''(a) = \frac{2}{L^2}\sum_{m=-\infty}^{\infty} \frac{\cos\left(\beta_m a\right)}{\beta_m L},
 \label{sa.8}
\end{equation}
which we choose to write as
\begin{equation}
	J_1''(a) = \frac{2a}{L^3}\sum_{m=-\infty}^{\infty} \frac{\cos\left(ka + \lambda m\right)}{ka + \lambda m},
 \label{sa.9}
\end{equation}
where we let $\lambda = 2\pi a/L$ so $0<\lambda<2\pi$.
Now consider the function
\begin{equation}
	F(z) = \frac{\pi \cos(ka + (\lambda-\pi)z)}{(ka + \lambda z)\sin(\pi z)}.
 \label{sa.10}
\end{equation}
Clearly $F(z)$ has poles at $z=n\in\mathbb{Z}$ and at 
$z=-ka/\lambda=-kL/2\pi$ where we have assumed $kL \neq 2\pi n$ so all 
poles are simple poles. If we let $z = R\exp^{\im\theta}$, it is clear 
that as $|z|\rightarrow\infty$ (i.e. $R\rightarrow\infty$) that we have
\begin{equation}
	|F(z)| \sim \frac{\pi\exp^{|\lambda-\pi||z|}}{\lambda|z|\exp^{\pi|z|}},
 \label{sa.11}
\end{equation}
which is clearly exponentially decaying and thus we can write
\begin{equation}
	0 = \lim_{N\rightarrow\infty}\oint_{C_N}F(z)\;\wrt z
 \label{sa.12}
\end{equation}
where $C_N$ is a circle of radius $N+\frac{1}{2}$ centred on the origin
of the complex plane. 
Computing the residues inside $C_N$ and letting $N \to \infty$
results in the identity
\begin{equation}
	\sum_{m=-\infty}^{\infty} \frac{\cos\left(ka + \lambda m\right)}{ka + \lambda m} = \frac{L}{2a}\cot\left(\frac{kL}{2}\right).
 \label{sa.13}
\end{equation}
Thus, we have established that
\begin{equation}
 J_1''(a) = \frac{1}{L^2}\cot\left(\frac{kL}{2}\right).
 \label{sa.14}
\end{equation}
Integrating twice with respect to $a$ and applying the conditions 
$J_1'(0) = J_1(0) = 0$ gives
\begin{equation}
 J_1(a) = \frac{a^2}{2L^2}\cot\left(\frac{kL}{2}\right)
 \label{sa.15}
\end{equation}
and (i) follows.

In order to calculate result (ii) start with the definition
\begin{equation}
 J_2(a) = \sum_{m=-\infty}^{\infty}\frac{\cos^2\left(\beta_m a/2\right)}{(\beta_m L)^3}
 \label{sa.16}
\end{equation}
and so we can write
\begin{equation}
 J_2(a) = \sum_{m=-\infty}^{\infty}\frac{1}{(\beta_m L)^3} - \sum_{m=-\infty}^{\infty}\frac{\sin^2\left(\beta_m a/2\right)}{(\beta_m L)^3}=  
\frac{1}{8}\cot\left(\frac{kL}{2}\right)\text{cosec}^2\left(\frac{kL}{2}\right)
 - \frac{1}{4}J_1(a),
 \label{sa.17}
\end{equation}
 using \citet{graandryz81}.
Additionally, let us define
\begin{equation}
 J_3(a) = \sum_{m=-\infty}^{\infty} \frac{\sin(\beta_m a/2)\cos(\beta_m a/2)}{(\beta_m L)^4}
 \label{sa.18}
\end{equation}
so that
\begin{equation}
 J_3'(a) = \frac{1}{2L}\sum_{m=-\infty}^{\infty}\frac{1}{(\beta_m L)^3} - \frac{1}{4L} J_1(a) = \frac{1}{16L}\cot\left(\frac{kL}{2}\right)\left(\text{cosec}^2\left(\frac{kL}{2}\right)-\frac{2a^2}{L^2}\right).
 \label{sa.19}
\end{equation}
Integrating with respect to $a$ and applying $J_3(a) = 0$ gives
\begin{equation}
 J_3(a) = \frac{a}{16L}\cot\left(\frac{kL}{2}\right)\left(\text{cosec}^2\left(\frac{kL}{2}\right)-\frac{2}{3}\frac{a^2}{L^2}\right).
 \label{sa.20}
\end{equation}
Finally, letting
\begin{equation}
	J_4(a) = \sum_{m=-\infty}^{\infty}\frac{\sin^2\left(\beta_m a/2\right)}{(\beta_m L)^5}
 \label{sa.21}
\end{equation}
we see that $J_4'(a) = J_3(a)/L$ and $J_4(a) = 0$ so that we can deduce
\begin{equation}
 J_4(a) = \frac{a^2}{32L}\cot\left(\frac{kL}{2}\right)\left(\text{cosec}^2\left(\frac{kL}{2}\right)-\frac{1}{3}\frac{a^2}{L^2}\right).
 \label{sa.22}
\end{equation}
From the definition (\ref{sa.4}) we find that
\begin{equation}
	C^{(p,p)} = \sum_{m=-\infty}^{\infty} \frac{\left|{g_m}^{(p)}\right|^2}{\beta_m L} = \frac{aL^2}{\ell^2}J_2(a) - \frac{4L^3}{\ell^2}J_3(a) + \frac{4L^4}{\ell^2a}J_4(a)
 \label{sa.23}
	\end{equation}
and (iii) follows.

Moving on,  we define
\begin{equation}
 J_5(a) = \sum_{m=-\infty}^{\infty}\frac{\cos(\beta_ma)}{\left(\beta_mL\right)^4} 
 \label{sa.24}
\end{equation}
so that $J_5(0) = 0$ and
\begin{equation}
	J_5'(a) = \frac{1}{L}\sum_{m=-\infty}^{\infty}\frac{\cos(\beta_ma)}{\left(\beta_mL\right)^2}
 \label{sa.25}
\end{equation}
so that $J_5'(0) = (1/4L)\text{cosec}^2(kL/2)$. Finally,
\begin{equation}
	J_5''(a) = \frac{1}{L^2}\sum_{m=-\infty}^{\infty}\frac{\sin(\beta_ma)}{\beta_mL}
 \label{sa.26}
\end{equation}
or
\begin{equation}
	J_5''(a) = -\frac{a}{L^3}\sum_{m=-\infty}^{\infty} \frac{\sin\left(ka + \lambda m\right)}{ka + \lambda m}
 \label{sa.27}
\end{equation}
where we have let $\lambda = 2\pi a/L$ so $0<\lambda<2\pi$.
Similarly to before we define
\begin{equation}
	G(z) = \frac{\pi \sin(ka + (\lambda-\pi)z)}{(ka + \lambda z)\sin(\pi z)}
 \label{sa.28}
\end{equation}
so that $G(z)$ has poles at $z=n\in\mathbb{Z}$ and at $z=-ka/\lambda=-kL/2\pi$ and we assume $kL \neq 2\pi n$ so all poles are simple poles. Again,
it can be shown that $|G(z)| \to 0$ as $|z| \to \infty$
and so
\begin{equation}
	0 = \lim_{N\rightarrow\infty}\oint_{C_N}G(z)\;\wrt z
 \label{sa.29}
\end{equation}
where $C_N$ is a circle of radius $N+\frac{1}{2}$.
Calculating the residues from the poles inside $C_N$ and letting
$N \to \infty$ results in
\begin{equation}
	\sum_{m=-\infty}^{\infty} \frac{\sin\left(ka + \lambda m\right)}{ka + \lambda m} = \frac{\pi}{\lambda} = \frac{L}{2a}.
 \label{sa.30}
\end{equation}
and hence
\begin{equation}
	J_5''(a) = -\frac{1}{2L^2}.
 \label{sa.31}
\end{equation}
Integrating twice with respect to $a$ and applying our the initial
conditions gives
\begin{equation}
	J_5(a) = -\frac{a}{4L}\left(\frac{a}{L} + \text{cosec}^2\left(\frac{kL}{2}\right)\right).
 \label{sa.32}
\end{equation}
Finally consider
\begin{equation}
	J_6(a) = \sum_{m=-\infty}^{\infty}\frac{\sin(\beta_ma)}{\left(\beta_mL\right)^3} 
 \label{sa.33}
\end{equation}
from which it follows that
\begin{equation}
	J_6'(a) = -\frac{1}{L}\sum_{m=-\infty}^{\infty}\frac{\sin(\beta_ma)}{\beta_mL} = -\frac{1}{L}R(a) = \frac{a}{4L^2}\left(\frac{a}{L} + \text{cosec}^2\left(\frac{kL}{2}\right)\right)
 \label{sa.34}
\end{equation}
and so
\begin{equation}
	J_6(a) = \frac{a^2}{24L^2}\left(\frac{2a}{L} + 3 \;\text{cosec}^2\left(\frac{kL}{2}\right)\right) + J_6(0),
 \label{sa.35}
\end{equation}
\begin{equation}
	C^{(h,p)} = \frac{\im L^2}{da}\left(P_5(a) + 2\frac{L}{a}\left(P_6(a)-P_6(0)\right)\right).
 \label{sa.36}
\end{equation}
and the result in (iii) follows.

\refstepcounter{section}\setcounter{equation}{0}
\section*{Appendix \Alph{section}: Another infinite series}
\addcontentsline{toc}{section}{Appendix \Alph{section}: Another infinite series}

We start with
\begin{equation}
	\sum_{m=-\infty}^{\infty}\frac{\exp^{\im \beta_m (x-x')}}{\beta_m L} = \sum_{m=-\infty}^{\infty} \frac{\cos(kX + \hat{\lambda}m)}{kX + \hat{\lambda}m} + \im\sum_{m=-\infty}^{\infty} \frac{\sin(kX + \hat{\lambda}m)}{kX + \hat{\lambda}m},
 \label{sb.1}
\end{equation}
where we have let $X=(x-x')>0$ (without loss of generality)
 and $\hat{\lambda} = 2\pi X/L$ such that $0 < X \leq \ell$ and $0 < \hat{\lambda} < 2\pi$. Using (\ref{sa.13}) and (\ref{sa.30}) from Appendix A we have
\begin{equation}
	\sum_{m=-\infty}^{\infty}\frac{\exp^{\im \beta_m X}}{\beta_m L} = \frac{1}{2}\left(\cot\left(\frac{kL}{2}\right) + \im\right).
 \label{sb.2}
\end{equation}
Conjugating this result is the same as reversing the sign of $X$ and if $X=0$ the series is real. Then the general result is
\begin{equation}
		\sum_{m=-\infty}^{\infty}\frac{\exp^{\im \beta_m (x-x')}}{\beta_m L} = \frac{1}{2}\cot\left(\frac{kL}{2}\right) + \frac{\im}{2}\;\text{sgn}(x-x').
 \label{sb.3}
\end{equation}
where $\text{sgn}(X) = -1,0,1$ when $X < 0, X= 0, X> 0$ respectively.
